\documentclass[useAMS,usenatbib,usegraphicx]{mn2e}
\usepackage{url,makeidx,xcolor,amsmath,amssymb,times}
\usepackage{enumerate}
\graphicspath{{imgpdf/}}
\DeclareGraphicsExtensions{.pdf}

\newcommand\vect[1]{\ensuremath{\boldsymbol{#1}}}
\newcommand\electron{\ensuremath{\mathrm{e}^{-}}}
\newcommand\cemga{{\sc CEMGA}}

\newcommand\gaia{\textit{Gaia}}
\newcommand\hst{\textit{HST}}
\newcommand\chandra{\textit{Chandra}}

\title[Image location in the presence of radiation damage.]{The impact of CCD radiation damage on {\gaia} astrometry:\\I. Image location estimation in the presence of radiation damage}

\author[T. Prod'homme, B. Holl, L. Lindegren,  A.G.A. Brown]{T. Prod'homme$^{1}$\thanks{E-mail:
prodhomme@strw.leidenuniv.nl}, B. Holl$^{2}$, L. Lindegren$^{2}$, and A.G.A. Brown$^{1}$\\
$^{1}$Leiden Observatory, Leiden University, P.O.\ Box 9513, 2300 RA, Leiden, The Netherlands\\
$^{2}$Lund Observatory, Lund University, Box 43, 22100 Lund, Sweden}
\begin{document}

\date{Accepted 2011 October 4. Received 2011 April 6; in original form 2011 April 6}

\pagerange{\pageref{firstpage}--\pageref{lastpage}} \pubyear{2011}

\maketitle
\label{firstpage}
\begin{abstract}
The {\gaia} mission has been designed to perform absolute astrometric measurements with unprecedented accuracy; the end-of-mission parallax standard error is required to be of the order of 10 micro-arcseconds for the brightest stars ($V \le 10$) and 30 micro-arcseconds for a G2V type star of magnitude~15. These requirements set a stringent constraint on the accuracy of the estimation of the location of the stellar image on the CCD for each observation: e.g., 0.3~milli-arseconds (mas) or 0.005~pixels for the same $V=15$ G2V star. However the {\gaia} CCDs will suffer from charge transfer inefficiency (CTI) caused by radiation damage that will degrade the stellar image quality and may degrade the astrometric performance of {\gaia} if not properly addressed. For the first time at this level of detail, the potential impact of radiation damage on the performance of {\gaia} is investigated. In this first paper we focus on the evaluation of the  CTI impact on the image location accuracy using a large set of CTI-free and damaged synthetic {\gaia} observations supported by experimental test results. We show that CTI decreases the stellar image signal-to-noise ratio and irreversibly degrades the image location estimation precision. As a consequence the location estimation standard errors increase by up to 6\% in the {\gaia} operating conditions for a radiation damage level equivalent to the end-of-mission accumulated dose. We confirm that in addition the CTI-induced image distortion introduces a systematic bias in the image location estimation (up to 0.05~pixels or 3~mas in the {\gaia} operating conditions). Hence a CTI mitigation procedure is critical to achieve the {\gaia} requirements. We present a novel approach to CTI mitigation that enables, without correction of the raw data, the unbiased estimation of the image location and flux from damaged observations. We show that its current implementation reduces the maximum measured location bias for the faintest magnitude to 0.005~pixels ($\sim$4$\times $10$^{-4}$~pixels at magnitude 15) and that the {\gaia} image location estimation accuracy is preserved.
In a second paper we will investigate how the CTI effects and CTI mitigation scheme affect the final astrometric accuracy of {\gaia} by propagating the residual errors through the astrometric solution.
\end{abstract}
\begin{keywords}
instrumentation: detectors -- space vehicles -- astrometry -- methods: numerical -- methods: analytical -- methods: statistical
\end{keywords}

\section{Introduction}\label{sect:introduction}

{\gaia} is a European Space Agency mission that aims to create the most complete and accurate stereoscopic map to date of the Milky Way, containing parallaxes, proper motions, radial velocities, and astrophysical parameters for one billion stars, one percent of the estimated stellar population in our galaxy \citep{Gaia2001,Gaia2008}.
Due to the satellite's constant spinning motion, the determination of the astrometric parameters ultimately comes down to measuring very precisely the time $t_{\mathrm{obs}}$ at which a particular star crosses a fiducial line on the focal plane \citep{lindegren2011, bastian2005}. The required astrometric precision is extreme, e.g., the end of mission parallax uncertainty for a star of magnitude $V=15$ is required to be better than 25 micro-arcseconds ($\mu$as)\footnote{A list of acronyms is provided in Table~\ref{t:acronyms} at the end of the paper.}. In order to determine $t_{\mathrm{obs}}$, one needs to measure the image location on the Charge-Coupled Device (CCD) relative to the instrument axes.
As a consequence the required astrometric accuracy sets a direct and stringent requirement on the residual image location uncertainty per CCD star transit. In the left part of Table~\ref{tab:requirement} we detail the end-of-mission parallax standard error, $\sigma_\varpi$, as function of stellar magnitude and type\footnote{Updated estimates of the science performance are given on:
\url{http://www.rssd.esa.int/index.php?project=GAIA&page=Science_Performance}} computed using \cite{JdB055}. These predicted standard errors do include the increased photon noise due to the  radiation damage induced charge loss, but not the residual bias-calibration errors considered in the present paper (except for a general contingency margin of 20\%). For this paper we are interested in the mean image location uncertainty per CCD star transit $\sigma_{\kappa}$ that would be needed to reach a given targeted parallax accuracy.
Based on \cite{JdB2005Acc} we estimate the corresponding `requirement' on the image location uncertainty shown in the right part of Table~\ref{tab:requirement} and computed as: $\sigma_{\kappa} = \sqrt{N_{\mathrm{obs}}} \; \sigma_{\varpi} / (m \, g_{\varpi} )$, with $N_{\mathrm{obs}}$ the average number of astrometric observations per star  (662), $m$ the end-of-mission scientific contingency margin which is 1.2, and $g_{\varpi}$ the geometrical parallax factor which is 2.08 for the {\gaia} solar aspect angle $\xi=45^{\circ}$.
This formula has also been used  to compute the `requirement' curve as function of {\gaia} $G$-band\footnote{The {\gaia} G-band magnitude is a broad-band, white-light magnitude in the wavelength range 300 -- 1000 nm defined by the telescope transmission and CCD quantum efficiency. $G=V$ for an un-reddened A0V star \citep{jordi2010, perryman2001}.} magnitude shown in several figures throughout this paper.
No spectral type distinction is needed when  these uncertainties are expressed in $G$ because they are virtually independent of spectral type. Note that the computed location uncertainties do not contain the 20\% contingency margin, making them very stringent.

During the 5 year mission life time, solar wind protons will collide with {\gaia}'s focal plane and create electron traps in the CCDs by displacement damage. These radiation-induced traps drastically increase the CCD charge transfer inefficiency (CTI) and will lead to a significant loss of signal for all {\gaia} measurements by stochastically capturing and releasing signal electrons. The resulting electron redistribution will also distort each stellar image. The CTI effects are expected to significantly contribute to the error budget of all the {\gaia} measurements (astrometric, photometric, and spectroscopic), especially if not properly taken into account in the data processing.

We present here the first part of a detailed evaluation of the impact of the radiation damage effects on the final accuracy of the {\gaia} astrometric measurements. This paper focuses on the effect of CTI on the image location accuracy. Studying the accuracy of a measurement is a rather complex enterprise. Hence we present in the following section the overall applied methodology and the different steps of this study. In a second paper, we will investigate how the final {\gaia} astrometric accuracy is affected by the CTI-induced errors at the image processing level as characterized in this study.

\begin{table}
\centering
\begin{tabular}{l|r|r|r|c|r|r|r|}
\hline
 & \multicolumn{3}{|c|}{Parallax accuracy target} &  \multicolumn{4}{|c|}{Corresponding CCD image } \\
 & \multicolumn{3}{|c|}{standard error $\sigma_{\varpi}$ [$\mu$as] } & \multicolumn{4}{|c|}{location uncertainty $\sigma_{\kappa}$ [mas]} \\
\hline
	Type & B1V & G2V & M6V & & B1V & G2V & M6V \\
	$V\!-\!G$ & $0.03$ & $0.16$ & $2.18$ & & $0.03$ & $0.16$ & $2.18$ \\
\hline
 	$V\!=\!10$		&6.9		                 &7.0	                  &7.3                       && 0.072                   & 0.073                     & 0.076 \\
 	$V\!=\!15$		&25\phantom{0}	 &24\phantom{0}	  &10\phantom{0}	&& 0.26\phantom{0} & 0.25\phantom{0}    & 0.11\phantom{0} \\
	$V\!=\!20$		&322\phantom{0} &300\phantom{0} &102\phantom{0}	&& 3.3\phantom{0}\phantom{0}  & 3.1\phantom{0}\phantom{0}   & 1.1\phantom{0}\phantom{0} \\
\hline
\end{tabular}
\caption{On the left side we tabulate the {\gaia} target performance of the sky averaged end-of-mission parallax standard error, $\sigma_{\varpi}$ (in $\mu$as), as function of spectral type and Johnson $V$-band magnitude \citep[see][]{JdB055}. $V\!-\!G$ allows for the conversion from $V$ to the {\gaia } $G$-band magnitude. 
On the right we give the corresponding mean CCD image location uncertainty, $\sigma_{\kappa}$ (in mas), that would result in the parallax standard error on the left when observed 662 times (the average number of astrometric observations per star). See Section~\ref{sect:introduction} for more details.}
\label{tab:requirement}
\end{table}

\section{Overall methodology\label{methodology}}

\paragraph*{The use of synthetic data}
To evaluate the impact of the CTI effects on the image location accuracy, we apply the {\gaia} image parameter estimation procedure (Section \ref{sect:imageParam}) to a large data set of simulated CTI-free and CTI-affected observations (from here on the latter are referred to as `damaged observations'). The use of synthetic data presents several fundamental advantages compared to the use of experimental data: while in experimental studies, the true image parameters, the instrument model or Point Spread Function (PSF), and the different noise contributions need to be estimated, in the simulation, these are known parameters. Hence the uncertainties related to the estimation of such parameters cannot bias the result of our study. Furthermore only simulation can allow the determination of the absolute image location bias and the associated standard errors as this requires the knowledge of the true image location. Finally, by using synthetic data one can compute the statistical uncertainties on the measured image location bias and standard errors, and this at any precision level just by increasing the number of simulated observations for a particular set of conditions. In Section \ref{sect:observations}, we detail the simulation of {\gaia}-like observations in the absence and presence of radiation damage for different stellar magnitudes, image widths, background and readout noise levels.

The bias allows for the quantification of the trueness of an estimation, and the standard errors for the quantification of the estimator precision. If an estimator delivers bias-free estimations, then its standard errors can also be regarded as a means to quantify the accuracy of this estimation. In the following we will thus make the important distinction between precision and accuracy when it is justified.

\paragraph*{The {\gaia} image parameter estimation procedure}
In the {\gaia} data processing, the image location and flux are estimated or `self-calibrated' through the use of an iterative procedure, that allows for the successive determination and improvement of the PSF and the image parameters without prior knowledge. A detailed description of this procedure is provided in Section \ref{sect:imageParam}. In Sections \ref{sect:locBiasUndamaged} and \ref{sect:locAccUndamaged}, the procedure is applied to the data set of simulated CTI-free observations, in order to verify that, in the absence of CTI, the {\gaia} image parameter estimation procedure performs efficiently, according to expectations.

\paragraph*{The Cram\'{e}r-Rao bound}
Assessing the efficiency of the {\gaia} image parameter estimation procedure necessitates the computation of the theoretical limit to the image location accuracy in the {\gaia} observing conditions. This theoretical limit corresponds to the ultimate accuracy achievable by any bias-free estimator. It is set by the Cram\'{e}r-Rao bound, described in Section~\ref{sect:cramerrao}. We thus compute the Cram\'{e}r-Rao bound as a function of magnitude ($G$), image width, background, and readout noise level (Section~\ref{sect:locAccUndamaged}) and subsequently use it as a reference in a comparison with the standard errors of the estimated image parameters.
The Cram\'{e}r-Rao bound is also required to assess the impact of the CTI effects independently from any estimation procedure.

\paragraph*{The radiation damage impact on the image location estimation}
In Section \ref{sect:damagedCR}, we use the set of damaged observations to demonstrate that the image distortion and the charge loss induced by CTI imply an irrevocable loss of accuracy in the image location determination. This loss of accuracy, which directly affects the performance budget of {\gaia}, is independent from any estimation method and can only be avoided by physically preventing charge trapping. This is done by optimizing the hardware (e.g., the CCD operating temperature) and using hardware countermeasures such as the periodic injection of artificial charges, or the use of a supplementary buried channel, an extra doping implant in each pixel that confines small charge packets.
Taking into account these countermeasures in the simulation of the damaged observations allows us to verify that they indeed substantially contribute to diminish the CTI effects on the theoretical image location accuracy. However, Sections~\ref{sect:bias} and \ref{sect:damagedLocStdErr} show that it is not possible to rely solely on these countermeasures: indeed we find that if the CTI effects are not properly taken into account in the image parameter determination the image location bias can be as large as $10$ mas for a star of magnitude 15 (to be compared to the requirement of $\sim$0.08 mas, see Table~\ref{tab:requirement}). This is in agreement with the experimental tests performed on {\gaia} irradiated CCDs (see Section \ref{sect:comparison}) and confirms that the CTI effects need to be addressed by the {\gaia} data processing in order to achieve the mission requirements.

\paragraph*{Mitigating the CTI effects}

The software mitigation of CTI effects is a complicated task. Several schemes have been discussed in the literature; they usually imply the direct correction of the raw data in the context of photometry-based measurements. In Section \ref{sect:approaches}, we review the different potential schemes, and where they intervene in the data processing chain. Then, in Section \ref{sect:fwd}, we present and motivate a novel approach to CTI mitigation that does not involve a direct correction, so that the noise properties of the raw observations remain unchanged. This approach, developed by the {\gaia} Data Processing and Analysis Consortium \citep[DPAC,][]{mignard2008}, relies on the forward modelling of each observation including the CTI distortion, such that the true image parameters can be directly estimated from the damaged observations. The modelling of the distortion of the stellar image is performed thanks to a fast analytical CTI effects model, a so-called Charge Distortion Model (CDM). The success of this CTI mitigation approach depends on the performance of such a model. In Section \ref{sect:idealFwdTest}, the potential accuracy of this approach is assessed in the case of an ideally calibrated CDM. Finally, using the current best CDM candidate \citep[][and Section \ref{sect:cdm}]{short2010, prodhomme2010a}, we apply the image parameter estimation procedure and the DPAC CTI mitigation scheme to the set of damaged observations and show that one can recover the CTI-induced image location and flux bias (see Sections \ref{sect:locBiasRecovery}, \ref{sect:fluxBiasRecovery}). Only then are we able to answer the question: does the current {\gaia} image location procedure combined with the presented CTI mitigation scheme allow an unbiased estimation of the image location with a sufficient accuracy in the presence of radiation damage?

\paragraph*{Methodology summary}
\begin{enumerate}
\setcounter{enumi}{0}
\setlength{\itemsep}{3pt}
\item Generation of CTI-free and damaged {\gaia}-like observations.  
\item Determination of the theoretical limit to the image location accuracy in the absence of CTI by computing the Cram\'{e}r-Rao bound.
\item Performance assessment of the {\gaia} image parameter estimation procedure in the absence of CTI.
\item Evaluation of the intrinsic loss of accuracy in the image location estimation induced by radiation damage by computing the Cram\'{e}r-Rao bound for a damaged LSF.
\item Characterization of the CTI effects on the image parameter estimation procedure.
\item Performance assessment of the {\gaia} image parameter estimation procedure in the presence of CTI and including a forward modelling approach to mitigate the CTI effects.
\end{enumerate}

\section{Generating \textit{GAIA}-like observations}\label{sect:observations}

In the following, we first describe the main principles of the {\gaia} observations, and explain how we generate {\gaia}-like reference images. These images are used to simulate thousands of observations for different stellar magnitudes and operating conditions, and thus constitute the basis of our study. 
Then we detail how we simulate the stellar transits over a CCD. To achieve a high level of realism, we use a physically motivated Monte Carlo model that simulates the CCD charge collection and transfer, as well as the trapping processes, at the pixel-electrode level \citep{prodhomme2011}.
Finally we summarize the expected CTI effects on the stellar images in the {\gaia} operating conditions and explain the choices we made regarding the radiation damage parameters of the simulation (trap species, level of radiation).

\subsection{How {\gaia} observes}\label{sect:gaiaObs}

The {\gaia} spacecraft will orbit around the second Lagrange point (L2) and constantly spin around its own axis such that its two telescopes scan a great circle on the sky several times a day.  The precession of the spin axis changes the orientation of the consecutive great circles, allowing for the coverage of the whole sky in about six months. The measurements are recorded in a single focal plane consisting of 106 CCDs. Due to the satellite spinning motion, the star projections will not remain stationary during an observation but will transit the focal plane in the along-scan (AL) direction.
The orthogonal direction is called across-scan (AC). To integrate the stellar images along the star transits, the CCDs are operated in time-delayed integration (TDI) mode. In this mode the CCD is constantly readout and the satellite
scanning rate (and induced light source motion) has been synchronized with the charge transfer
period, so that the charge profile continues to build up as the image travels across the CCD
avoiding as much as possible image smearing. The charge transfer period is $0.9892$ ms and the
integration time $4.4$ s.
The observing principle of {\gaia} relies on differential positional measurements among the stars simultaneously visible in the two superposed fields of view. In particular the differential measurements \emph{between} the two fields of view (covering arcs of about $106.5^\circ$ on the sky) are essential for the construction of a global reference frame and the determination of absolute parallaxes. For these measurements the AC component of the differential positions is largely degenerate with the instrument pointing, and mainly the AL component matters \citep{lindegren2011}. Gaia is therefore primarily optimized for AL measurements and the image location accuracy in that direction is the most critical one for the performance.

Because of limitations on the amount of data that can be sent to ground only a small window around each source is read out. 
For sources fainter than magnitude 13 the windowing scheme is simple. In the AL direction, the window size is 12 pixels for stars brighter than magnitude 16, and 6 pixels for fainter stars. In the AC direction the size of the on board readout windows is 12 pixels (the pixel size is 58.9 mas AL and 176.8 mas AC). The observations are then binned across-scan before being sent to the ground. Hence, for a particular star, a {\gaia} observation results in a one-dimensional along-scan set of electron counts that corresponds to the sampling of a one-dimensional point spread function or Line Spread Function (LSF). 
Note that {\gaia} will observe sources as bright as $G=5.7$, but due to the relatively low number of stars between magnitude 5.7 and 13 ($\sim$1\% of the expected $10^9$ sources) and because of the use of a complicated gating scheme (to avoid pixel saturation) these magnitudes are ignored in this study.

\subsection{Construcing a {\gaia}-like reference image}\label{sect:imageRefModel}
\begin{figure}
\centering
\includegraphics[width=0.49\textwidth]{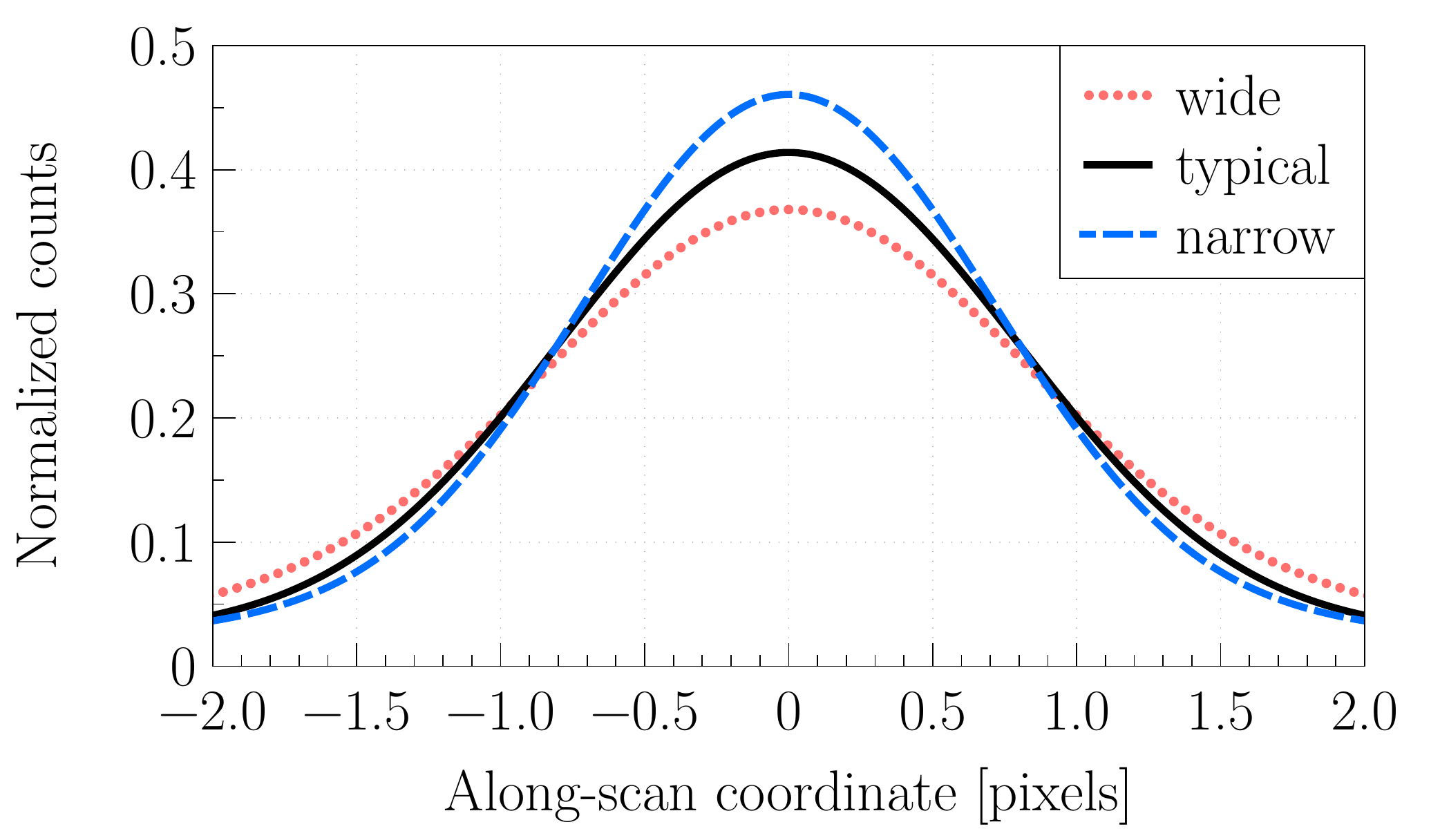}
\includegraphics[width=0.49\textwidth]{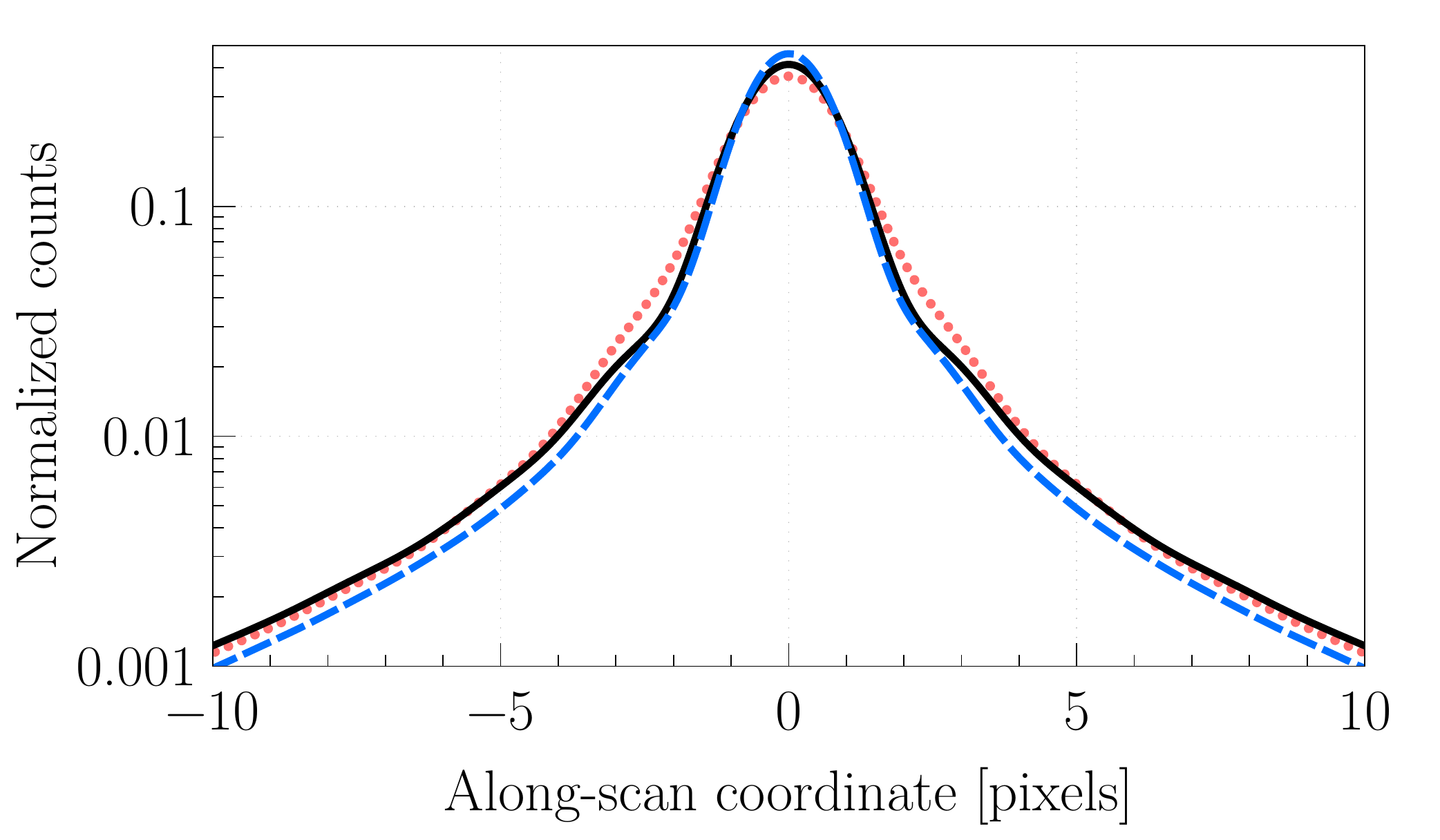}
\caption{One-dimensional {\gaia}-like reference images generated from a symmetrized subset of realistic {\gaia}-like line spread functions (LSFs).
These reference images are used to reflect the range of narrow, typical, and wide profiles that results from different stellar types and wave front errors.
The bottom figure shows the same profiles as the top figure for a wider AL coordinate range and with a logarithmic ordinate scale.}
\label{fig:refLsf}
\centering
\end{figure}
In this study we only analyze observations of point sources because these will be used in the Astrometric Global Iterative Solution \citep[AGIS,][]{lindegrenAgis2011}
to calibrate the instrument and satellite attitude, which are then used to estimate the astrometric parameters for all sources. The PSF, the actual two-dimensional flux distribution of an unresolved star that illuminates a CCD, depends on the spectral energy distribution of the star, the CCD properties, and the optics and its associated wavefront errors. Because the spectral energy distribution depends on the type of star, and the wavefront errors depend on the position in the focal plane, we cannot construct a single PSF that would be representative of the whole range of possible profile shapes.

To construct a set of PSFs that samples the range of possible profile shapes we make use of a study by \cite{lindegren2009LL084} in which a set of 20,000 one-dimensional line spread functions (LSFs) were generated that are representative of the {\gaia} optics and wave front errors, stellar spectral energy distributions, and CCD effects (e.g., smearing due to TDI). A set of basis functions was extracted from this dataset using a Principle Component Analysis (PCA) in order to describe the data with a minimum number of parameters. 
A set of ten basis functions was found to be sufficient to represent any of the LSFs with an RMS error of $10^{-4}$. 
To be readily usable the basis functions were subsequently fitted by a special quartic spline\footnote{The special quartic spline used to represent the LSF can be described as the convolution of an ordinary cubic spline, defined on a regular knot sequence with a knot separation of half a pixel, with a rectangular function of width equal to one pixel. This spline has the property that the sum of points sampled at one pixel separation is independent of the sub-pixel phase of the sampled points. Any `effective' LSF, being the result of an optical LSF convolved with the pixel response function \citep{anderson+king2000}, should have this property. Details of the special quartic spline are found in a technical note by \cite{lindegren2003LL046}.}
 that is flexible enough to fit the data while being smooth enough to avoid overfitting that could result in sub-pixel position bias when the function is used for location estimation on pixel sampled data.

For our study we decided to use three reference images. First of all we construct a reference profile based on a selection of all profiles that have a FWHM within $\pm$ 1\% of the mean FWHM of 1.958 pixels. These profiles are subsequently symmetrized (i.e. $L_{symm}=(L(x)+L(-x))/2$), averaged, and then fitted with the first four even basis functions, resulting in a symmetric mean profile with FWHM 1.957 pixels, from hereon referred to as the `typical' reference profile. Four components are used to get a profile that is sufficiently close to the target mean FWHM. To represent the extremes we introduce a `narrow' and a `wide' reference profile. The narrow and wide profiles are constructed in exactly the same way as the typical reference profile, only differing in the selection of LSF samples, which are: 90 $\pm$ 1\% and 110 $\pm$ 1\% of the mean FWHM respectively. This results in a FWHM of 1.767 pixels for the narrow, and 2.161 pixels for the wide reference profile. All three reference images are normalized as shown in Fig.~\ref{fig:refLsf}.

Although the two-dimensional PSF in the {\gaia} focal plane is wider in the across-scan direction than in the along-scan direction, the pixels are shaped such that in pixel units the PSF is nearly identical in both directions. Because we are only interested in how the pixels are illuminated we can therefore construct the two-dimensional reference image from simply multiplying the one-dimensional reference image $L(x)$ in two dimensions:
\begin{equation}\label{eq:2dRefModel}
P(x,y) = L\left( x - \kappa\right) \times L\left( y - \mu\right)
\end{equation}
In all our further analyses we assume $\mu=0$. Because we defined the zero-point of the symmetric profile $L$ to be at the symmetry point, this means that the PSF is always in the centre of the window in the across-scan direction.

\subsection{Monte-Carlo simulations of observations}\label{sect:mcModel}

The approach we have chosen for this study is to simulate a fully synthetic dataset using a detailed physical simulation of the photo-electron collection and transfer in CCDs at the pixel-electrode level \citep{prodhomme2011}, available through the {\cemga} software package\footnote{\url{www.strw.leidenuniv.nl/~prodhomme/cemga.php}} \citep{prodhomme2010b}. The model also allows for a detailed treatment of radiation induced traps that capture and release electrons and thereby distort the charge profile transferred through the CCD (see Section~\ref{sect:damageSim}).
The observations are simulated in two dimensions: 4494 pixels in along-scan and 12 pixels in the across-scan direction. In the software we illuminate the CCD with a two-dimensional reference image $P\left( x,y\right)$ described in Section~\ref{sect:imageRefModel}. The normalized reference image is scaled to produce an illumination that corresponds to a particular stellar magnitude.
The photon detection is modelled as a Poisson process: at each transfer step, the photo-electrons are generated using a random generator with a Poisson distribution and a mean equal to the expected number of collected photons (given by the reference image) within the integration time ($\sim$1ms $\times$ 4494 pixels in the case of {\gaia}).
Note that we control the exact along-scan location of the reference image, therefore allowing us to determine the exact error when a location estimate from the observation has been made. In the simulations we can optionally include a constant background. The electron packet transfer in the readout register is not simulated.

The raw two-dimensional observation counts are cropped in along-scan direction to $80$ pixels (centred around the signal) and the resulting $80\times 12$ pixels are stored. All used reference images are zero for $|x|>20$, therefore any relevant signal is always contained in the cropped raw observation data.

When processing an observation we load the raw two-dimensional pixel counts. Depending on the windowing scheme for this particular magnitude and CCD we crop the data around the signal to the relevant window size and optionally bin the pixel counts in the across-scan direction resulting in a one-dimensional sample of the transit photo-electron counts $\{N_k\}$. 
Readout noise can be added to the counts using a normally distributed random generator $\mathrm{Normal}(0,r^2)$: having zero mean and standard deviation $r$ (the readout noise value). 

Our total synthetic observational dataset consists of:
\begin{enumerate}
\setcounter{enumi}{0}
\setlength{\itemsep}{3pt}
\item 
3 different CCD states: CTI-free, damaged with 1~trap~pixel$^{-1}$ and damaged with 4~traps pixel$^{-1}$ (see Section~\ref{sect:damageSim} for details about the damaged cases),
\item
3 different reference images: narrow, typical and wide (see Section~\ref{sect:imageRefModel}),
\item 2 different levels of sky background: 0 and 0.44698 \electron pixel$^{-1}$s$^{-1}$ (the latter corresponding to the average sky surface brightness),
\item
9 different magnitudes: G= 13.3, 14.15, 15.0, 15.875, 16.75, 17.625, 18.5, 19.25, 20.0,
\item the (two-dimensional) photo-electron counts of 250 CCD transits, each with the reference image incrementially shifted by $1/250^{\mathrm{th}}$ of a pixel in the along-scan direction.
\end{enumerate}
In almost all of the processing we select a unique combination of (i), (ii), (iii) and magnitude (iv), containing all 250 transits. The selection of all across-scan binned transits for a given magnitude is denoted as: $\{\{N_k\}\}_{G}$.

\subsection{Simulation of the CTI effects}\label{sect:damageSim}
\begin{figure*}
\includegraphics[width=0.49\textwidth]{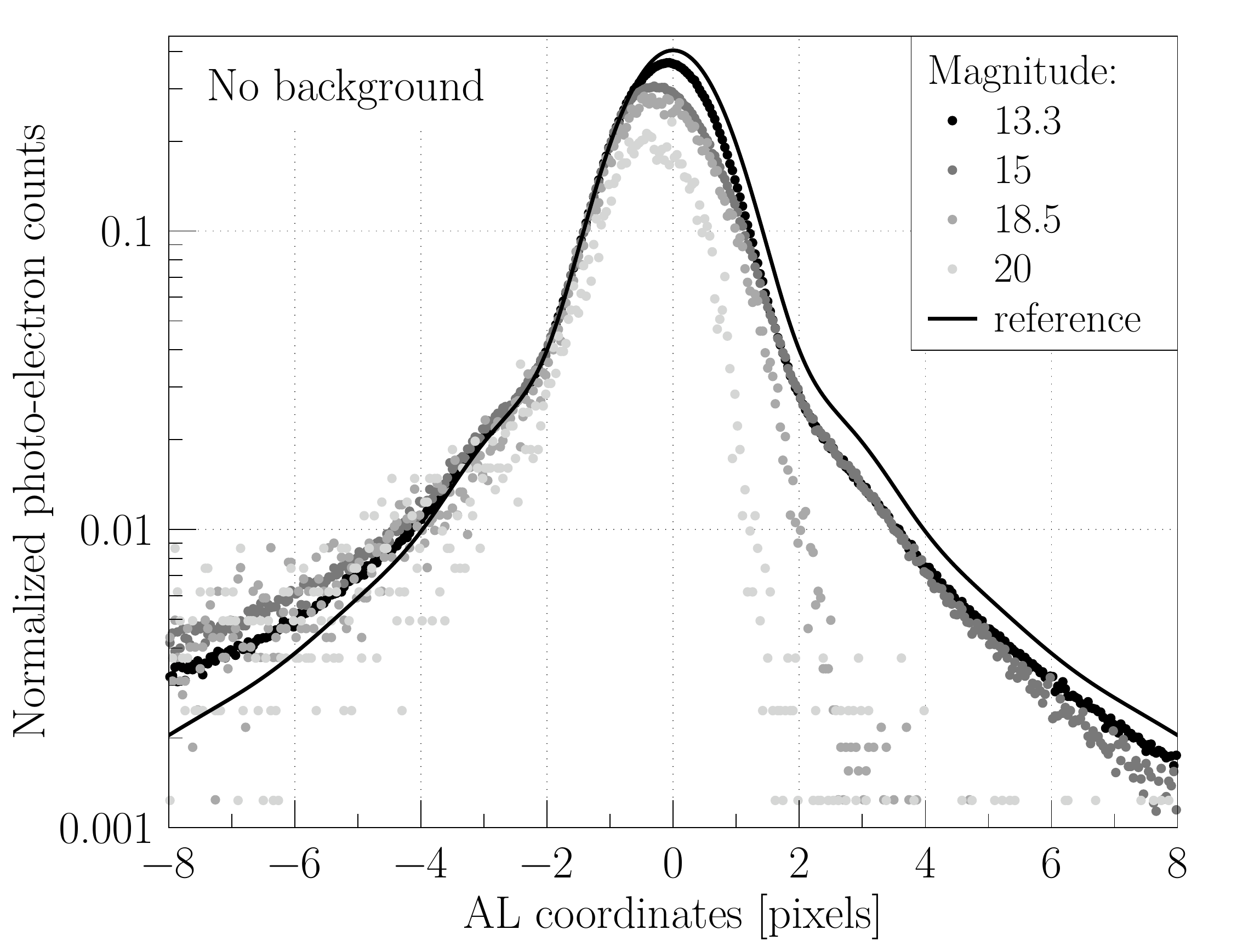}
\includegraphics[width=0.49\textwidth]{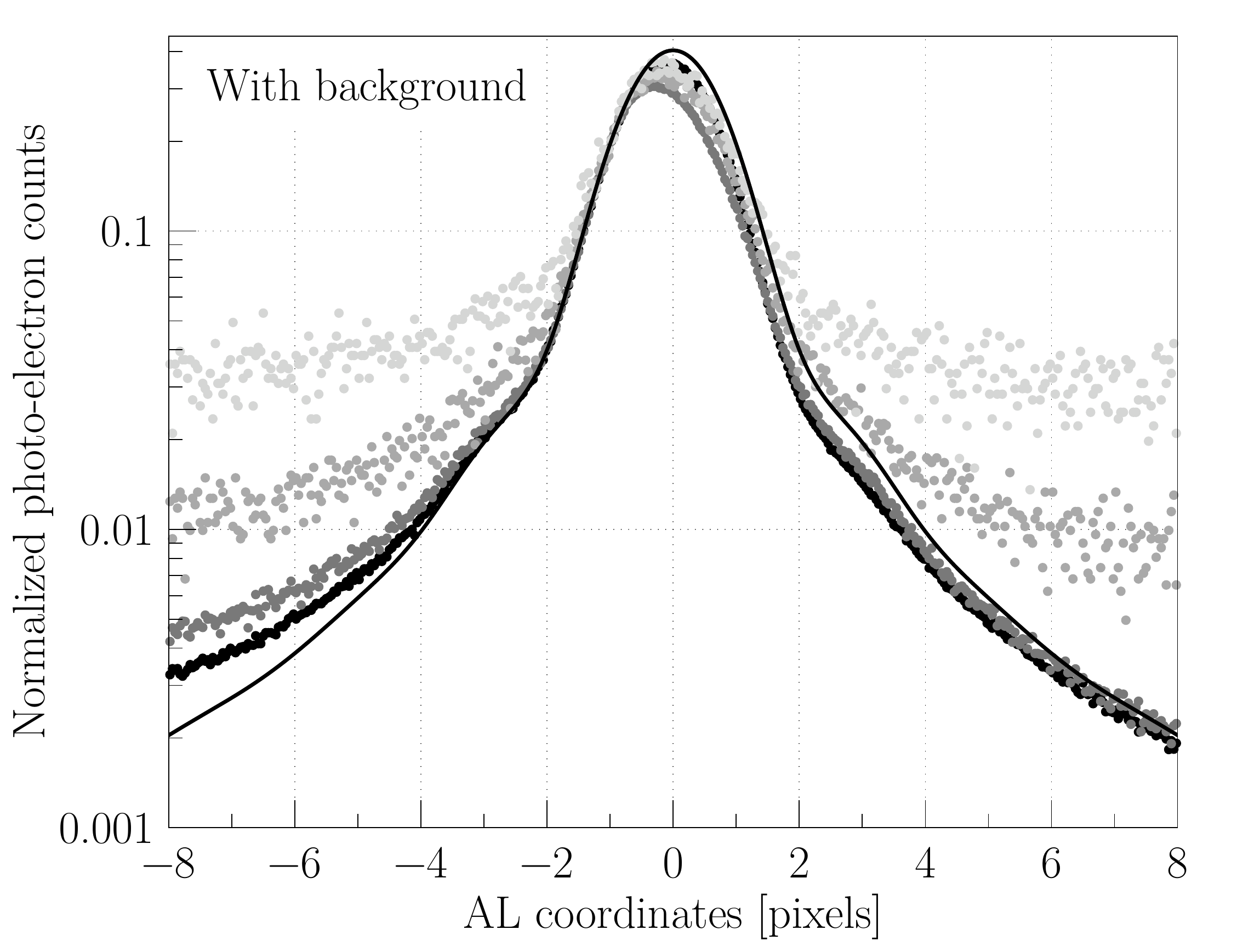}
\caption{The two plots show the simulated damaged observation resulting from multiple transits over an irradiated CCD containing 4~traps pixel$^{-1}$ at different magnitudes with no background (left) and a background set to the sky brightness (right). In both cases the typical reference image is used and no readout noise has been added. The transfer direction is from left to right. The ordinate scale corresponds to the photo-electron counts normalized by the total expected CTI-free flux at each magnitude. Only 25 observations per magnitude (out of 250) are shown. The reference image (line) enables us to appreciate the variation of amplitude in the CTI induced distortion and charge loss as a function of the signal level. Also the overall profile centroid shift is clearly visible. Note that, as one can observe from the left plot, a radiation dose leading to 4~traps pixel$^{-1}$ translates into a severe distortion of the {\gaia} measurements, particularly for fainter stars. The expected end-of-mission dose, based on the latest prediction of the next solar cycle, is however three quarters of the one we chose to simulate. Moreover the right plot shows that even a low level of background ($\sim$1 \electron pixel$^{-1}$) strongly mitigates the CTI effects by filling the traps prior to the star transit.}
\label{fig:damagedObservations}
\end{figure*}
At L2 the radiation environment is dominated by energetic protons emitted during solar flares. The proton fluence is thus governed by the cyclic activity of the Sun
which is usually monitored through sunspot counts. According to the latest predictions \citep[see e.g,][]{sidc}, the next peak of activity will occur during 2013 coinciding with the launch of {\gaia}. Using the JPL 1991 model, the reference interplanetary proton fluence model by \cite{feynman1993}, taking
into account the satellite design, and assuming 4 years of operation during the solar maximum (and one year during minimum), the average accumulated radiation dose received by a CCD of the astrometric instrument is predicted to be $\sim3\times$10$^9$ (10~MeV equivalent) protons~cm$^{-2}$.
These protons will collide with and displace atoms in the {\gaia} CCD silicon lattice, and lead to the creation of electron traps. These traps stochastically capture and release the electrons transferred in the CCD. For more information concerning the trapping processes see \cite{prodhomme2011} and references therein. The traps originate from different chemical complexes generally referred to as a trap species: a summary of the expected trap species in the {\gaia} CCDs is provided by \cite{seabroke2008} and \cite{hopkinson2005}. One usually distinguishes between trap species with short and long release time constants relative to the characteristic trap-electron interaction time ($\sim$1 ms for the {\gaia} CCDs), as they have different effects on the measurements. The traps with short release time constants capture electrons from the image leading edge and redistribute them within the telemetry window, which induces a distortion of the charge profile. The traps with longer release time constants capture electrons from the stellar profile and release them outside the telemetry window, which implies a charge loss that reduces the signal to noise ratio (see Fig.~\ref{fig:damagedObservations}). The {\gaia} CCDs comprise two hardware CTI mitigation tools: a charge injection (CI) structure and a supplementary buried channel (SBC). The CI structure is located all along the first CCD pixel row; it is composed by a diode capable of generating artificial charges and a gate that controls the number of electrons to be injected in the first pixel row and subsequently transferred across the whole CCD. Charge injections temporarily fill a large fraction of the traps present in the CCD and effectively prevent the trapping of the following generated and transfered photo-electrons. The SBC is a second and narrower doping implant on top of the buried channel. It creates a deeper potential that disappears into the shallower but wider buried channel for charge packets larger than 1500~\electron. By concentrating the electron distribution into a smaller volume it minimizes the electron-trap interactions in the rest of the pixel volume, effectively reducing the fraction of trapped electrons at low signal levels ($\leq$ 1500~\electron or $G > 15$).

In order to obtain representative results from our study, it is critical to achieve a high level of realism in the simulation of the CTI effects on each observations. This is why we make use of the most detailed CTI effects model to date \citep{prodhomme2011} verified against experimental tests performed on {\gaia} irradiated CCDs. At each transfer step, this model simulates the capture and release of electrons by computing for each trap the capture and release probabilities according to the trap characteristics and the local electron density distribution taking into account the {\gaia} pixel architecture and in particular the presence of the SBC. In the following we detail the considerations that led us to choose to simulate a unique trap species and two different radiation levels.

During the mission CI will be performed at periodic intervals. This means that most of the traps with release time constants greater than the injection period will be permanently filled, as only a very small fraction of them will have the time to release an electron. The current most likely value to be selected for the injection period is $1$~s. If one neglects the serial CTI effects (occurring during the charge transfer in the CCD readout register), and take for reference the trap species as presented in \cite{seabroke2008}, the only trap species that remains significantly active corresponds to the so-called `unknown' with a release time constant $\tau \sim90$ ms at the {\gaia} operational temperature and a capture cross-section $\sigma = 5\times 10^{-20}$ m$^2$. We thus decided to generate the damaged observations using a virtual irradiated CCD containing a unique trap species, with these parameters. Note that the release and capture time constants vary exponentially with the temperature. The temperature over the entire {\gaia} focal plane is expected to deviate at most 5~K from the nominal operating temperature. This means that for different CCDs the effect of a single trap species will be different. However the temperature variation over a single device is expected to be negligible, hence our assumption regarding a single trap species with a unique release time constant still holds.

The traps filled by the CI, release their electrons and induce a characteristic `release' trail after the CI. This trail changes the uniform nature of the background and must be carefully taken into account in the background estimation procedure. In order to prevent the CI background estimation and removal from affecting the results of our analysis, no CI was performed before the stellar transit during the simulations. Hence, the trap density has to be carefully selected to reproduce the trap density as it is perceived by a star after a CI delay (or time since last CI) comparable to the CI period. In this way, the simulated amplitude of the CTI effects corresponds to the one observed in the experimental tests with CI performed in similar conditions.

In a series of four different campaigns of experimental tests carried out on irradiated {\gaia} CCDs, the prime contractor for {\gaia}, EADS Astrium, investigated the performance of potential hardware mitigation tools and characterized the trend and amplitude of CTI effects on {\gaia}-like measurements. These campaigns are referred to as radiation campaigns (RC). The RCs were performed in simulated {\gaia} operating conditions: a CCD operated in TDI mode at a temperature of 163~K with a low level of background light. The devices were irradiated at room temperature with a radiation dose of $4\times10^9$ protons cm$^{-2}$ ($10$ MeV equivalent) that corresponds to an upper limit to the predicted {\gaia} end-of-life accumulated radiation dose. A trap density of 4~traps pixel$^{-1}$ is necessary to reproduce the amplitude of the CTI effects, in particular the fractional charge loss as observed in the second RC (RC2) from first pixel response measurements \citep{prodhomme2011}. This test was performed with charge injections occurring every $\sim 27$~s. The relative image location bias was measured in similar conditions during the same campaign for a star transit occurring 1 and 27~s after the last CI. These results are summarized in Fig. \ref{fig:damagedBias} along with the absolute location bias computed in this study. For a CI delay of 1 s the location bias is clearly smaller than for a longer CI delays (e.g., CI period of $\sim$27 s), this is due to the fact that shorter CI periods maintain a larger portion of the traps constantly filled. 
As a consequence, our simulations were performed for two different active trap densities (or level of radiation damage), 1 and 4~traps pixel$^{-1}$. By active we mean empty before the transit of the star of interest over the CCD. These densities reproduce the amplitude of the CTI effects as observed for short and long CI periods in the experimental tests.

Figure \ref{fig:damagedObservations} shows the resulting simulated CTI-induced distortion and charge loss by comparing, for different illumination levels, the simulated CTI-free observations and damaged observations (4~traps pixel$^{-1}$) after a normalization. Note that for a unique damaged CCD containing a single trap species with fixed parameters, the distortion varies significantly from one signal level to the other and not linearly. This is due in particular to the SBC, which mitigates the CTI effects at low signal level.

\section{The \textit{Gaia} image location estimation procedure}\label{sect:imageParam}

\subsection{Observation model: scene}\label{sect:observationModel}
To model the flux distribution that illuminates a CCD we need a model of the instrument response to a point-like source, and a model of the actual distribution of (point) sources on the sky. The former has already been parameterized in Section~\ref{sect:imageRefModel}: it is given by the line spread function $L$ when considering one dimension, or the PSF when considering two dimensions. Because we will mainly deal with one-dimensional data in this study we will hereafter only refer to the one-dimensional LSF. 
For the purpose of this study a simple observation model is sufficient: $E(N_k)$ is the expected number of photo-electron counts in pixel $k$, $\lambda_k$ is the modelled photo-electron count given by a flat background $\beta$ plus a single point source with flux $\alpha$ at location $\kappa$:
\begin{equation}\label{eq:model}
E(N_k) \, \equiv \, \lambda_k = \, \alpha L\left( k - \kappa\right) + \beta
\end{equation}
Here $\alpha$, $\kappa$, and $\beta$ are called the scene- or image parameters.

\subsection{Maximum-likelihood estimation of the image parameters}\label{sect:estimator}
In the {\gaia} data processing the image location $\kappa$ and flux $\alpha$ will be estimated by fitting the modelled photo-electron counts $\{\lambda_k\}$ (Eq.~\ref{eq:model}) to the observed photo-electron counts $\{N_k\}$ using a Maximum-Likelihood (ML) algorithm, and this for each observation.The image background $\beta$ is not determined using the ML algorithm but beforehand by a more adequate method. This method makes use of empty telemetry windows to estimate separately the different main components of the background: astrophysical background (zodiacal light, faint stars and galaxies i.e. $G > 20$), CI trails, and the CCD electronic offset. In this study we always assume that $\beta$ is known. Another parameter that is considered to be known beforehand is the CCD readout noise $r$. Therefore, when estimating any of the image parameters, the true values of $\beta$ and $r$ will be used. The ML algorithm is comprehensively described in \cite{lindegren2008}, therefore only the main assumptions and equations are detailed in this paper.

According to the ML principle, the best estimate of the parameter vector $\vect{\theta}$ (here $\theta_1=\kappa$
and $\theta_2=\alpha$) maximizes the likelihood function or equivalently the log-likelihood function:
\begin{equation}\label{eq:logl}
\ell\left(\vect{\theta}| \{N_k\}\right) = \sum\limits_{k} \ln p\left( N_k | \lambda_k, r \right)
\end{equation}
with $p$ the probability density function of the sample value given the modelled count and readout noise.
We hence need to adopt a probability model for the sample values. To do so we assume
(i) that the noise is not correlated from a sample to another (already implicit in the sum in Eq.~\ref{eq:logl});
(ii) that the variance of the noise is $E[(N_k-\lambda_k )^2]=\lambda_k+r^2$;
and (iii) that the sample value including the readout noise can be modelled as Poissonian random variables,
$N_k \sim \mathrm{Poisson}(\lambda_k+r^2)-r^2$ \citep[see][]{lindegren2008}. That the Poisson distribution
is discrete, while $N_k$ (obtained by correcting the digitized values for bias and gain) are in general non-integer,
is not a problem as long as $N_k+r^2\ge 0$. The continuous probability density function derived from the
Poisson distribution is:
\begin{equation}\label{eq:loglpoisson}
p\left( N_k | \lambda_k, r \right) = \mbox{const} \times
\frac{(\lambda_k + r^2)^{N_k+r^2}}{\Gamma(N_k+r^2+1)}e^{-\lambda_k-r^2}
\end{equation}
and Eq. \ref{eq:logl} can then be re-written:
\begin{equation}\label{eq:loglpoisson}
\ell\left(\vect{\theta} | \{N_k\}\right) = \mbox{const} + \sum\limits_{k}[(N_k + r^2)
\ln (\lambda_k (\vect{\theta})+r^2)-\lambda_k(\vect{\theta})]
\end{equation}
which is maximized by solving the following system of equations:
\begin{equation}
\frac{\partial \ell\left( \vect{\theta} | \{N_k\}\right) }{\partial \vect{\theta}}
= \sum\limits_{k} \frac{N_k-\lambda_k(\vect{\theta})}{\lambda_k(\vect{\theta})+r^2}
\frac{\partial \lambda_k}{\partial\vect{\theta}}= \vect{0}
\end{equation}
These equations are non-linear and must be solved by iteration. Given an initial estimate $\vect{\theta}^{(0)}$,
the linear system to be solved in iteration $m$ is:
\begin{equation}
\vect{A}^{(m)} \Delta \vect{\theta}^{(m)} = \vect{\delta}^{(m)}
\label{eq:linearSys}
\end{equation}
whereupon
\begin{equation}
\vect{\theta}^{(m+1)} = \vect{\theta}^{(m)} + \Delta\vect{\theta}^{(m)}
\label{eq:linearSys2}
\end{equation}
$\vect{A}$ is a symmetric positive definite matrix computed from the expectation of the Hessian matrix;
its elements are:
\begin{equation}
A_{ij} = \sum\limits_{k}\frac{1}{\lambda_k(\vect{\theta})+r^2}
\frac{\partial\lambda_k}{\partial\theta_i}\frac{\partial\lambda_k}{\partial\theta_j}
\end{equation}
and
\begin{equation}
\delta_{i} = \sum\limits_{k}\frac{1}{\lambda_k(\vect{\theta})+r^2}\frac{\partial\lambda_k}{\partial\theta_i}
\end{equation}
The iterations converge quickly if the initial estimate is reasonably close to the ML solution.

\subsection{First image parameter estimates and LSF model \label{sect:lsfModel}}

The ideal image model $L$ (the true underlying flux distribution for each observation) corresponds to the reference image that is used to generate the data. During the mission, $L$ will not be known. Therefore we have to estimate an image model $\tilde{L}$ using the observations themselves.
This estimation is an iterative process (Section~\ref{sect:iterativeProcedure}), and successive iterations are denoted with the superscript ($n$) for $n = 0, 1, \ldots$ (not to be confused with the iterations in Eqs.~\ref{eq:linearSys}~and~\ref{eq:linearSys2}).

Given a set of transits for a certain reference image, background and $G$, denoted by $\{\{N_k\}\}_{G}$, how do we make the first estimate of the image parameters and generate the first LSF model, $L^{(0)}$?
As mentioned in Section~\ref{sect:estimator}, the background $\beta$ and readout noise $r$ are assumed to be known already. The most straightforward initial flux estimate $\alpha^{(0)}$ can be made by simply taking the sum of the observed counts after subtracting the background. The initial estimate for the image location $\kappa^{(0)}$ is determined using Tukey's Biweight centroiding algorithm \citep{press92, lindegrenLL68}.

\begin{figure}
\includegraphics[width=0.49\textwidth]{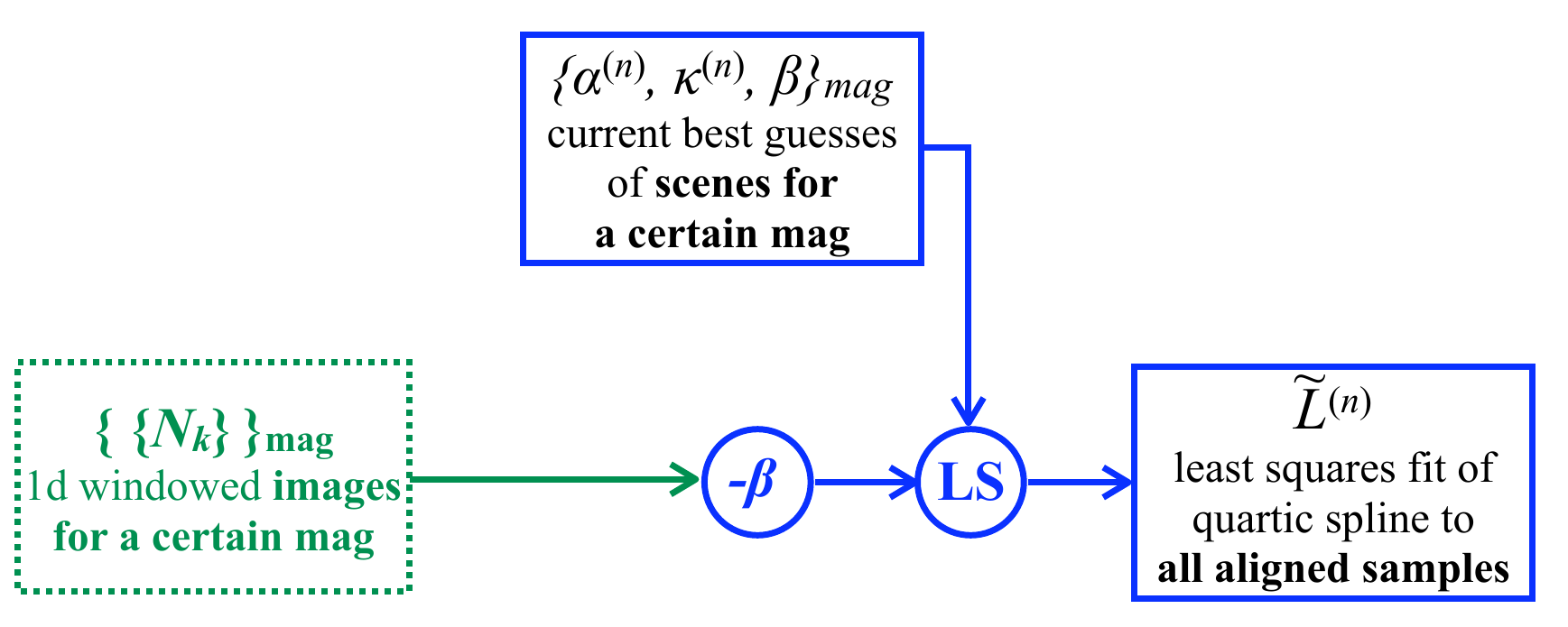}
\caption{Diagram of the construction of the estimated LSF model $\tilde{L}$. For a particular reference image, background level and $G$, the photo-electron counts of all selected profiles are aligned relative to each other using the current best estimates of the scene parameters to create an oversampled profile, which is then fitted by the special quartic spline using a Least Squares algorithm, after removal of the background.
}\label{fig:model_update}
\includegraphics[width=0.49\textwidth]{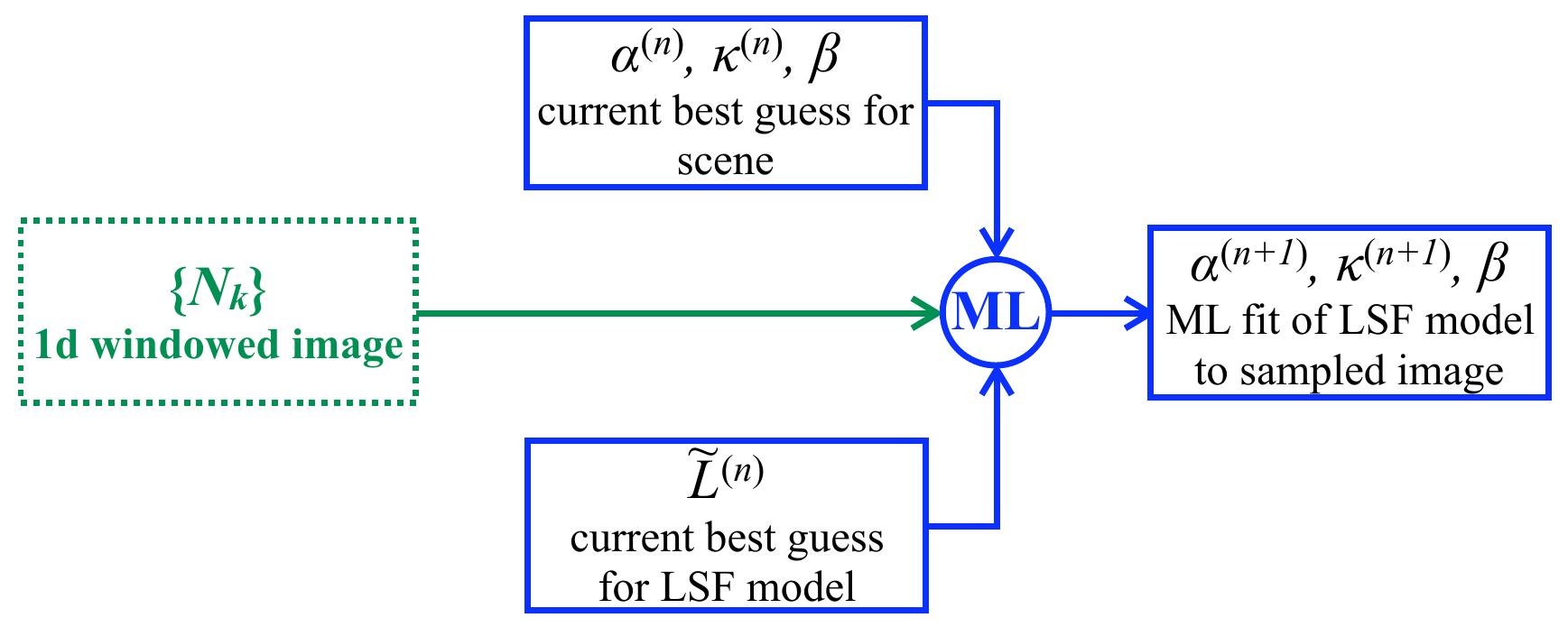}
\caption{Diagram summarizing the estimation of the scene (or image) parameters from a single observation. Modelled counts, computed using the latest LSF model, are fitted to the observed counts using our Maximum Likelihood (ML) algorithm (see Section~\ref{sect:estimator}). Note that the background is not determined by the ML algorithm but by a dedicated procedure not detailed in this paper. In our study, the background $\beta$ is assumed to be known.
}\label{fig:scene_update}
\includegraphics[width=0.49\textwidth]{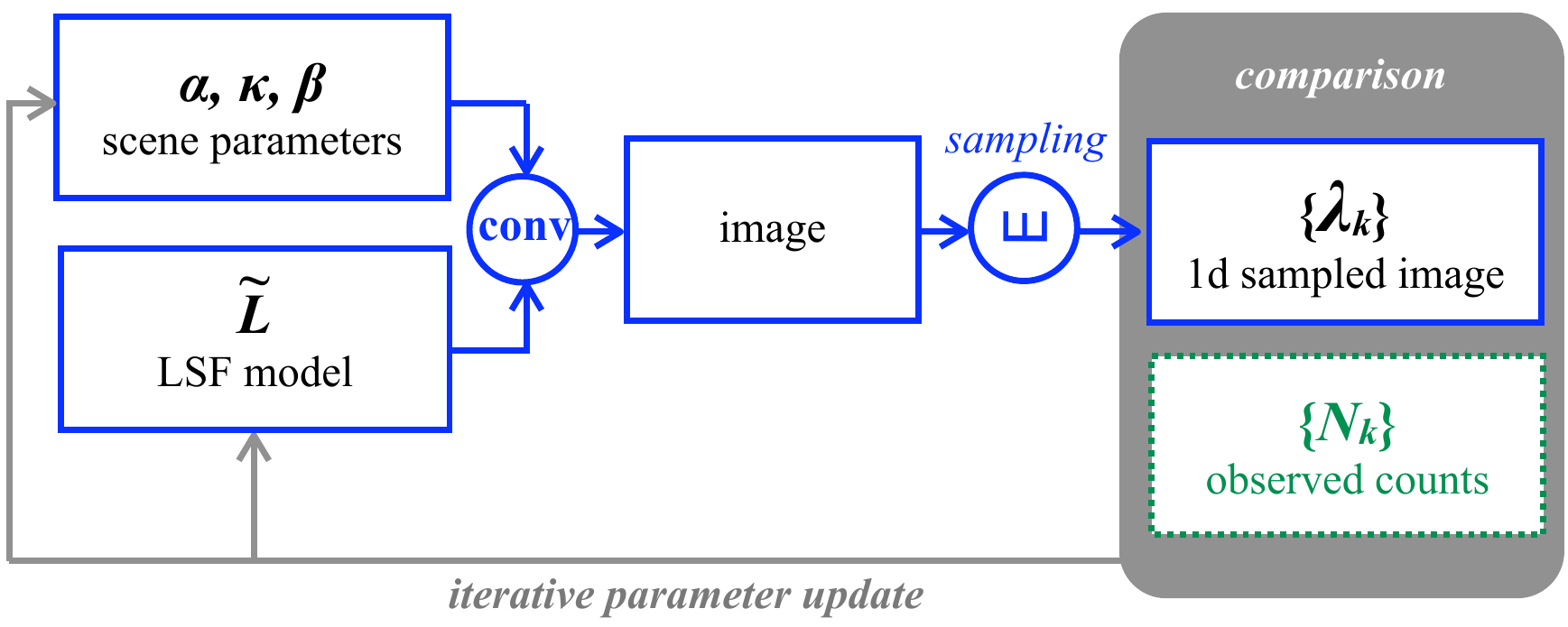}
\caption{Resulting top level diagram of the {\gaia} image parameter estimation iterative procedure. The LSF model, $\tilde{L}$, and the scene parameter estimates, $\kappa$, $\alpha$ (respectively the image location and flux), are iteratively improved by fitting the modelled counts $\{ \lambda_k \}$  to the observed counts $\{ N_k \}$. The modelled counts are predicted by our observation model (Eq. \ref{eq:model}) based on the current best scene parameter estimates. In our study the background $\beta$ is assumed to be known with a high accuracy (we use the true value).
}\label{fig:iterative-procedure}
\end{figure}

To generate the first estimate $\tilde{L}^{(0)}$ we use the initial location estimates $\{ \kappa^{(0)} \}_{G}$ to relatively align the photo-electron counts of all selected profiles and create an oversampled profile. The creation of the oversampled profile is possible because each count results from the sampling of the reference image at a different sub-pixel position (Section~\ref{sect:mcModel}).
After a background subtraction the oversampled profile is fitted by the special quartic spline to obtain $\tilde{L}^{(0)}$. This profile estimation procedure is illustrated in Fig.~\ref{fig:model_update}.

\subsection{Iterative image parameter and LSF model improvement\label{sect:iterativeProcedure}}
Once the first image model $\tilde{L}^{(0)}$ is available, an improved estimate of the image parameters of each individual transit can be made using the ML algorithm (described in Section~{\ref{sect:estimator}}). This is illustrated in Fig.~\ref{fig:scene_update}. Based on these improved image parameters an improved image model can be constructed, leading to the iterative scheme shown in Fig.~\ref{fig:iterative-procedure} where the image parameters and image model are improved one after the other. Note that in the whole procedure we have not used any prior knowledge: everything is estimated from the observed photo-electron counts (i.e. `self-calibrating'). 

After each iteration the residuals between the modelled and the observed photo-electron counts are monitored through the computation of the $\chi^2$:
\begin{equation} \label{eq:sampleChiSquare}
\chi^{2}_{G} = \sum_{t=0}^{T-1} \sum_{k=0}^{K-1} {\frac {\left( \lambda_{tk} - N_{tk} \right)^{2}}{\sigma_{tk}^2}}\ \ \mathrm{with}\ \ \sigma_{tk}^2 = N_{tk} + r^2
\end{equation}
where $T$ is the total number of transit profiles for a certain $G$, and $K$ the number of along-scan pixels in each profile. For transit $t$ and pixel $k$, $\lambda_{tk}$ and $N_{tk}$ are the predicted and observed photo-electron counts respectively. The uncertainty $\sigma_{tk}$ is considered to be equivalent to the quadratic sum of the photon noise and the readout noise~$r$.

\begin{figure}
\centering
\includegraphics[width=0.49\textwidth]{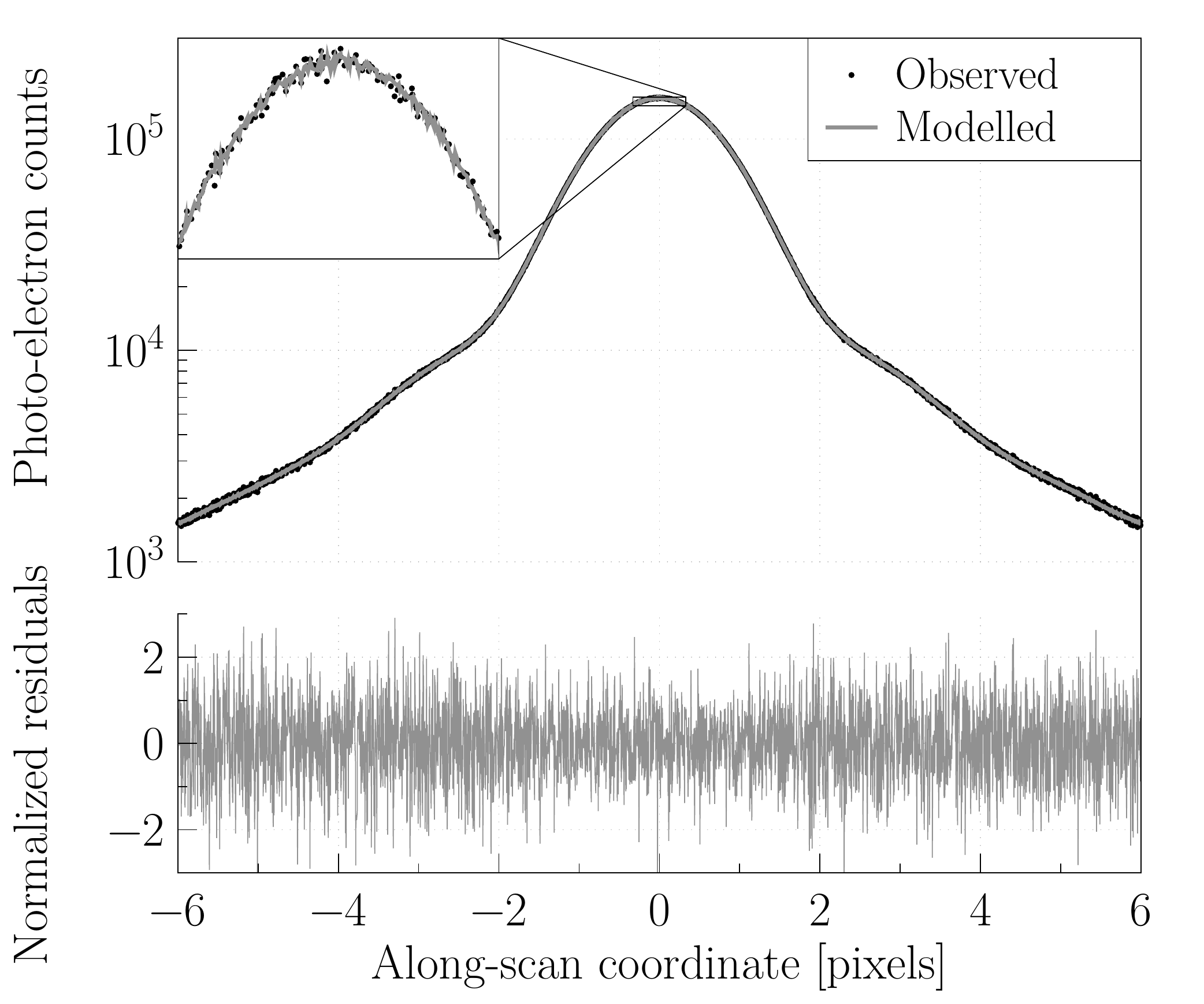}
\caption{Top: Comparison between the observed CTI-free observations $\{\{N_k\}\}_{13.3}$ (black dots), and the modelled counts $\{\{\lambda_k\}\}_{13.3}$ (grey line) computed following the presented iterative procedure. The transits were generated from the typical reference image, at $G= 13.3$, with readout noise and background.
\newline Bottom: Residuals normalized by the noise: $(N_{tk} - \lambda_{tk})/ \sigma_{tk}$, at the last stage of the image parameter estimation iterative procedure.} \label{fig:residuals}
\includegraphics[width=0.49\textwidth]{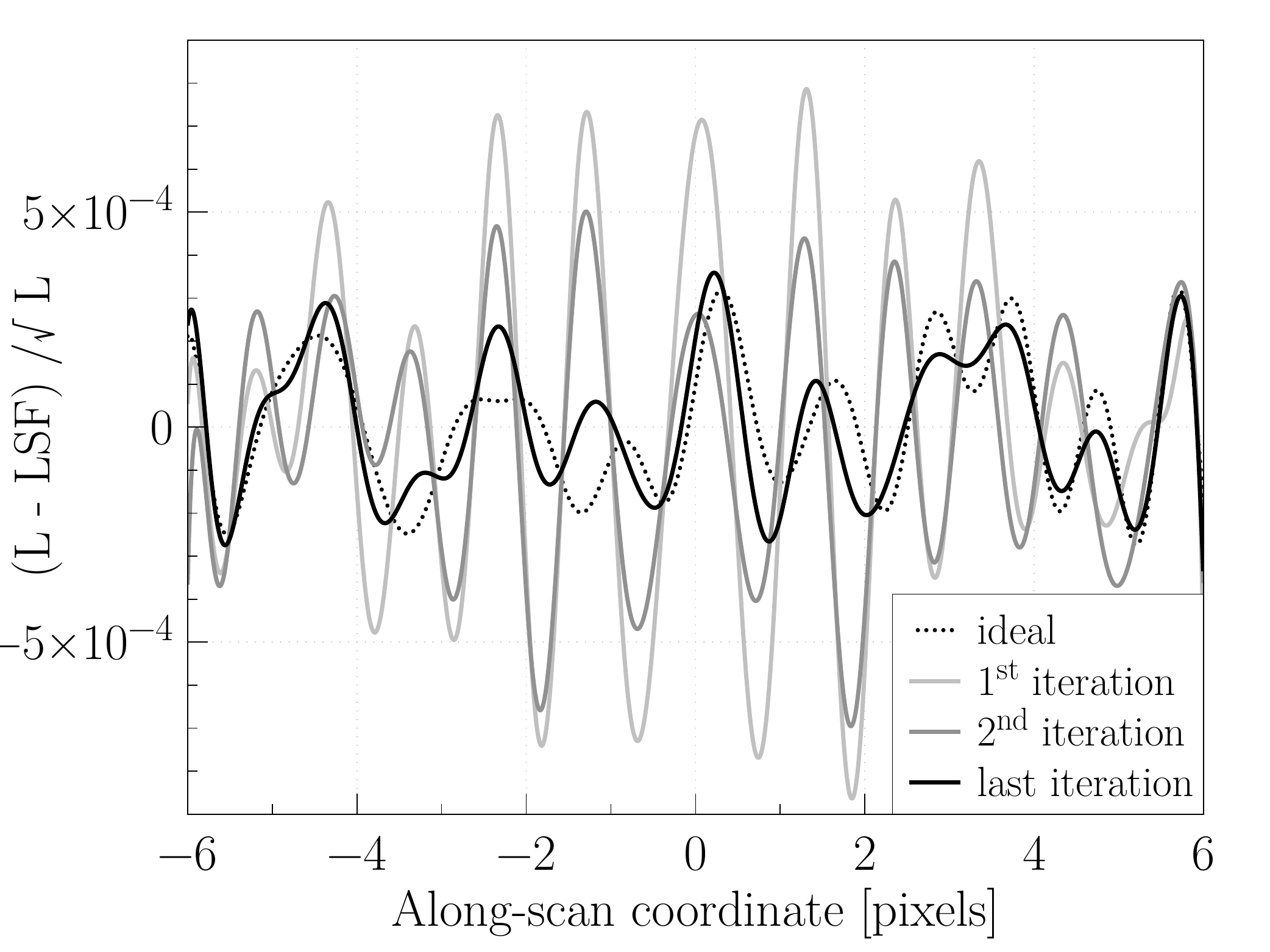}
\caption{Normalized difference between the LSF model and the true underlying flux distribution at different stages of the image parameter estimation iterative procedure ($G = 13.3$, readout noise and background, and typical reference image). The ideal case corresponds to a LSF model constructed from the observed data and the true image location.} \label{fig:difference}
\end{figure}

The agreement between observed and modelled counts (Fig.~\ref{fig:residuals}) does not significantly improve after 2 iterations, however the agreement between the LSF model and the reference image (see Fig.~\ref{fig:difference}), and the average image location bias as well as the location estimator standard errors (Sections~\ref{sect:locBiasUndamaged} and \ref{sect:locAccUndamaged}) can still improve after a certain number of iterations that essentially depends on the stellar magnitude. As a consequence, we stop the iterative procedure after a particular number of iterations that is determined for each magnitude beforehand.

\section{Theoretical and actual limit to the image location accuracy}\label{sect:limits}
To be able to evaluate the accuracy of the image location estimation procedure, we first need to determine what the theoretical limit of any image location estimator is. This is done by computing the Cram\'{e}r-Rao bound (Section~\ref{sect:cramerrao}), which shows that it depends uniquely on the image shape, flux, background and noise.
Subsequently, and first in the absence of CTI, we verify that any potential bias of the {\gaia} image location estimator does not depend on the image location (Section~\ref{sect:locationBias}). And then the estimator bias  and standard errors are calculated as a function of $G$, image reference width, and for different operating conditions (Sections~\ref{sect:locBiasUndamaged} and \ref{sect:locAccUndamaged}). We compare the latter to the theoretical limit and evaluate the efficiency of our estimation procedure in the absence of radiation damage.
Then we estimate the irreversible loss of accuracy intrinsic to radiation damage, by computing the Cram\'{e}r-Rao bound for a `damaged LSF' generated from the data set of damaged observations (Section~\ref{sect:damagedCR}).
Ultimately, we apply the {\gaia} image location estimator to the damaged observations without any CTI mitigation. This allows us to characterize the radiation damage induced location bias (Section~\ref{sect:bias}), and check the consistency of our CTI effects simulation by comparing our results to experimental test results (Section~\ref{sect:comparison}).

\subsection{Definition of the astrometric Cram\'{e}r-Rao bound}\label{sect:cramerrao}
For a dataset with a known underlying probability density function the Cram\'{e}r-Rao minimum variance bound theorem gives the minimum reachable variance of a free parameter using any estimation procedure.
In the case of estimating the location $\kappa$ of a one-dimensional image containing $N_p$ detected photons, the Cram\'{e}r-Rao bound $\sigma_{\kappa}^2$ can be expressed as follows \citep{lindegren78}: 
\begin{equation}\label{eq:cr}
\sigma_{\kappa}^2 = \frac{1}{N_p} \left( \int{ \frac{|I'(x)|^2}{I(x)+(\beta + r^2)/N_p}dx} \right)^{-1}
\end{equation}
with $I(x)$ a normalized one-dimensional flux distribution of the image along $x$, $\beta$ the background and $r$ the CCD readout noise.

\subsection{Location independent error and standard deviation}\label{sect:locationBias}
\begin{figure}
\centering
\includegraphics[width=0.49\textwidth]{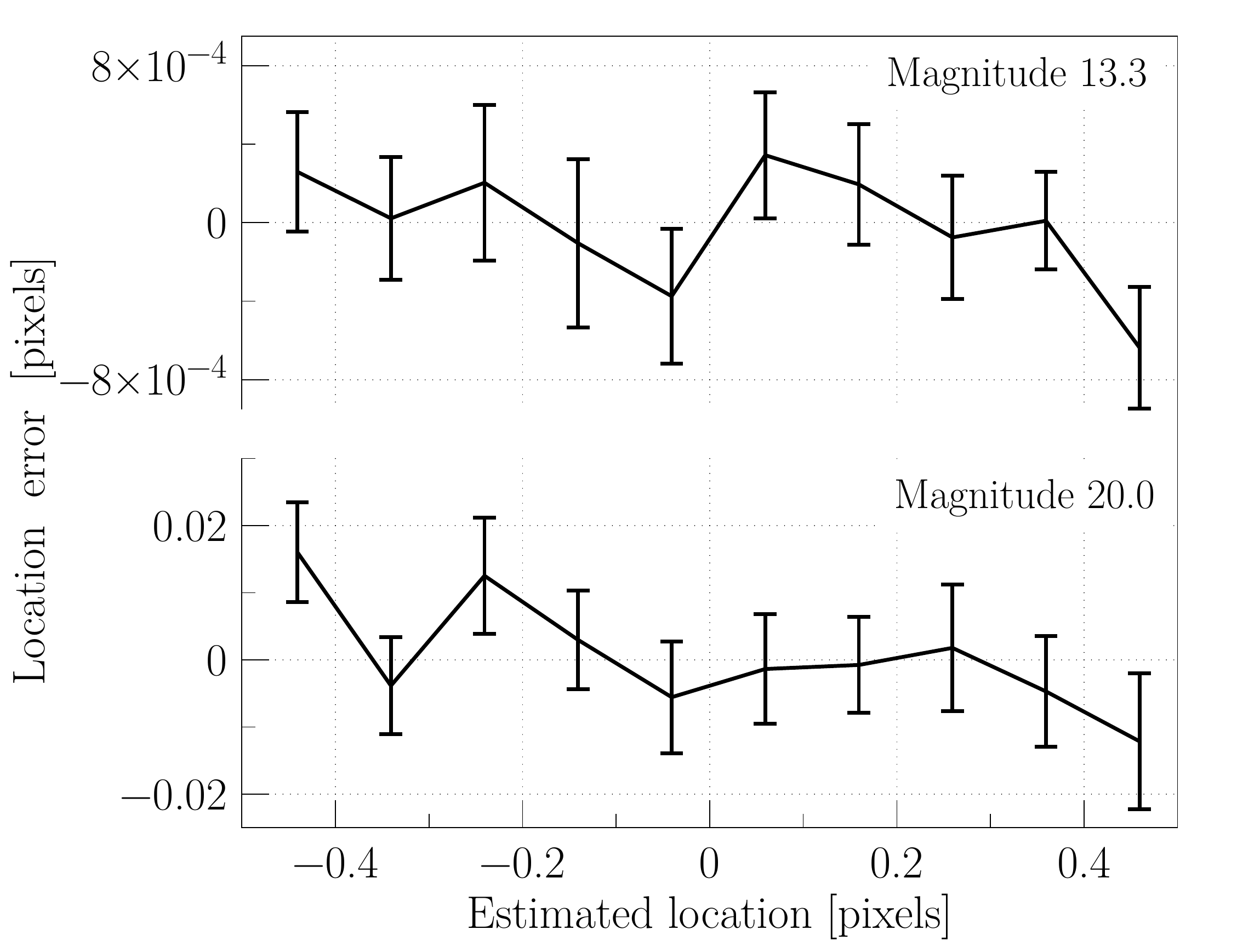}
\caption{Image location error $\{ \delta_\kappa^{(n)} \}_{G}$, as a function of the location offset from the pixel grid $\{ \kappa^{(n)} \}_{G}$, for the brightest (top) and the faintest magnitude (bottom) at the last stage of the iterative procedure (respectively 7 and 2 iterations).}
\label{fig:location_bias-vs-subpix}
\end{figure}

The iterative procedure on a transit (described in Section~\ref{sect:imageParam}) provides us with an image location estimation $\kappa^{(n)}$ for iteration $n$.
One can compute the image location error $\delta_{\kappa}$ by directly comparing $\kappa$ to the true image location $\kappa_{\mathrm{true}}$:
\begin{equation}
\delta_{\kappa}= \kappa - \kappa_{\mathrm{true}}
\end{equation}
Before averaging over all the image location estimates (or the corresponding errors) of a particular magnitude (for a particular reference image, window size, background level and readout noise value), one first needs to check that the error does not significantly fluctuate as a function of the relative location offset from the pixel grid. The latter is simply given by the collection of estimated image locations $\{ \kappa \}_{G}$. Fig.~\ref{fig:location_bias-vs-subpix} shows an example of this variation: each point corresponds to the average location error over 25 adjacent sub-pixel positions, the error bars represent the standard deviation of the points with respect to this mean. 
At the last stage of the iterative procedure, the set of estimated locations $\{ \kappa^{(n)} \}$ shows that there is virtually no significant systematic variation across the relative location offsets. A certain number of iterations (7 for the brightest and 2 for the faintest) is needed to remove the variation introduced during the procedure initialization by the Tukey's Biweight centroiding algorithm.

Having established that there is no significant error as function of the relative location offset from the pixel grid it is allowed to average over all the transits within a particular magnitude (for a particular reference image, window size, background level and readout noise value) to find the bias:
\begin{equation}
\langle\delta_{\kappa}\rangle =  \frac{1}{T}\sum_{t=0}^{T-1} \delta_{\kappa_t}
\label{eq:bias}
\end{equation}
with $t$ going through all transits of the transit selection. We will indicate the average over all transits of a particular magnitude as $\langle\delta_{\kappa}\rangle_{G}$. For this subset of transits we can now also compute the corresponding standard deviation:
\begin{equation}
\sigma_{\kappa} = \sqrt{\frac{1}{T-1} \sum_{t=0}^{T-1} (\delta_{\kappa_t} - \langle\delta_\kappa\rangle_{G})^2}
\label{eq:stdevBias}
\end{equation}
The statistical uncertainty of this standard deviation is:
\begin{equation}
\upsilon_{\sigma_{\kappa}} = \frac{\sigma_{\kappa}}{\sqrt{2T}}
\label{eq:uncertStdev}
\end{equation}
and the statistical uncertainty of the bias is:
\begin{equation}
\upsilon_{\langle\delta_{\kappa}\rangle} = \frac{\sigma_{\kappa}}{\sqrt{T}}
\label{eq:uncertBias}
\end{equation}
Summarizing, we can for all transits of a particular magnitude quantify the location bias as: $\langle\delta_{\kappa}\rangle_{G} \pm \upsilon_{\langle\delta_{\kappa}\rangle}$, and the location standard deviation, hereafter called location precision, as: $\sigma_{\kappa,G} \pm \upsilon_{\sigma_{\kappa}}$. In the absence of any significant bias, the latter can be referred to as the location accuracy. 

\subsection{CTI-free location bias results per magnitude}\label{sect:locBiasUndamaged}

\begin{figure}
\centering
\includegraphics[width=0.49\textwidth]{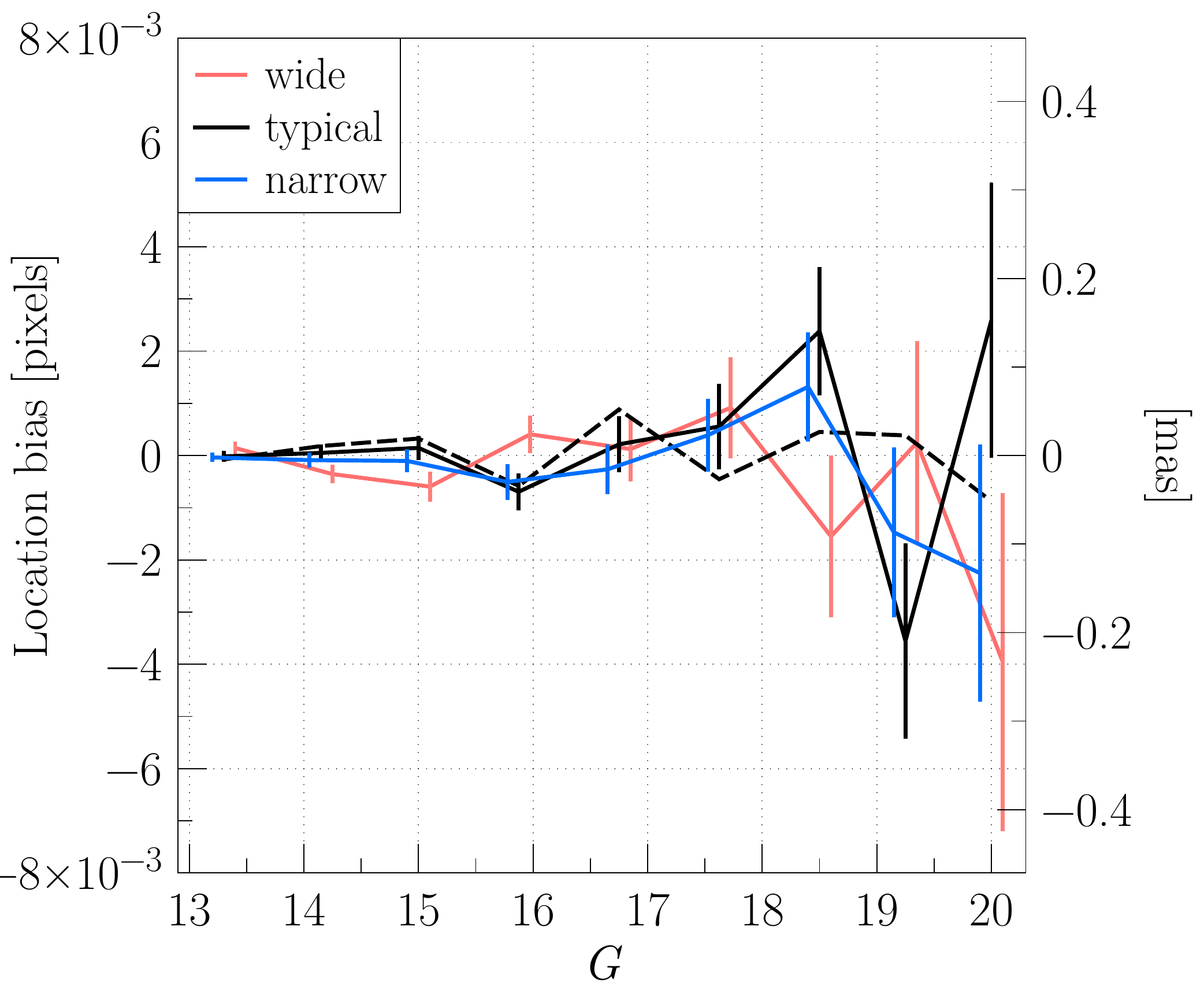}
\caption{Comparison of the location bias $\langle\delta_{\kappa }\rangle \pm \upsilon_{\langle\delta_{\kappa}\rangle}$ for different reference image widths as a function of $G$ measured in the {\gaia} operating conditions (cf. Section \ref{sect:locBiasUndamaged}). The dashed line corresponds to a measurement realized in the same conditions for the typical image width but without background. The error bars correspond to $\upsilon_{\langle\delta_{\kappa}\rangle}$, the statistical uncertainty on the location bias (see Eq.\ \ref{eq:uncertBias}). Note that for readability a slight offset has been introduced on the $G$ axis for the narrow and wide reference image results.
There are two ordinate axes, left is the location bias in units of pixels and the right in units of mas. The same holds for the following figures.
}
\label{fig:location_bias-vs-magnitude}
\end{figure}

Figure \ref{fig:location_bias-vs-magnitude} shows the image location bias as a function of $G$ for the three different reference image widths, a sky background level set to the average sky brightness and the {\gaia} CCD operating conditions regarding the readout noise value (4.35 $\electron$) and the size of the telemetry windows in the along-scan direction (12 pixels for $G<$16 then 6 AL pixels $G>$16). This set of conditions, hereafter referred as to {\gaia} operating conditions, constitutes the most realistic case of our study and also the most unfavourable case for the image parameter estimation procedure. Yet, one can observe from Fig.~\ref{fig:location_bias-vs-magnitude} that the location bias, $\langle\delta_{\kappa}\rangle_{G}$, for none of the magnitudes exceeds the level of 5 milli-pixels ($\sim0.3$ mas). Moreover: within the uncertainty of our measurement, and for the three different image widths, $\langle\delta_{\kappa}\rangle_{G}$ does not significantly deviate from zero. Hence we can establish that the {\gaia} image location estimator is a bias-free estimator in the absence of radiation damage. Increasing the window size in the along-scan direction or reducing the readout noise has no significant effect on $\langle\delta_{\kappa}\rangle_{G}$. Only setting the background level to zero seems to slightly decrease the bias for the faintest magnitudes. This effect is illustrated by the dashed line in Fig.~\ref{fig:location_bias-vs-magnitude}.

\subsection{CTI-free location accuracy per magnitude}\label{sect:locAccUndamaged}
\begin{figure}
\centering
\includegraphics[width=0.49\textwidth]{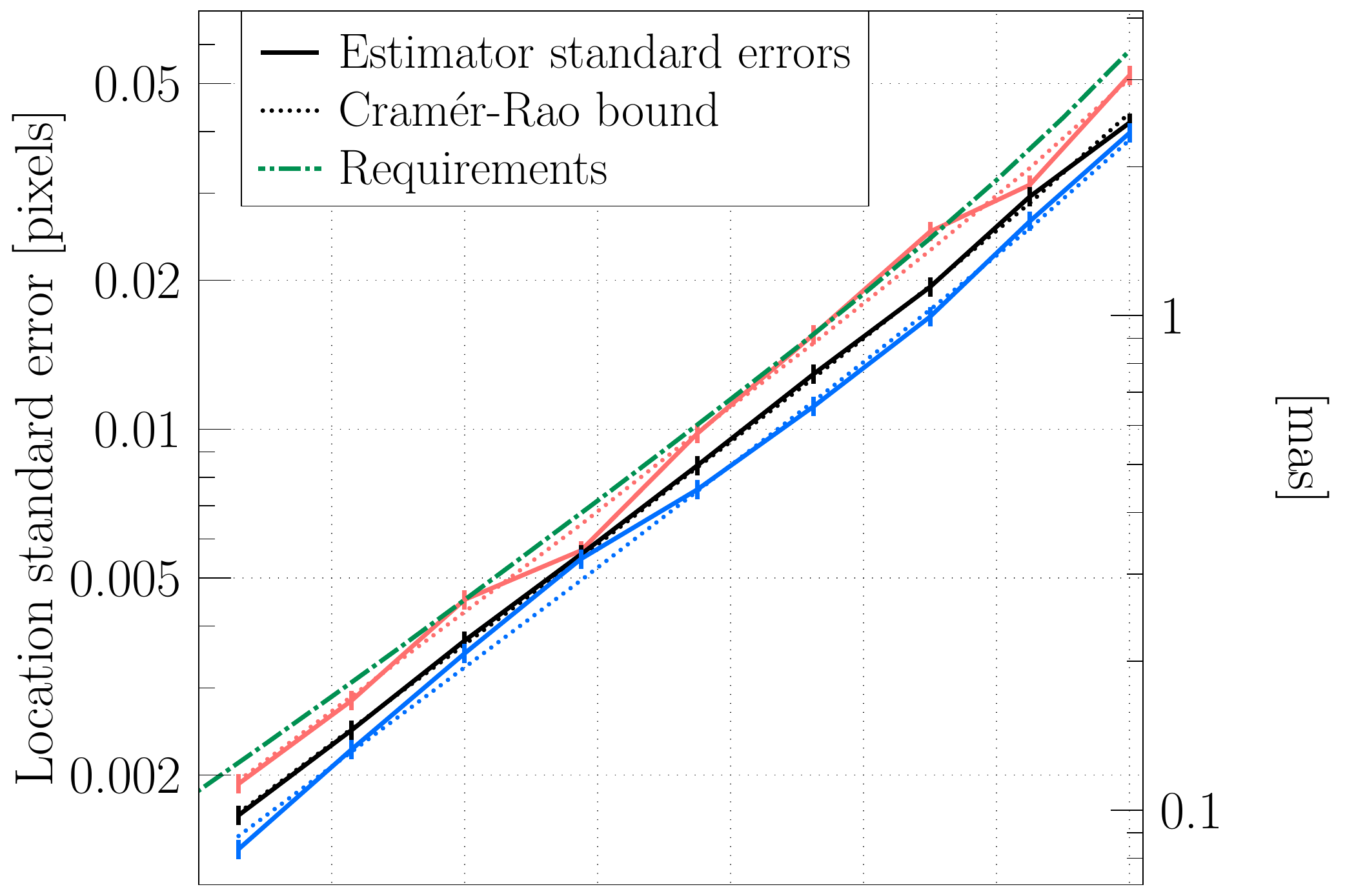}
\includegraphics[width=0.49\textwidth]{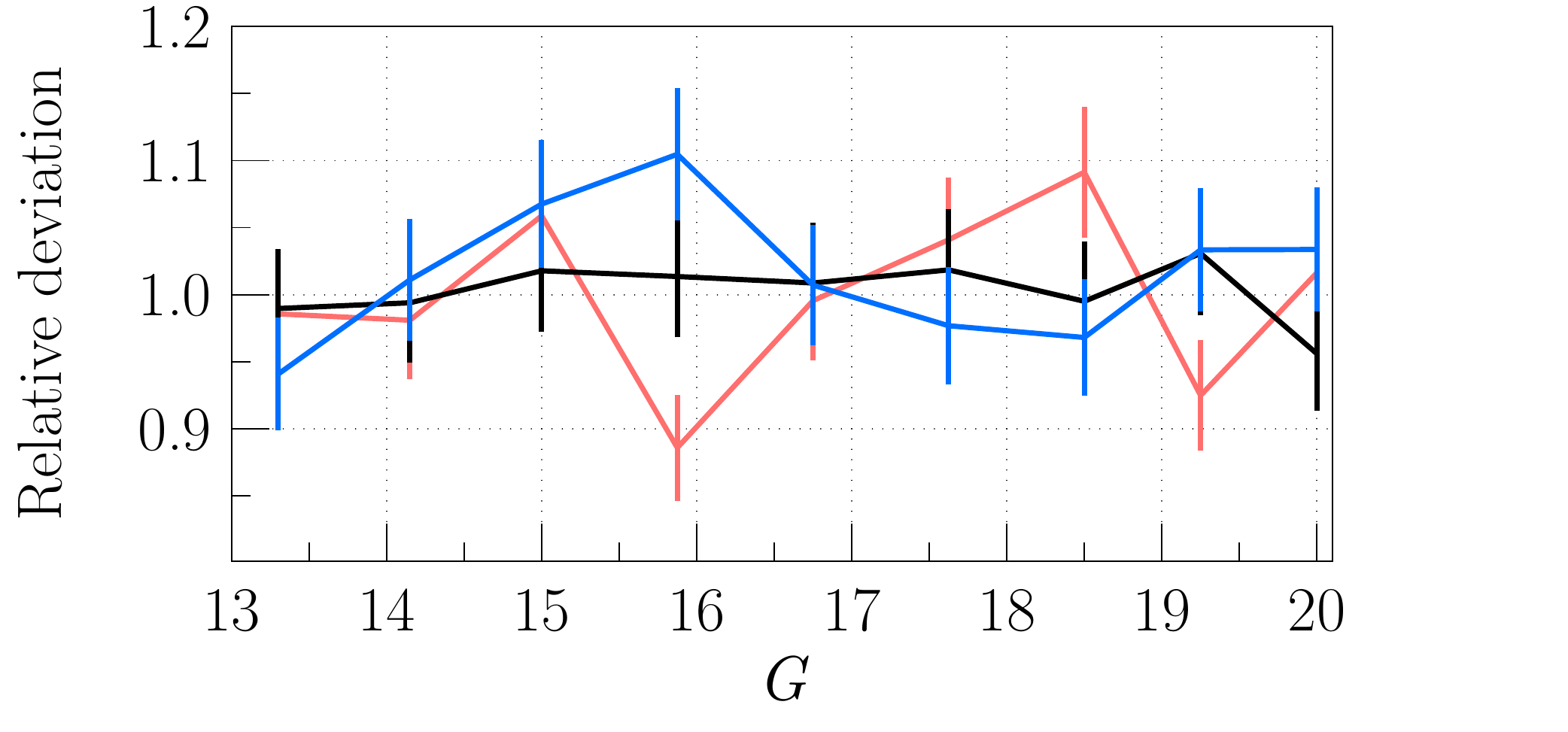}
\caption{Comparison between the actual and theoretical limit to the image location accuracy, as a function of $G$. Top: The continuous lines correspond to the {\gaia} image location estimator accuracy standard errors, $\sigma_{\kappa,G} \pm \upsilon_{\sigma_{\kappa}}$, measured in the {\gaia} operating conditions for the three different reference images: narrow (blue), typical (black) and wide (red). The associated error bars corresponds to the statistical uncertainty on the location accuracy (see Eq.\ \ref{eq:uncertStdev}). The dotted line represents the Cram\'{e}r-Rao bounds computed for the same reference image widths (same colour coding), window size, background level, and readout noise value. Finally, the dashed and dotted (green) line shows the {\gaia} required image location accuracy.
\newline Bottom: The ratio between $\sigma_{\kappa,G}$ and the Cram\'{e}r-Rao bounds and the associated error bars are depicted as a function $G$. For the three reference image widths, the relative deviation does not exceed 10\%. The {\gaia} image location estimator can thus be considered efficient in the absence of radiation damage.
}
\label{fig:standard_errors-vs-magnitude}
\end{figure}
To evaluate the efficiency of our estimator, we compare the measured standard errors, $\sigma_{\kappa}$, to the astrometric Cram\'{e}r-Rao bound, the theoretical limit to the image location accuracy of any bias-free estimator (see Section \ref{sect:cramerrao}). The comparison results are summarized in Table \ref{tab:limits} for different values of $G$, image widths, window size, background levels, and values of CCD readout noise. In Fig.~\ref{fig:standard_errors-vs-magnitude}, we compare the accuracy of the {\gaia} image location estimator in the {\gaia} operating conditions, the Cram\'{e}r-Rao bound computed for the same reference image width and level of readout noise, and the requirements as presented in Table \ref{tab:requirement}. As one can see, the {\gaia} image location estimator performs remarkably well. The estimator standard errors are always below the requirements and this for any reference image width. Also the standard errors, within the measurement statistical uncertainty $\upsilon_{\delta_{\kappa}}$, strictly follow the Cram\'{e}r-Rao bounds at every signal level. Note how stringent the {\gaia} requirements are: for the wide reference image, the actual and theoretical limits to the image location accuracy are very close to the required accuracy. 

As mentioned in Section~\ref{sect:introduction}, the targeted performance predictions of {\gaia} contain a margin of 20\% to take into account unmodelled  on-ground calibration errors including for instance residual bias.
In this context we consider an estimator efficient if its standard errors are within 10\% of the Cram\'{e}r-Rao bound and thus not consuming more than half the margin. The {\gaia} estimator rigorously fulfills this criteria. This is illustrated in Fig.~\ref{fig:standard_errors-vs-magnitude} (bottom) for the three reference image profiles: the ratio between the estimator standard errors and the Cram\'{e}r-Rao bounds remain below 1.1 (i.e. 10\% relative deviation). As expected, in both the theoretical and actual cases, an increase in the image width is directly translated into a loss in location accuracy. This loss varies linearly with the image FWHM. Table~\ref{tab:limits} shows that increasing the readout noise, the background level, and/or decreasing the window size also increases the Cram\'{e}r-Rao bound and the {\gaia} estimator standard errors.

In the absence of radiation damage, we established that in realistic operating conditions and from bright to faint magnitudes, the {\gaia} image location estimator is bias-free, efficient, and performs within the requirements, with a high accuracy close to the theoretical limit. It is now important to characterize in detail the impact of radiation damage on the image location uncertainty.

\subsection{Radiation damage intrinsic uncertainty increase}\label{sect:damagedCR}
\begin{figure}
\centering
\includegraphics[width=0.49\textwidth]{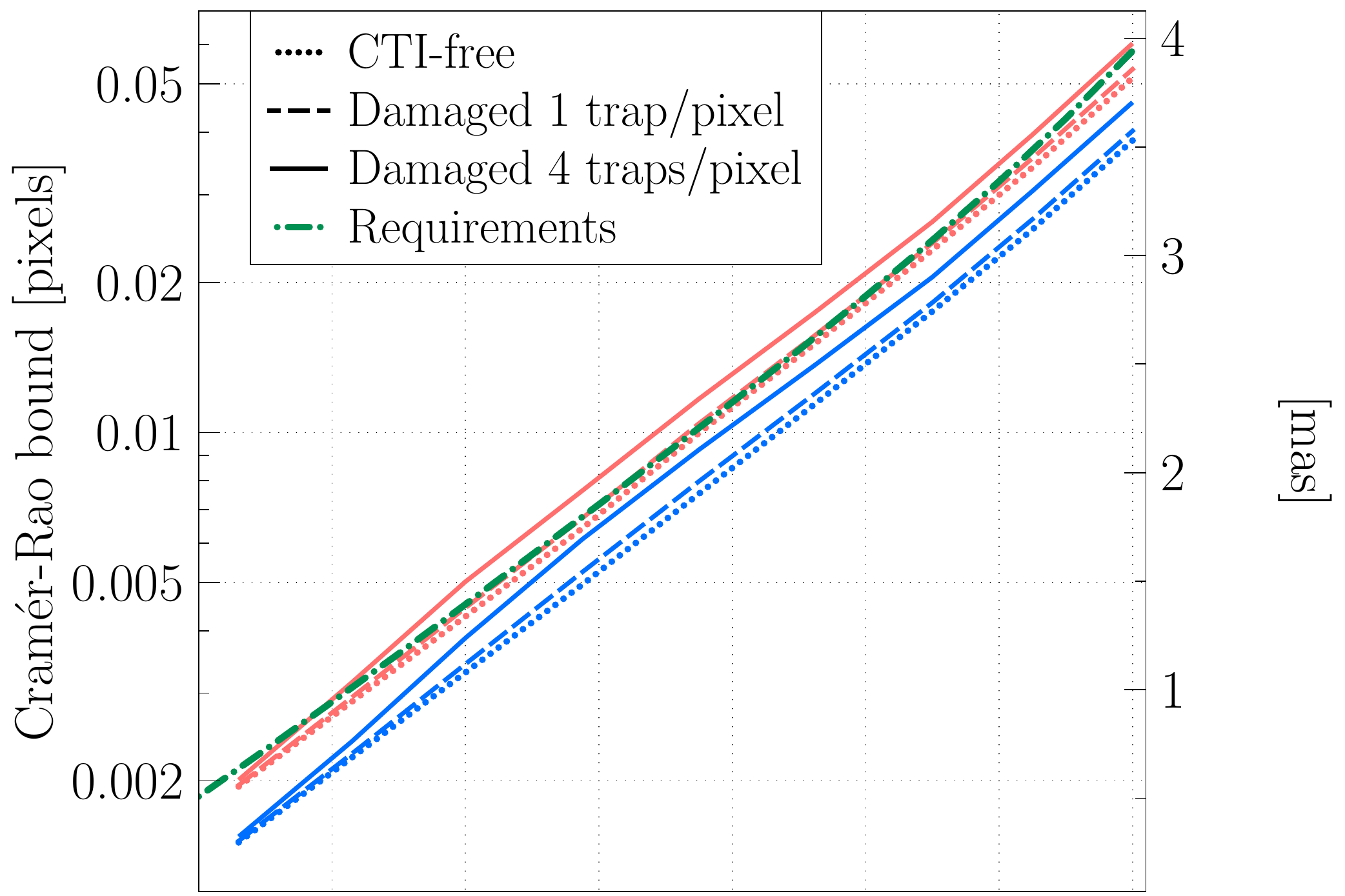}
\includegraphics[width=0.49\textwidth]{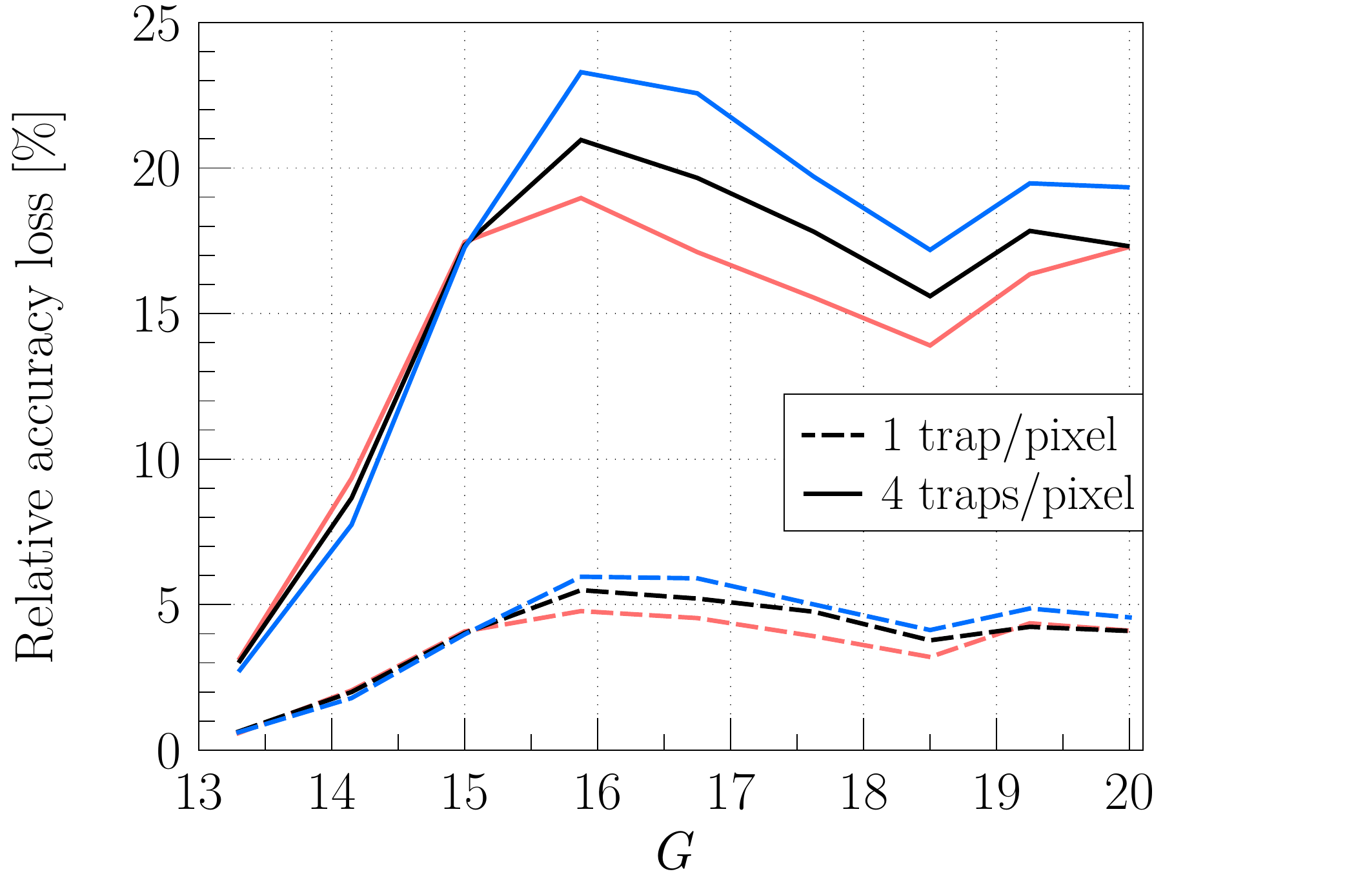}
\caption{Top: Comparison between the Cram\'{e}r-Rao bounds computed from the original flux distribution (dotted lines) and the constructed damaged flux distribution for the two trap densities: 1~trap~pixel$^{-1}$ (dashed line), 4~traps pixel$^{-1}$ (continuous line). Two reference images are here considered: the narrow (blue) and the wide (red). The window sizes, background level and readout noise value correspond to the {\gaia} operating conditions. The computed required image location accuracy is also shown (dashed-dotted green line).
\newline Bottom: The relative intrinsic loss of accuracy induced by radiation damage as a function of $G$ for the three reference images: narrow (blue), typical (black), wide (red), and the two trap densities: 1~trap~pixel$^{-1}$ (dashed line), 4~traps pixel$^{-1}$ (continuous line). The relative loss of accuracy corresponds to the relative difference between the Cram\'{e}r-Rao bound computed from the original flux distribution and the constructed damaged flux distribution. Note the important difference in loss amplitude, for the two different trap densities: a reduction in the active trap density (e.g., by the means of CI) is directly translated into a gain in location accuracy of a similar factor. Similarly, the flattening of the intrinsic loss for $G>15$ is due to the effect of the SBC.}
\label{fig:crdamaged}
\end{figure}
\begin{figure*}
\centering
\includegraphics[width=0.49\textwidth]{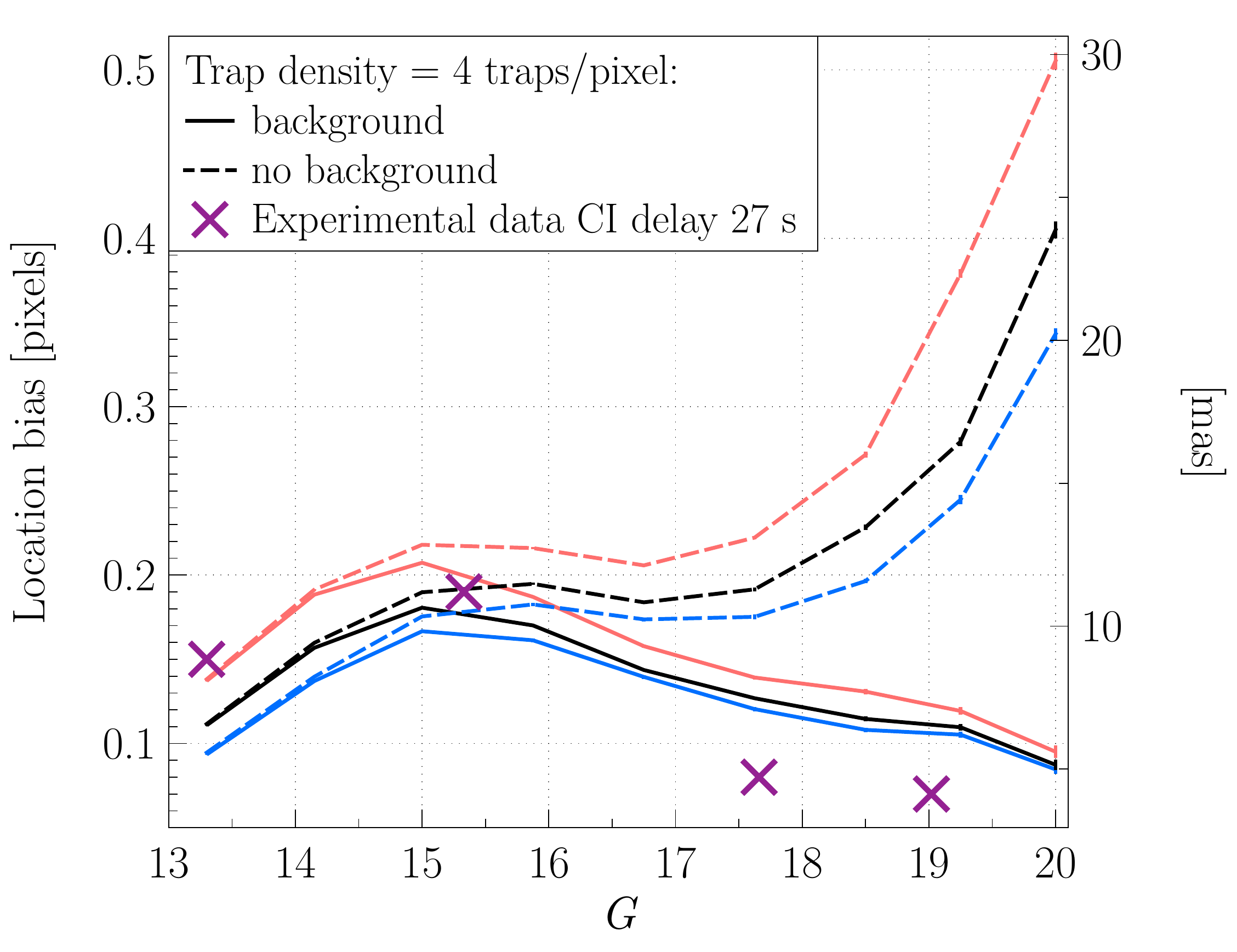}
\includegraphics[width=0.49\textwidth]{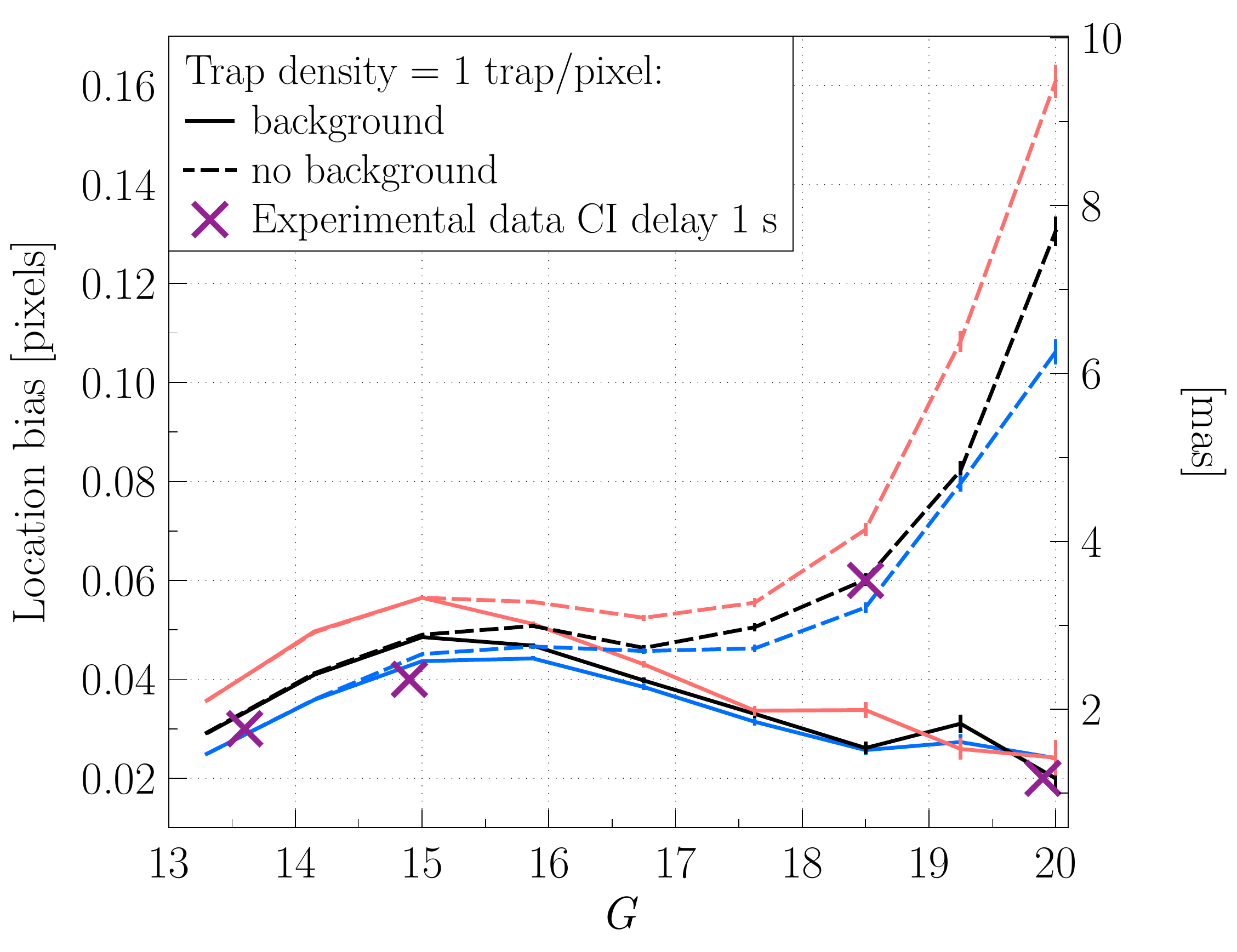}
\caption{Image location bias, $\langle\delta_{\kappa}\rangle_{G,D}$, resulting from the radiation damage effects on the stellar image and the use of an image location estimator that does not take into account these effects. $\langle\delta_{\kappa}\rangle_{G,D}$ corresponds to the mean location error for all transits at a particular $G$. It is measured in the {\gaia} operating conditions (window size and readout noise value) for the three reference image widths and with no background (dashed line) and a background level set to the averaged sky brightness (continuous lines). The left figure shows the results obtained for a trap density of 4~traps pixel$^{-1}$, representative of a CI delay of $\sim$27~s as demonstrated by the comparison to experimental results (crosses). The right figure shows the results obtained for a trap density of 1~trap~pixel$^{-1}$, representative of a CI delay of 1~s. The very small error bars correspond to the computed statistical uncertainty on the measured bias, $\upsilon_{\delta}$ (see Eq. \ref{eq:uncertBias}). The crosses represent the average relative location bias computed from experimental tests carried out during RC2, for a CI delay of $\sim$27 s and 1 s and the lowest background level. The large experimental uncertainties are not shown, but as can be noticed the overall trend and amplitude of the measured bias is well reproduced by our model.}
\label{fig:damagedBias}
\end{figure*}

Computing the Cram\'{e}r-Rao limit (Eq.\ \ref{eq:cr}) for a flux distribution including the CTI distortion and taking into account the charge loss allows to quantify this intrinsic uncertainty increase induced by the radiation damage.
As one can observe from Fig.~\ref{fig:damagedObservations} the CTI induced distortion sharpens the image profiles and renders them more asymmetric but the charge loss significantly decreases the signal-to-noise ratio. The latter effect prevails and, at a given $G$, causes an increase in the image location uncertainty. To generate $L_D$, the damaged flux distribution, we proceed in a similar fashion to the construction of $\tilde{L}$ (cf. Sections~\ref{sect:lsfModel} and \ref{sect:iterativeProcedure}). First we place each data point from the damaged observations at the right sub-pixel position to create an oversampled damaged profile.
Then the over-sampled profile is fitted by the special quartic spline so that we can use an analytical representation. 
The resulting minimum variances on the estimate of an image location acquired by a damaged CCD, and thus accounting for the CTI effects, are summarized in Table \ref{tab:cr} for different image widths, background levels and levels of radiation damage.

In the {\gaia} operating conditions, the relative intrinsic uncertainty increase (or accuracy loss) can be as large as 23\% (see Fig.~\ref{fig:crdamaged}) for the highest trap density and 6\% for the lowest. Here we recall that this drop in active trap density results from the use of a more frequent CI: from a CI period of 27~s to 1~s. 
In both cases, this increase is more pronounced for narrower stellar profiles and peaks at a signal level of $G = 15.875$. Then, due to the mitigating effects of the SBC at lower signal levels, one clearly observes a flattening of the uncertainty increase. This illustrates the critical importance of the two hardware mitigation tools (see Section~\ref{sect:damageSim}), which are the only mitigation countermeasures capable of reducing the CTI induced intrinsic loss of accuracy, by physically preventing the electron trapping and thus the image distortion and charge loss.

The Cram\'{e}r-Rao bound computed for the damaged flux distribution now constitutes the maximum achievable accuracy by any unbiased image location estimator in the presence of radiation damage. Although the loss of accuracy can be quite large, Fig.~\ref{fig:crdamaged} shows that the {\gaia} requirements would still be fulfilled, if an image location estimator that is bias-free and efficient enough can be elaborated (excluding the wide reference image and highest trap density case). In the next section, in order to assess the efficiency of the {\gaia} image location estimator without CTI effects mitigation (Section \ref{sect:imageParam}) in the presence of radiation damage, we directly apply it to the data set of damaged observations.

\subsection{Radiation induced image location bias}\label{sect:bias}

In this section we are interested in exploring the consequences of not accounting for the CTI effects during the image location estimation. We thus apply the {\gaia} image location estimator as presented in Section \ref{sect:imageParam} to the data set of damaged transits. In this case the image distortion shall not be accounted for in the LSF model construction. This is achieved by using $\tilde{L}_U$, the LSF model generated from the CTI-free transits. After applying the procedure, one eventually obtains an estimated location $\kappa_D$ for each transit, which after subtraction of the true image location $\kappa_{\mathrm{true}}$, gives us the error $\delta_{\kappa}$. After averaging for a particular magnitude and CCD operating conditions, we obtain the image location bias induced by the CTI effects as a function of signal level, $\langle\delta_{\kappa}\rangle_{G,D}$. The location bias results from the mismatch between the observed profile shape and the modelled LSF used to
estimate the location. Only one iteration of the scheme from Fig.~\ref{fig:iterative-procedure} is performed since the LSF model cannot be improved using the damaged counts without taking into account the CTI effects.

The results for different image widths, window sizes, and background levels are summarized in Table \ref{tab:bias} and depicted in Fig.~\ref{fig:damagedBias} (left) for a trap density of 4~traps pixel$^{-1}$ and in Fig.~\ref{fig:damagedBias} (right) for 1~trap~pixel$^{-1}$. The bias strongly varies as a function of $G$. In the {\gaia} operating conditions including background, the location bias reaches a maximum for $G =$~15. The mitigation effects of the SBC is clearly noticed for $15 < G < 18$ as the bias is either reduced or levels off. For $G > 18$, the background plays an important role in limiting the image distortion and reducing the bias as can be seen by comparing the dashed and solid lines. From these results we can conclude that in the presence of radiation damage, and without any attempt at any stage to correct or mitigate the CTI effects, the estimator is strongly biased. Indeed the image location can be shifted from a tenth of a pixel up to half a pixel for the fainter stars in the no-background case. When the estimator is applied to the damaged observations simulated with a background level set to the average sky brightness, the location bias is not as dramatic at low signal level. Nevertheless the image location bias for any image width and any signal level is constantly higher than $\sim$0.1 pixels in the 4~traps pixel$^{-1}$ case, which is not acceptable. Changing the telemetry window size has no significant effect on the location bias. As can be seen from Fig.~\ref{fig:damagedBias} (right), for a shorter CI delay (or CI period), and thus less active traps (here 1~trap~pixel$^{-1}$), the location bias is significantly lowered with a minimum level of $\sim$0.02 pixels. It is interesting to note that decrease in bias is scaled by the same factor ($\sim$ 4) as the decrease in trap density. Regarding the required performance, for the faintest magnitude this level of bias might be acceptable in a limited amount of cases (e.g., the bluest stars). However, in most cases, and especially for the bright stars this level of bias inevitably requires a software-based CTI mitigation scheme.

\subsection{Comparison with experimental data \label{sect:comparison}}

In order to check how representative the results obtained from synthetic data are in terms of the overall amplitude of the CTI effects and also fluctuation as a function of signal level, Fig.~\ref{fig:damagedBias} shows results obtained experimentally from RC2 \citep{astrium2008, swb2009}. In the experimental case the location bias does not correspond to an absolute image location bias since the true image location is by definition unknown. The presented bias is thus the relative location bias. It is computed by comparing the stellar transits over the irradiated part of the CCD and the same stellar transits over the non-irradiated part of the same CCD.
Taking into account the differences between real and synthetic data, as well as experimental uncertainties, the overall agreement between the results obtained from the RC2 and our simulations is remarkable. The combined mitigating effects of the SBC and the background are also noticeable in the test data at low signal levels. Hence not only the amplitude of the location bias for different CI delays (or densities of active traps) is reproduced by our model but also the overall bias evolution over a wide range of signal levels: 7 magnitudes. The simulations suggests that the illumination setup (and resulting PSF width) as well as slight differences in background light between experiments can have a significant impact on the measured CTI effects. This may explain observed discrepancies between the results from different RCs and within a RC.

\subsection{Damaged location estimation standard errors \label{sect:damagedLocStdErr}}
\begin{figure}
\centering
\includegraphics[width=0.49\textwidth]{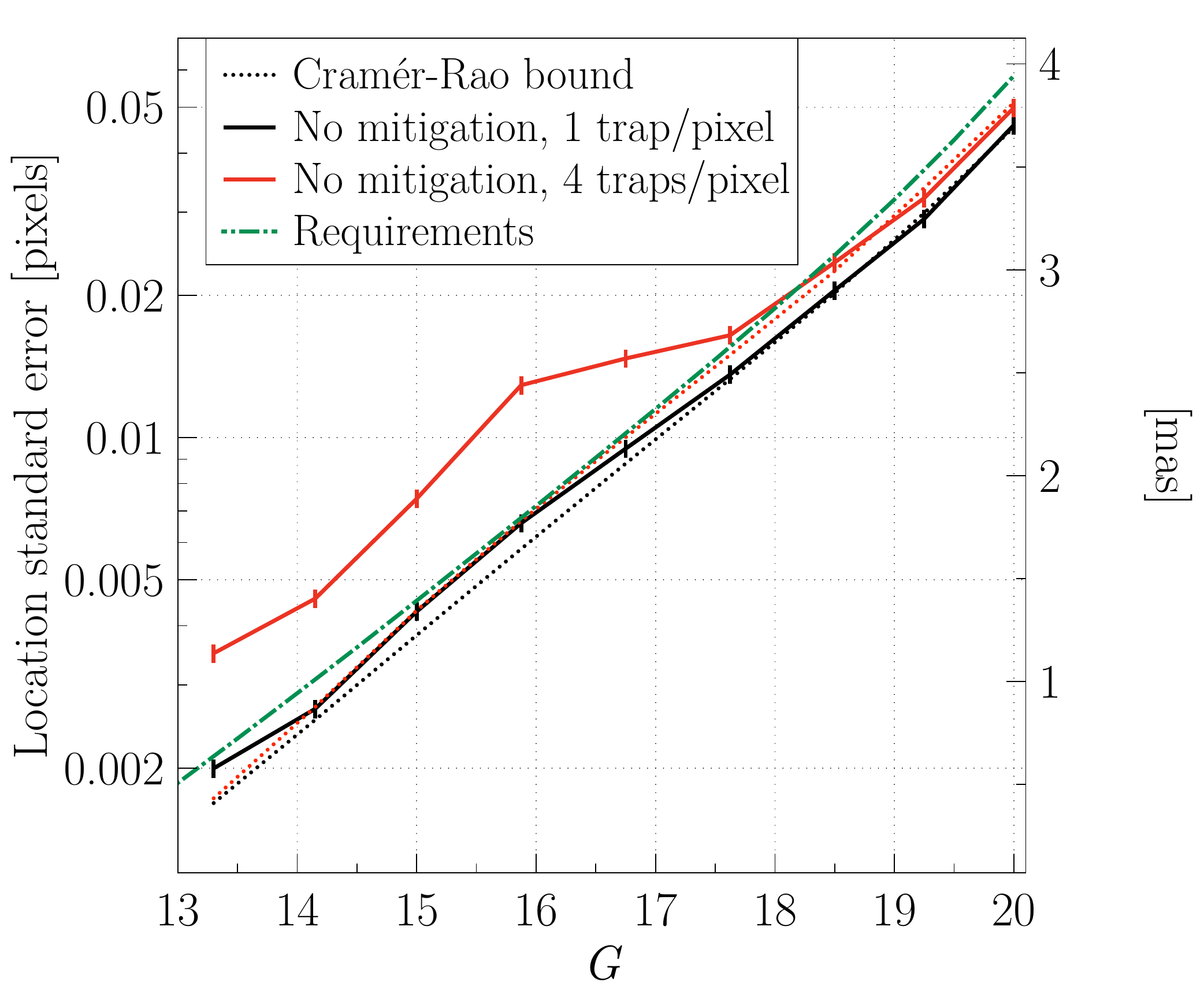}
\caption{Comparison between the Cram\'{e}r-Rao bound in the presence of radiation damage (dotted lines) and the standard errors (continuous lines) obtained by applying the {\gaia} image location estimation procedure to the set of damaged observations without any CTI mitigation (i.e. using the LSF model generated from the CTI-free observations). The Cram\'{e}r-Rao bounds were computed and the standard errors measured for the typical reference image considering the {\gaia} operating conditions and two different levels of radiation damage: 1~trap~pixel$^{-1}$ (black) and 4~traps pixel$^{-1}$ (red). In the presence of radiation damage, and without CTI effects mitigation, the {\gaia} image location estimator cannot be considered efficient anymore as its standard errors deviate significantly from the Cram\'{e}r-Rao bound (in addition to the estimator being biased c.f.\ Fig.~\ref{fig:damagedBias}).}
\label{fig:damagedStdErrors}
\end{figure}
Finally we show the resulting standard errors, $\sigma_{\kappa_D}$, as a function of $G$ in Fig.~\ref{fig:damagedStdErrors}: the standard errors are larger than the theoretical minimum variance, especially for intermediate magnitudes. For the most severe radiation level, the standard errors are larger than the requirements, and for the lowest radiation level the requirements are barely met; the mismatch between modelled and observed line spread function implies a broader spread in the locations estimated by the ML algorithm. This effect is less pronounced for the lowest level of radiation as the distortion, and thus the mismatch, is less important. The overall variance remains quite low as compared to the bias. Table \ref{tab:stdErr} summarizes these results for the three different image widths.

\section{CTI effects mitigation}\label{sect:mitigation}
Correcting for CTI is a complicated task and not only because the induced charge loss and distortion are considerable (Fig.~\ref{fig:damagedObservations}). The trapping probabilities \citep[e.g.,][]{prodhomme2011} depend on the electron density and thus the CTI effects vary with $G$. This variation is not linear, in particular due to the presence of the SBC that mitigates the CTI effects only at low signal levels. This can be clearly observed from Fig.~\ref{fig:damagedBias} from both simulations and experimental data. An important consequence is that the stellar core and wings (in the CCD serial direction) will not experience the same distortion. These different contributions to the global stellar image distortion will nevertheless be collapsed into a one-dimensional signal.
In addition, the location bias and charge loss will not be repeatable for a particular star or signal level as the CTI effects depend on the state (empty or filled) of the traps prior to the stellar transit. During the mission each star will transit on average $\sim$72 times over the focal plane of  {\gaia}.  For each of these stellar transits the scanning direction of the satellite will differ, and thus also the CCD illumination history that determines the trap state. It is also likely that the trap density will have increased between two consecutive transits. The CI will play an important role here, not only by decreasing the active trap density but also by simplifying the illumination history by reseting it every 1~s.
Finally, it is important to note that, as already mentioned, {\gaia}'s launch and first year of operations coincide with the predicted peak of the Sun's activity for the current solar cycle, and that none of the {\gaia} measurements will be free of radiation damage. This will strongly limit our knowledge of the exact instrument LSF/PSF in space.

\subsection{Potential alternative approaches}\label{sect:approaches}

\begin{figure}
\centering
\includegraphics[width=0.49\textwidth]{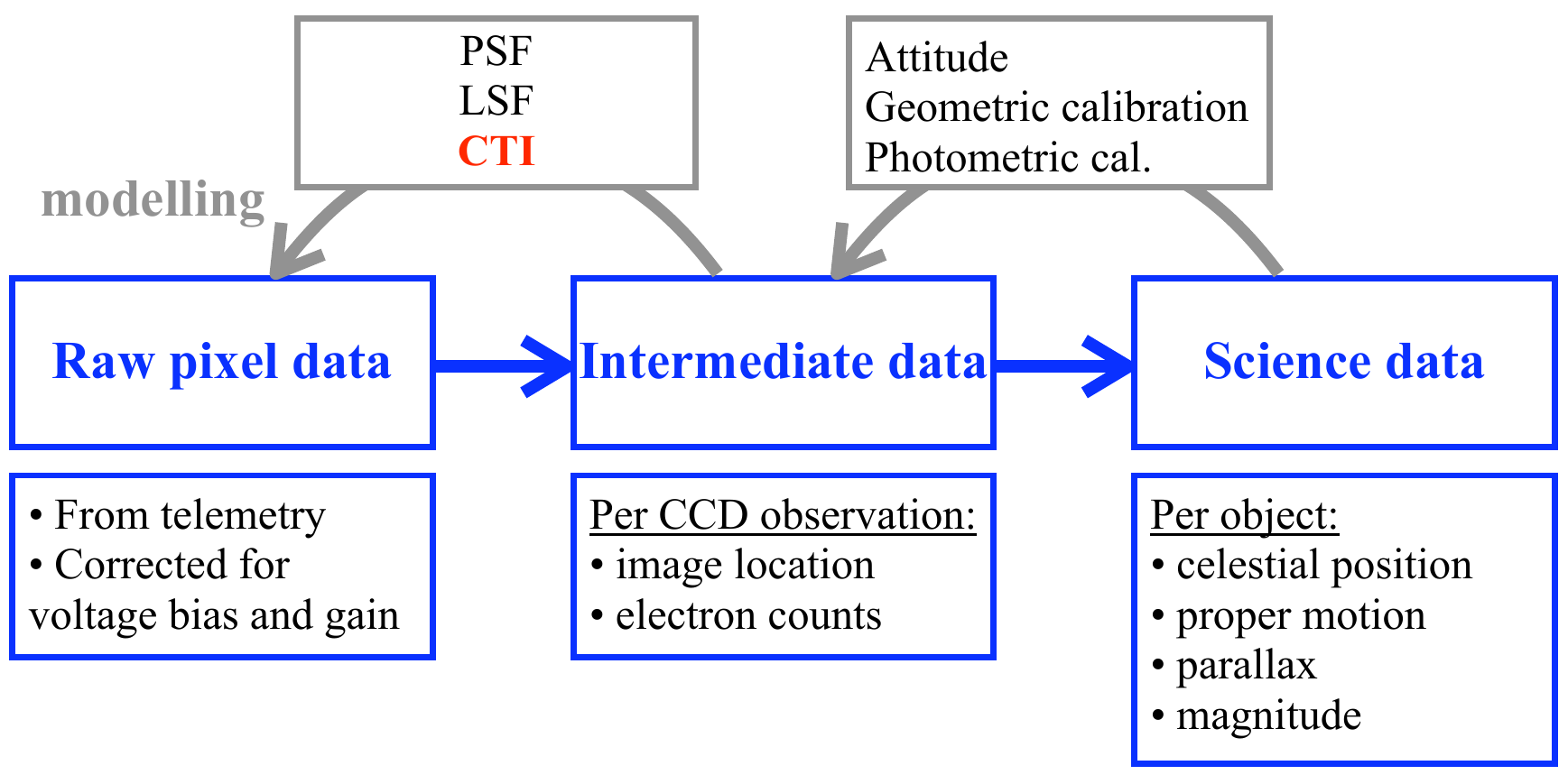}
\caption{From raw to science data: summary of the {\gaia} data processing chain in three stages (middle boxes). The top boxes contain the required models/assumptions to go from one stage to the other. The top arrows symbolize the feedback occurring at each stage which enables the iterative improvements of both models and data. The bottom boxes give some details about the data delivered/used at each stage. In principle, each set of data can be empirically corrected for CTI, however we choose not to perform any correction but to model the CTI distortion as part of the image parameter estimation procedure (see Section \ref{sect:fwd} and below).}
\label{fig:dataprocessing}
\end{figure}

Figure \ref{fig:dataprocessing} summarizes the {\gaia} data processing chain in three different stages at which a different set of data is available: (i) the raw data, (ii) the intermediate data, (iii) the science data. Each set of data is further explained in the figure. Different ways of handling the CTI effects in this chain are possible, and the literature provides us with a handful of correction procedures for photometric, spectroscopic, and (very rarely) astrometric measurements carried out in the optical or at X-Ray wavelengths.

\paragraph*{CTI correction at the level of the raw pixel data:}
one can correct the raw pixel data to obtain artificial CTI-free data and perform the rest of the data processing using the corrected raw data. This constitutes one of the most common approaches, and has been successfully used to correct the CTI effects on {\hst} data for instance. Its main advantage is that is minimizes the impact of CTI on the remaining data processing chain, the correction being performed very close to the source of the problem.
Either the photo-electron count correction is directly performed by means of a parametric empirical or semi-empirical formula \citep[e.g.,][]{goudfrooij2002, dolphin2009} that determines the CTI induced charge loss as a function of signal level, background, radiation dose, and source position on the CCD. Or it is performed by `comparing' the damaged observation to a simulated observation, for which the damage is simulated by an empirical or physically-motivated analytical forward model of the charge transfer and trapping \citep[e.g.,][]{bristow2003, massey2010, anderson2010}. \cite{bristow2003} provides a detailed comparison between direct empirical and model-based corrections: while the direct correction can only correct 
photometric and spectroscopic point source measurements, a model-based correction allows for astrometric correction of arbitrary complex sources (extended, binaries etc.). The latter is more complex, i.e. computationally intensive, but versatile and potentially more accurate. In principle, the model-based correction requires the generation of a synthetic undamaged observation to be subsequently distorted by the CTI model. However, as comprehensively described in \cite{massey2010}, and first proposed by \cite{bristow2005}, one can avoid this step and iteratively remove the CTI induced image distortion by subtracting actual and simulated observations, assuming, in a first step, that the actual damaged observation is the CTI-free input signal. This relies on the assumption that the CTI effects correspond to a slight perturbation around the true image. Although promising, the model-based correction of the raw data at the pixel level has only been tested against the empirical direct correction, and mostly for photometric and spectroscopic data. \cite{massey2010} go one step further and assess the astrometric correction induced shift as a function of signal level and distance from serial register. Although the correction performs as expected, the accuracy of such a method cannot be guaranteed yet due to the lack of reference or CTI-free data that prevents the measurement of the method absolute bias and standard errors. On top of this uncertainty regarding the final accuracy of this method, two other considerations preclude the direct use of this approach in the {\gaia} data processing before more investigations.
First, the noise properties of a corrected pixel value are no longer simple and may introduce hard-to-track effects
in the image location estimation procedure, and subsequently in the astrometric global iterative solution (AGIS)
that combines all observations to infer absolute astrometry for each observed object. In particular, the assumptions
on which the maximum likelihood estimation of the image parameters is based, namely that the individual samples
are statistically independent and described by the Poissonian model (Section~\ref{sect:estimator}), no longer hold
for the corrected samples. 
Secondly, the lack of full frame data and the binning of most telemetry windows implies that we lack the information required to perform a full pixel-based correction.

\paragraph*{CTI correction at the level of the intermediate data:}
at this level, the correction is performed thanks to a parametric ad-hoc model \citep[e.g.,][]{rhodes2007,schrabback2010}. It offers the advantage of being simple and fast to apply, and once formulated the model should be relatively simple to calibrate. However, the elaboration of such model is not trivial. It first requires a careful study of the CTI effects on the parameters extracted from the raw measurements as a function of a finite number of pre-selected variables. Subsequent to this study, the dependency of the CTI induced bias on the pre-selected variables must be mathematically described for each estimated parameter of interest. It is not guaranteed that such a mathematical formulation is possible and the resulting models have by definition no predictive power.
In the case of {\gaia}, the CTI-induced image location bias and charge loss could be parametrized as function of the signal level, background, radiation dose (or observation time), source position on the CCD, and illumination history (or time since last CI). Fig.~\ref{fig:damagedBias} shows an example of the image location bias dependence on the signal level and background. Comparison between Fig.~\ref{fig:damagedBias} left and right, also provides additional information about the dependence on the time since last CI. Such an approach was studied by EADS Astrium, but does not constitute the current baseline approach of the {\gaia} Data Processing and Analysis Consortium (DPAC) as it cannot handle complex scenes but only single stars. 

\paragraph*{CTI correction at the level of the science data:}
this last potential approach is the most impractical. It also requires a parametric ad-hoc model, most likely impossible to formulate as the CTI effects are too entangled at the level of the science data. Moreover, the calibration of such approach would require the use of reference data, which in the case of {\gaia} will be mostly not available.

\subsection{A complete forward modelling approach}\label{sect:fwd}
\begin{figure}
\includegraphics[width=0.49\textwidth]{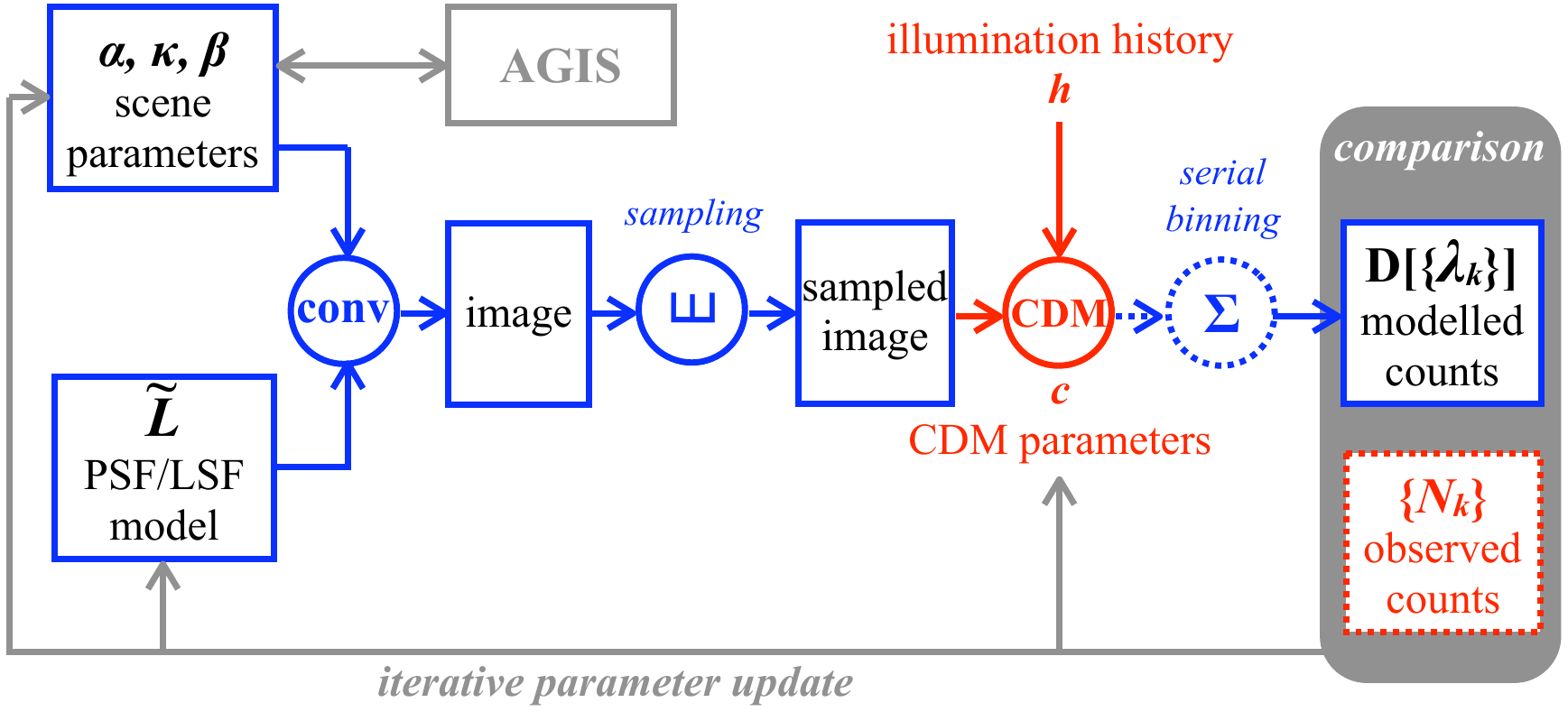}
\caption{Forward modelling approach to CTI mitigation \citep{lindegren2008}: a CTI-free sampled image is generated following the method explained in Section \ref{sect:imageParam} and is subsequently distorted by a fast analytical CTI effects model, so-called CDM (Charge Distortion Model). The distorted counts are then compared to the observed counts. In an iterative procedure, the scene, the LSF model, and the CDM parameters are successively improved. Note that if CDM takes as input a two-dimensional signal, the CDM output needs to be binned in the CCD serial direction before comparison to the observed counts. AGIS, the {\gaia} Astrometric Global Iterative Solution, uses the scene parameters to estimate the astrometric parameters. AGIS also provides an updated estimate of the scene parameters that is corrected for `nuisance' parameters such as the satellite attitude.}
\label{fig:forwardModelling}
\includegraphics[width=0.49\textwidth]{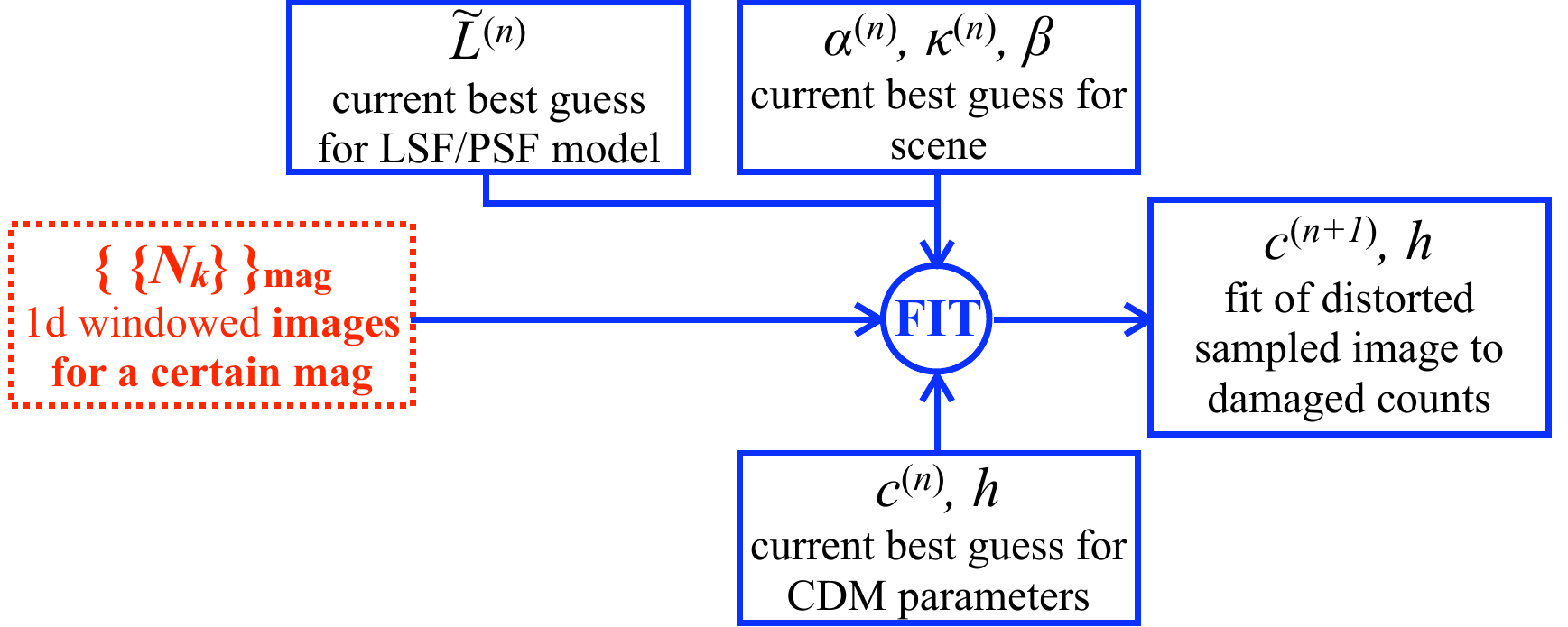}
\caption{CDM parameter estimation procedure: a set of CDM input signals is generated by sampling the current best LSF or PSF model using the best estimate of the scene parameters at a particular $G$. CDM simulates a set of damaged modelled observations that is subsequently compared to the set of observed counts corresponding to a particular $G$. A fitting algorithm provides us with a set of CDM parameters that optimizes the agreement between the CDM predictions and the observed counts.}
\label{fig:cdmparameter-update}
\end{figure}
Due to the complexity of the CTI effects and the extreme accuracy required in the image location estimation, as well as for the reasons mentioned above, the DPAC adopted a forward modelling approach. Thus in contrast to the solution applied to {\hst} data, no direct correction of the raw data shall be performed, essentially to preserve the simple noise properties and avoid arbitrary assumptions. Instead, the true image parameters are estimated in an iterative scheme, in which each observation is ultimately compared to a modelled charge profile for which the distortion has been simulated through an analytical CTI model, a so-called charge distortion model (CDM). This approach is illustrated by the schematic depicted in Fig. \ref{fig:forwardModelling}, where the modelled counts are now described as follows:
\begin{equation}
\lambda_k = D \left[ \alpha L\left( k - \kappa\right) + \beta \;|\; \vect{c}, \, \vect{h} \right]
\label{eq:distortion}
\end{equation}
with $D$ the CTI distortion applied to the sampled image using CDM, $\vect{c}$ a set of CDM parameters (e.g., trap species characteristics, electron density distribution parameter), and  $\vect{h}$ a set of parameters that describes the illumination history (the most obvious being the time since the last charge injection).

As illustrated in the Fig.~\ref{fig:forwardModelling}, the scene, the CDM, and the instrument (LSF/PSF) parameters are iteratively adjusted until the modelled counts $D \left[ \{\lambda_k\} \right]$ agree with the observation $\{N_k\}$. Fig.~\ref{fig:cdmparameter-update} gives the details of the CDM parameter update.
It is important to note that the model LSF cannot be directly generated from the observations anymore as they are now affected by CTI. During the mission, the LSF model will  thus be extracted from a LSF library composed partly by modelled LSFs and by a subset of observations: mostly the single bright stars that are the least affected by radiation damage (i.e. early mission data and/or observations close to a charge injection).
If the CDM and the instrument model are properly calibrated, the estimated scene parameters subsequently used to determine the stellar parallaxes should be unbiased and free of CTI. Since no direct correction is performed the noise properties of the observation should remain dominated by the photon and readout noise and thus a complex contamination of the rest of the data processing chain and its products is avoided. This approach can handle arbitrarily complex scenes and offers the advantage being in accordance with the general {\gaia} data processing principle of self-calibration.
A similar approach was successfully used to handle CTI effects on photometric and spectroscopic X-ray measurements performed by {\chandra} \citep{townsley2000, townsley2002, grant2004}.

In the following (Section \ref{sect:idealFwdTest}), we demonstrate the ability of the {\gaia} CTI mitigation approach to reach the best achievable image location estimation accuracy for damaged observations (Section \ref{sect:damagedCR}) in the case of an ideal CDM, and ideally calibrated LSF and CDM parameters. Then we assess the actual performance of this approach regarding the recovery the image location estimate bias (Section \ref{sect:locBiasRecovery}) and image flux estimate bias (Section \ref{sect:fluxBiasRecovery}), using the current best CDM candidate \citep{short2010}.

\subsection{Testing the forward modelling approach}\label{sect:idealFwdTest}
\begin{figure}
\centering
\includegraphics[width=0.49\textwidth]{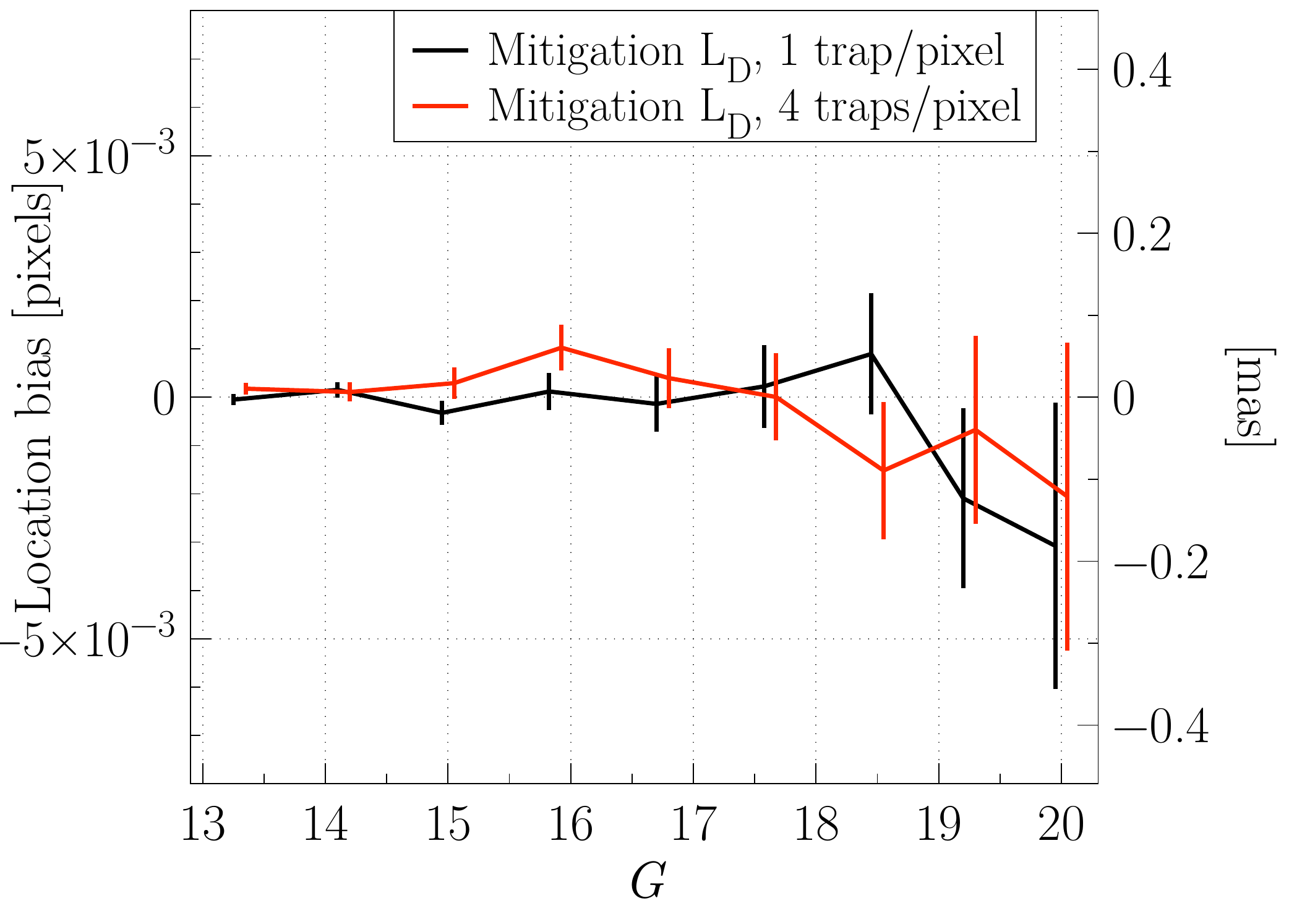}
\caption{Location bias $\langle\delta_{\kappa,D}\rangle \pm \upsilon_{\langle\delta_{\kappa}\rangle}$ after a CTI mitigation using the LSF model $\tilde{L}_D$, representative of an ideally calibrated forward modelling approach. These results are obtained in the {\gaia} operating conditions, considering the typical reference image only and for two different levels of radiation damage: 1~trap~pixel$^{-1}$ (black) and 4~traps pixel$^{-1}$ (red). As one can see, in these conditions, the forward modelling approach allows for a full recovery of the CTI-induced location bias. Note that for readability a slight offset on the $G$~axis as been introduced between the results for the two trap densities.}
\label{fig:idealMitigBias}
\includegraphics[width=0.49\textwidth]{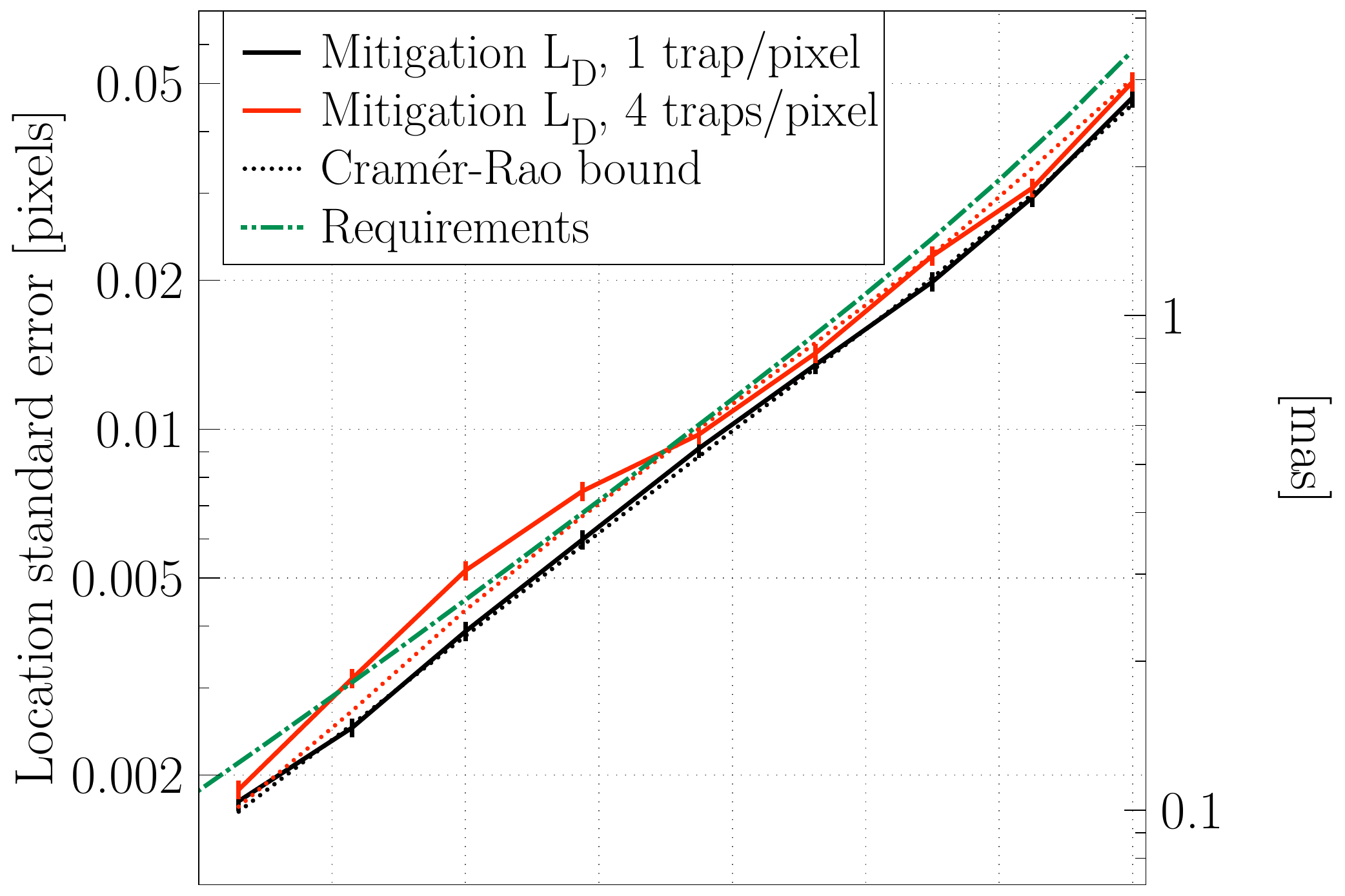}
\includegraphics[width=0.49\textwidth]{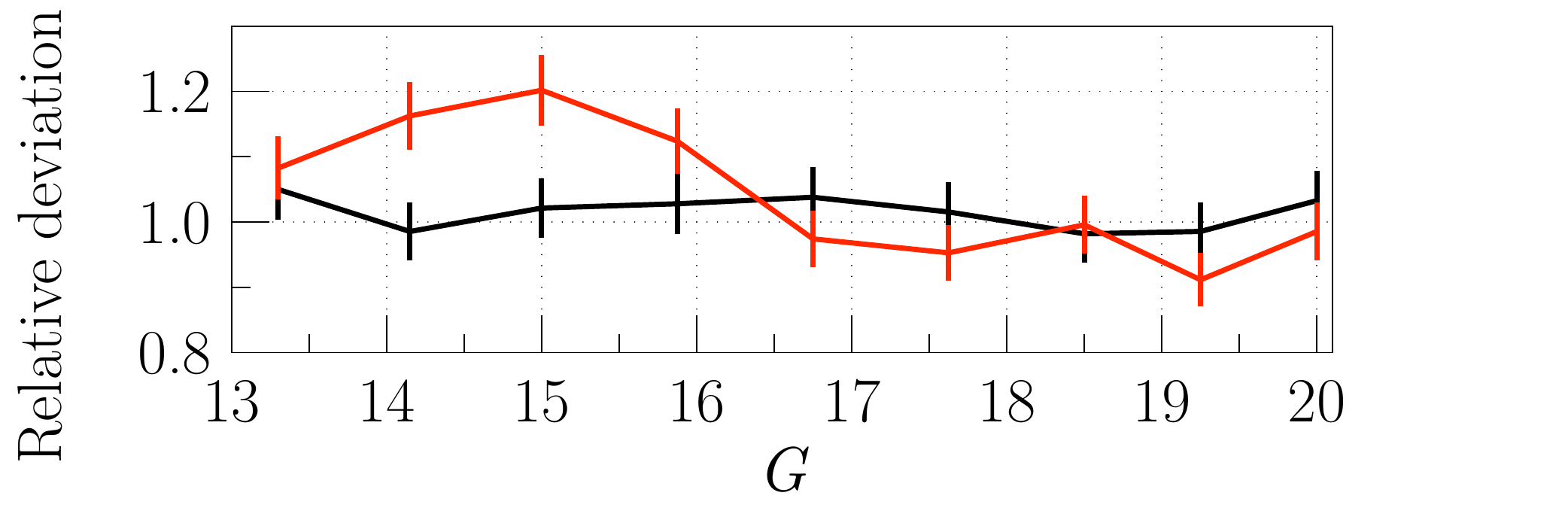}
\caption{Comparison between the Cram\'{e}r-Rao bound in the presence of radiation damage (dotted lines) and the measured standard errors (continuous lines) when applying the {\gaia} image location estimation procedure including CTI mitigation to the set of damaged observations. These results are obtained in the {\gaia} operating conditions, considering the typical reference image only and for two different levels of radiation damage: 1~trap~pixel$^{-1}$ (black) and 4~traps pixel$^{-1}$ (red). The CTI mitigation is performed following the forward modelling approach, using $\tilde{L}_D$ (i.e. and ideally calibrated CDM and LSF model). In these conditions the estimator standard errors are below the requirements (green dashed and dotted line) even for the most severe damage (top). The relative deviation from the best achievable accuracy (bottom) is reasonable (below 10\%) in the 1~trap~pixel$^{-1}$ case but can reach 20\% in the 4~traps pixel$^{-1}$ case for the intermediate magnitudes.}
\label{fig:idealMitigStdErr}
\end{figure}
In a first step towards a more complete validation of our approach, we would like to ensure that this approach, if perfectly calibrated, enables an unbiased estimation of the image location with high enough precision.
To do so we estimate the (unknown) scene parameters for the set of damaged observations, in the case of an ideal CDM and ideally calibrated LSF and CDM parameters. This ideal case is simulated by using $\tilde{L}_D$, the damaged LSF (cf. Section \ref{sect:damagedCR}).
This is allowed because in this scheme the true LSF and CDM parameters correspond to a model that is capable of fully explaining the image distortion and the charge loss in the damaged observations. 

Figures \ref{fig:idealMitigBias} and \ref{fig:idealMitigStdErr} show the location bias and the estimator standard errors obtained in these conditions, for the two different levels of radiation damage, and for the typical reference image. The results obtained for the two other reference images can be found in Tables \ref{tab:bias} and \ref{tab:stdErr}. As one can see, in the case of ideal CTI mitigation, the location bias in the presence of radiation damage is now comparable to the one obtained for the CTI-free observations (see Fig.~\ref{fig:location_bias-vs-magnitude}); the bias does not exceed 5 milli-pixels and does not significantly deviate from zero within the error bars ($\upsilon_{\langle\delta_{\kappa}\rangle}$, the statistical uncertainty), and this even for the most severe level of damage. Regarding the estimator standard errors, they comply with the {\gaia} requirements, even in the case of the most severe level of damage for most of the magnitudes. The bottom part of Fig.~\ref{fig:idealMitigStdErr} shows that the location estimator including CTI mitigation performs efficiently for the lowest level of damage (i.e. less than 10\% relative deviation). However it is interesting to note that even in this favourable case (ideally calibrated LSF model and CDM parameters), the relative deviation of the estimator precision from the best achievable one can reach 20\% for the intermediate magnitudes and the strongest level of damage.

From these results we can conclude that a forward modelling approach to CTI mitigation, as presented in the previous section, allows the recovery of the CTI-induced location bias and enables the bias-free estimation of the image location at the required precision, close to the theoretical limit. This level of performance is achieved in the favorable conditions of a very good LSF model and the CDM parameter calibration, but for the strongest expected image distortion in the {\gaia} operating conditions, i.e. stars located the furthest away from the last CI and a density of traps equivalent to the predicted upper limit to the {\gaia} end-of-life accumulated radiation dose.

\subsection{Current best CDM candidate}\label{sect:cdm}

The elaboration and calibration of a CDM that allows to reach the level of performance presented in Section \ref{sect:idealFwdTest} is challenging. The presented mitigation scheme requires a CDM that must be both accurate and fast, as the iterative procedure is performed for each observation, and the CDM distortion applied at each iteration. The DPAC strategy regarding the elaboration of such CDM is detailed by \cite{vanLeeuwen2007a} and \cite{vanLeeuwen2007b} and a short summary is given by \cite{prodhomme2010b}. In this study, to demonstrate the validity of our CTI mitigation approach including a CDM and thus assess its present actual performance, we use the current best CDM candidate (later referred as to CDM for simplicity) as it is described in \cite{short2010}, and for which a first comparison of its outcomes to experimental test data is presented in \cite{prodhomme2010a}.

CDM is based on the common Shockley Read Hall formalism \citep{shockley52, hall52} and describes the capture and release processes in a statistical way. To cope with the computational speed requirement, it suppresses the treatment of the numerous charge transfer steps required to transfer the signal from one CCD end to the other, but computes the signal transit in a single calculation making use of several assumptions \citep{short2010}. CDM is able to simulate the CTI effects in TDI and imaging mode for any kind of signal (single, double stars, spectrum etc.). The CDM free parameters are: $\gamma$ which determines how the volume of the electron packet grows as electrons are added, $\beta$ the background light (respectively denoted $\beta$ and $S_{dob}$ in \cite{short2010}, but changed herein for disambiguation), and three trap parameters per trap species, $\rho$, $\sigma$ and $\tau$, respectively the trap density, the capture cross-section, and the release time constant. 
It has to be noted that a more recent version of this model has been elaborated. This newer version incorporates a better handling of the charge injection modelling and the possibility of simulating the serial CTI that occurs in the readout register. As charge injections are not explicitly simulated in our synthetic data set and the serial CTI was not simulated, the hereafter demonstrated performances remain representative of the current performances of our mitigation scheme.

\subsection{The forward modelling approach initialization} \label{sect:initialization}

The iterative image parameter estimation procedure including CTI mitigation now involves three different sets of parameters to be successively improved: the scene, the LSF/PSF, and the CDM parameters (see Fig.~\ref{fig:forwardModelling}). Reaching a stable solution in these conditions is complex; each set of parameters needs to be initialized with values not too far off from the `true' ones for the iterative procedure to converge. 

\paragraph*{LSF model:} as already mentioned, the LSF model cannot be generated from the damaged observations, as they are not directly representative of the instrument anymore. During the mission the LSF model will partly be generated using the least damaged observations of single bright stars. In the following we thus use $\tilde{L}_U$, the LSF model generated from the CTI-free observations at a particular $G$. This constitutes a favourable yet realistic case.

\paragraph*{Scene parameters:} the initial estimate for the image location $\kappa^{(0)}$ is determined using the Tukey's Biweight centroiding algorithm, the initial flux estimate $\alpha^{(0)}$ corresponds to the sum of the observed counts after background subtraction (and as in the rest of the study the background $\beta$ is considered to be known).
Hence the initial location and flux estimates are biased by the CTI effects. However one should note that due to its construction the LSF model contains some information about the true location of the observations in its zero-point. This is still reasonable as we have so far ignored that during the mission the astrometric solution (AGIS) will provide extra information about the true location of each observation through a feedback mechanism (see Fig.~\ref{fig:forwardModelling}).

\paragraph*{CDM parameters:} it is first important to realize that several fundamental differences exist between CDM and the detailed Monte-Carlo CTI effects model that we used to simulate the damaged observations: the most important ones being related to the charge transfer simulation, the computation of the capture and release probabilities, and the modelling of the electron density distribution. Hence, in this context, no `true' CDM parameters exist but only CDM parameters that allow the reproduction of the simulated damaged observations.
This actually constitutes a similar situation to the one that will be experienced during the operation of {\gaia}. Indeed, due to the simplifications intrinsic to the elaboration of a fast analytical model of a complex phenomenon, even with the right parameterization, the agreement between the damaged observations and the CDM predictions will not be perfect.

Furthermore: although the trapping occurs during the transfer of two-dimensional stellar images, only one-dimensional information is accessible from the binned observations. The CDM distortion can be applied to a one- or a two-dimensional CTI-free signal. In the latter case, one needs to reconstruct a PSF and the resulting modelled counts must be binned prior to a comparison with the observed damaged counts. In our study we generally obtained a significantly better agreement between CDM predictions and the damaged observations by applying the CDM distortion to a one-dimensional signal. In the following we thus only present results obtained in this case. In reality this might be different, in particular due to the serial CTI that was not taken into account here. Yet, if a comparable performance level can be achieved, the one-dimensional option would still be preferred during the mission for the 1D binned data as it presents the advantage of saving a significant number of computations.
\begin{figure}
\centering
\includegraphics[width=0.49\textwidth]{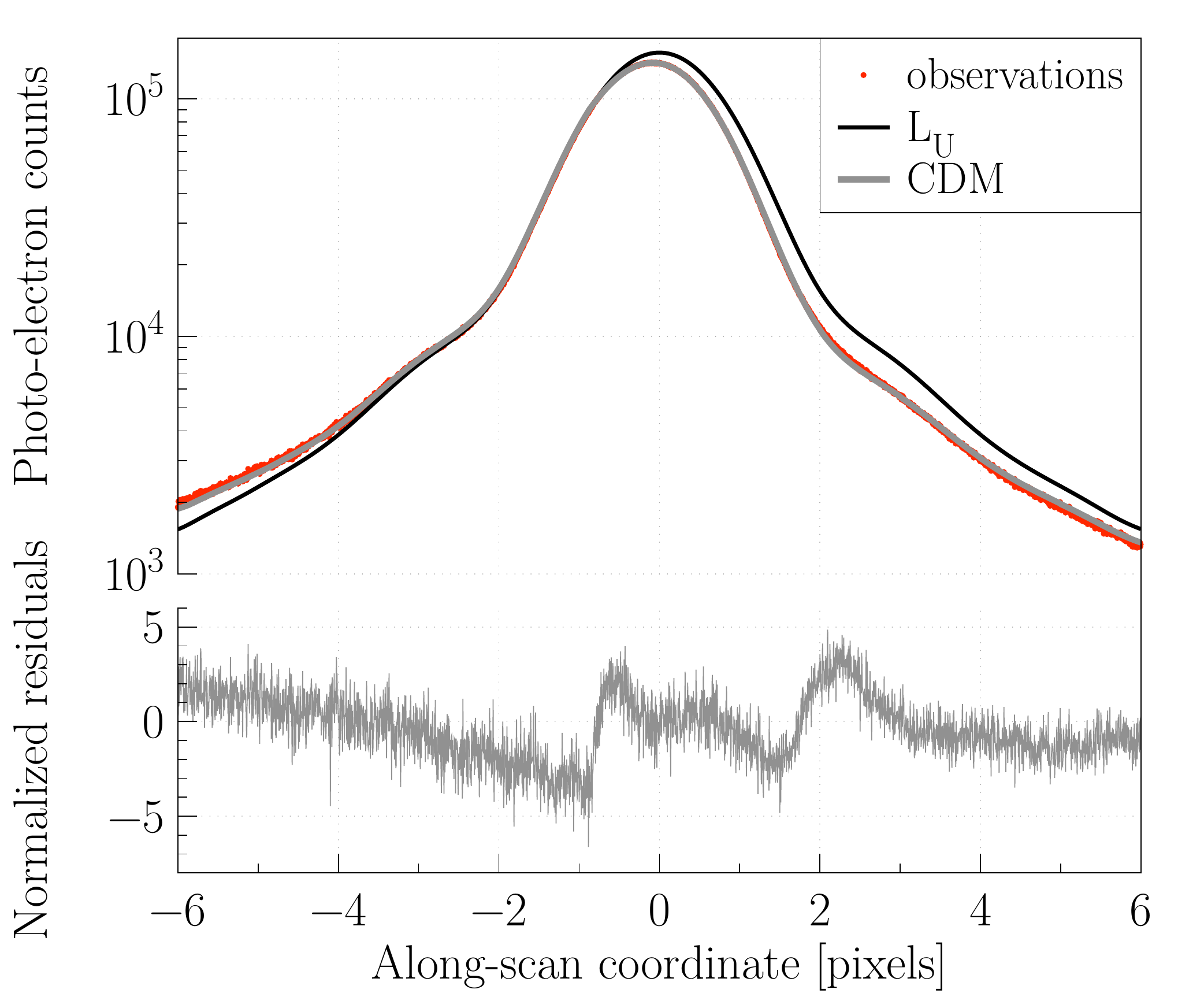}
\caption{Top: Comparison between the CDM predictions after initialization of the CDM parameters (grey) and the damaged observations $\{\{N_k\}\}_{13.3}$ (red dots) under the {\gaia} operating conditions and for a radiation level of 4~traps pixel$^{-1}$. To enable the observation of the CTI induced distortion the input signal $\tilde{L}_U$ is also shown (black line).
\newline Bottom: Residuals normalized by the photon noise, the reduced $\chi^2$ is $\sim$3.0.}
\label{fig:cdmResiduals}
\end{figure}
To obtain an initial set of CDM parameters that describes reasonably well the damaged observations, we use $\tilde{L}_U$ as input signal, and fit the CDM predictions to the damaged observations $\{\{N_k\}\}_{G}$ for a particular $G$ and set of operating conditions (i.e. windowing scheme, background and readout noise level). The fitting procedure minimizes the $\chi^2$ (Eq. \ref{eq:sampleChiSquare}) between the CDM predictions and the damaged observations. The fitted parameters are $\gamma$, $\rho$, $\sigma$, and $\tau$ (see Section~\ref{sect:cdm}), $\beta$ is fixed to the true value.
At this stage, the fitting procedure is an evolutionary algorithm\footnote[1]{\url{http://watchmaker.uncommons.org/}} that uses two mechanisms, mutation and cross-over. It is applied on an initial population of 100,000 parameter sets and evolves towards smaller $\chi^2$ generation after generation. After 10 generations, we select the set of parameters with the smallest $\chi^2$. This set of parameters can be further improved by using the downhill simplex minimization method \citep{nelder1965}. 
Fig.~\ref{fig:cdmResiduals} gives an example of the obtained agreement between the CDM outcomes and the damaged `observations' (generated with the Monte Carlo model described in Section~\ref{sect:observations}) at a particular value of $G$. This example is representative of the best level of agreement achieved after applying the described initialization procedure.
The illumination history parameters, $\vect{h}$, will be fixed to the reconstructed illumination history. Here $\vect{h}$ is only the time since last CI that is set to infinity as no CI has been explicitly simulated. The effect of not calibrating for disturbing stars, i.e. stars located between the last CI and the star of interest, will be studied in the second part of this study. It can however already be mentioned that stars located between a CI and the star of interest are only disturbing if they are located in the same pixel column (or an adjacent one) and that, for a CI period of 1~s, the number of disturbing stars is expected to be very low even for the densest parts of the sky \citep{holl2011}.

\subsection{Image location bias and accuracy recovery}\label{sect:locBiasRecovery}

\begin{figure}
\includegraphics[width=0.49\textwidth]{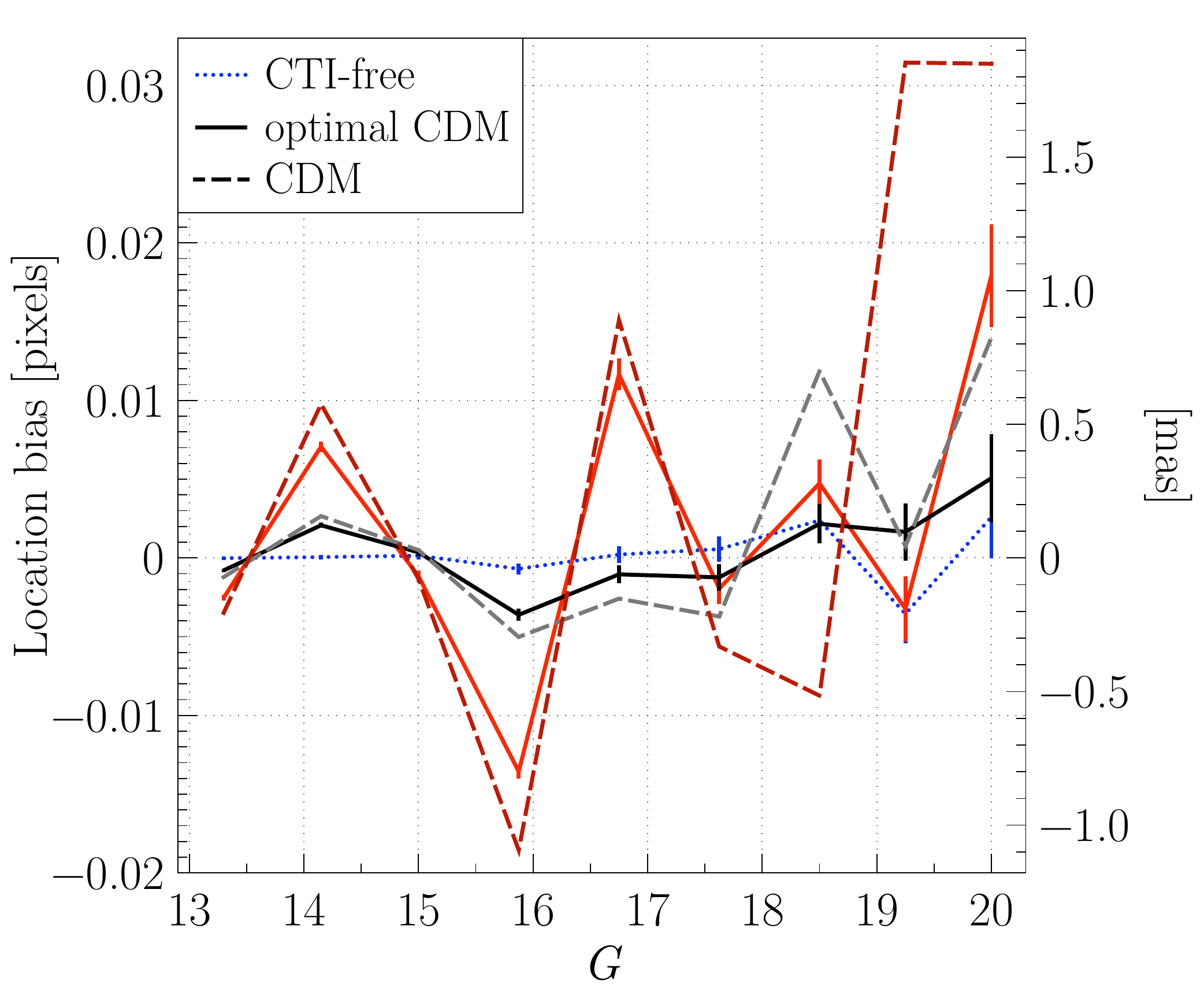}
\caption{Image location bias recovery using the CTI mitigation forward modelling approach including the current best CDM candidate. These results are obtained in the {\gaia} operating conditions, considering the typical reference image only, for two different levels of radiation damage (1~trap~pixel$^{-1}$ in black, and 4~traps pixel$^{-1}$ in red), and for two different initial optimizations of the CDM parameters: fully optimized (continuous line) and iteratively improved (dashed line). For comparison the average location bias measured for the CTI-free observations is also shown (blue dotted line). The CDM parameters are calibrated per magnitude and as a result there is a different set of CDM parameters for each magnitude and the resulting final agreement between modelled and observed counts varies from one signal level to the other. This explains the bias oscillations (in particular for the highest trap density).}
\label{fig:biasRecovery}
\includegraphics[width=0.49\textwidth]{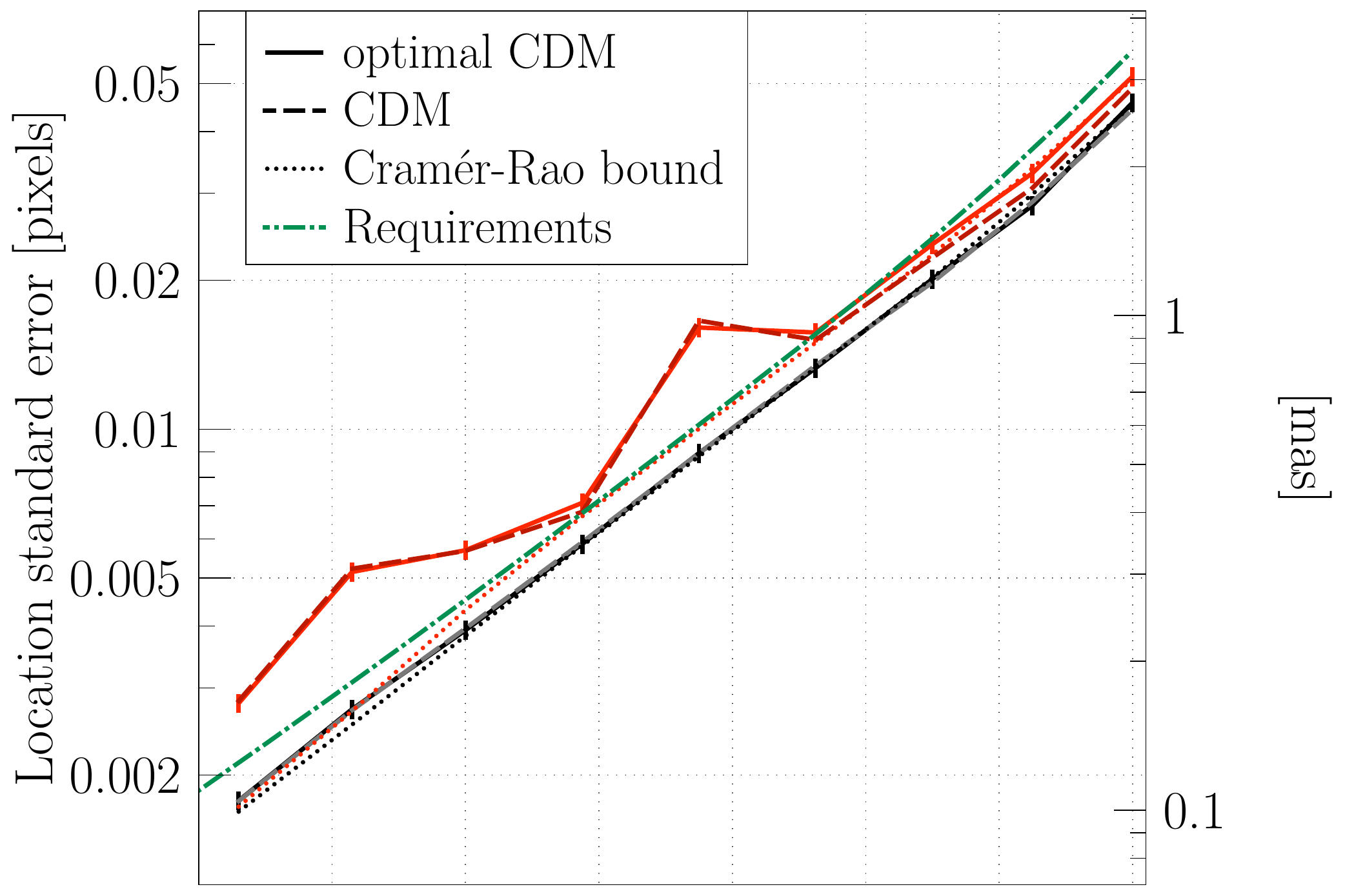}
\includegraphics[width=0.49\textwidth]{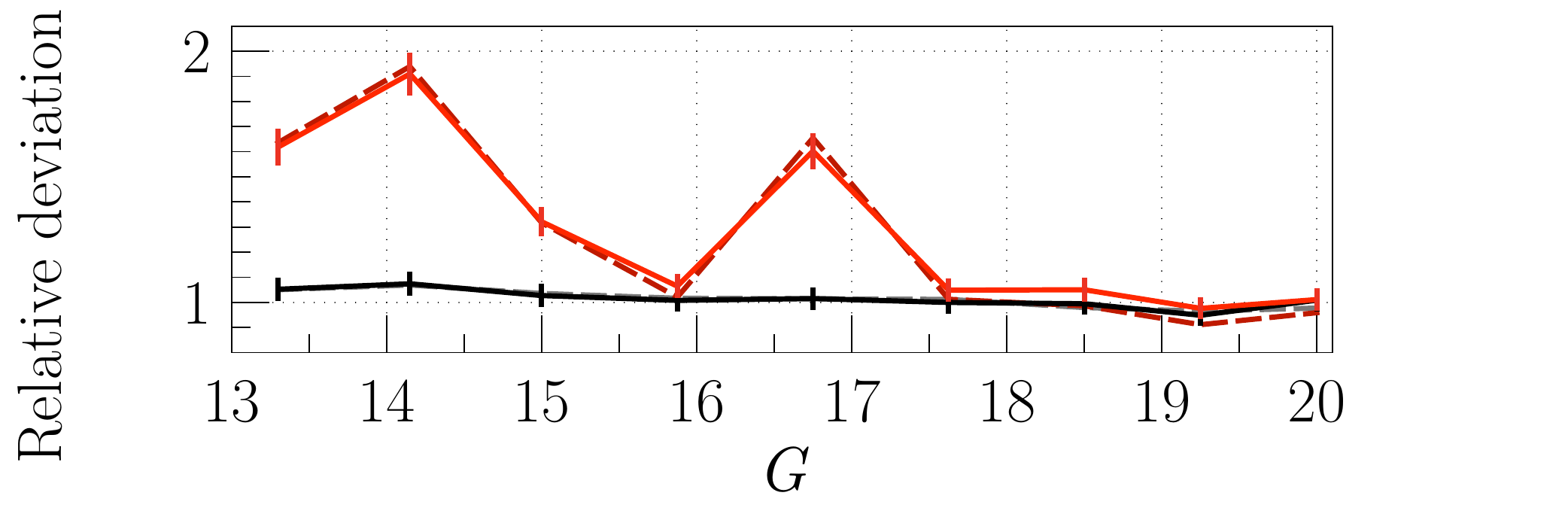}
\caption{Image location standard errors resulting from the use of the CTI mitigation forward modelling approach including the current best CDM candidate for two different levels of radiation damage (1~trap~pixel$^{-1}$ in black, and 4~traps pixel$^{-1}$ in red), and for two different initial optimization of the CDM parameters (fully optimized as a continuous line, and iteratively improved as a dashed line). The lower panel shows the relative deviation from the best achievable accuracy. The Cram\'{e}r-Rao bound in the presence of radiation damage and the {\gaia} requirements are also shown (respectively the dotted lines and the green dashed and dotted line).}
\label{fig:accuracyRecovery}
\end{figure}

As mentioned in the previous section, after initialization, the CDM parameters can still be further improved by the use of the downhill simplex method. In the following we thus distinguish between two different cases: (\textit{a}) the CDM parameters have been fully optimized and the scene parameters are then estimated, no more iterations are performed, (\textit{b}) the CDM parameters have not yet been fully optimized and are refined as part of the image location iterative procedure (Fig.~\ref{fig:forwardModelling}). 
In the latter case, once each set of parameters is initialized, the iterative procedure is performed as follow: (i) the scene parameters are first estimated using $\tilde{L}_U$ and the CDM parameters $\gamma^{(0)}$, $\rho^{(0)}$, $\sigma^{(0)}$, $\tau^{(0)}$ and $\beta$, then (ii) the CDM parameters are updated as presented in Fig.~\ref{fig:cdmparameter-update} using the newly estimated scene parameters, and (iii) a new scene parameter update is performed.
In our study, the CDM parameter update (Fig.~\ref{fig:cdmparameter-update}) is performed using the downhill simplex minimization method \citep{nelder1965} only; as we shall see it proves to be quite inefficient at this stage of the procedure. A maximum likelihood based procedure would be better suited and is currently being developed to perform this task in the {\gaia} data processing.  

Figure~\ref{fig:biasRecovery} shows the remaining image location bias after using the CTI mitigation forward modelling approach including the current best CDM candidate for the two different initial optimizations of the CDM parameters. These results should be compared to Fig.~\ref{fig:damagedBias} that shows the location bias when no CTI mitigation is applied, and to Fig.~\ref{fig:idealMitigBias} that shows the ideal performance of the presented mitigation scheme. All these results are summarized in Table \ref{tab:bias} for the three different reference images. The current best CDM candidate does not allow for a total recovery of the location bias at each magnitude, however this bias is considerably reduced for both levels of radiation damage. For instance, in the case of the lowest trap density and an optimal CDM optimization, the bias does not exceed the level of 0.005 pixels, while without any mitigation and in the same conditions the bias reaches 0.05 pixels (see Fig.~\ref{fig:damagedBias}). Fig.~\ref{fig:biasRecovery} shows that at the faint end the bias is significantly lowered by a better optimization of the CDM parameters, and that the CDM parameter update performed during the iterative procedure in these conditions is too limited to recover the full potential of CDM. This means that the CDM initialization is a crucial step of the CTI mitigation scheme, especially at faint magnitudes, and if the CDM parameters are iteratively refined, the simplex method seems not to be efficient enough. The bias oscillations as a function of $G$ are due to the fact that the CDM parameters have been calibrated per magnitude. As a result a different level of agreement between observed and modelled counts is achieved for each signal level.

Figure~\ref{fig:accuracyRecovery} shows the measured location standard errors for the typical reference image (see Table \ref{tab:stdErr} for the other reference images). As expected (see Fig.~\ref{fig:idealMitigStdErr}) for the most severe level of radiation damage, the {\gaia} requirements are not met for the bright magnitudes, and the relative deviation from the best achievable accuracy is large: it almost reaches 100\% at $G=14.15$. However when considering the more realistic trap density of 1~trap~pixel$^{-1}$, the standard errors are safely below the requirements, and the relative deviation from the Cram\'{e}r-Rao bound remains below 10\%. In these conditions our estimator thus remains efficient, even if, as already explained, biased. Finally, it has to be noted that the overall precision of the location estimation seems to be quite insensitive to the fine tuning of the CDM initial parameters.

\begin{figure}
\includegraphics[width=0.49\textwidth]{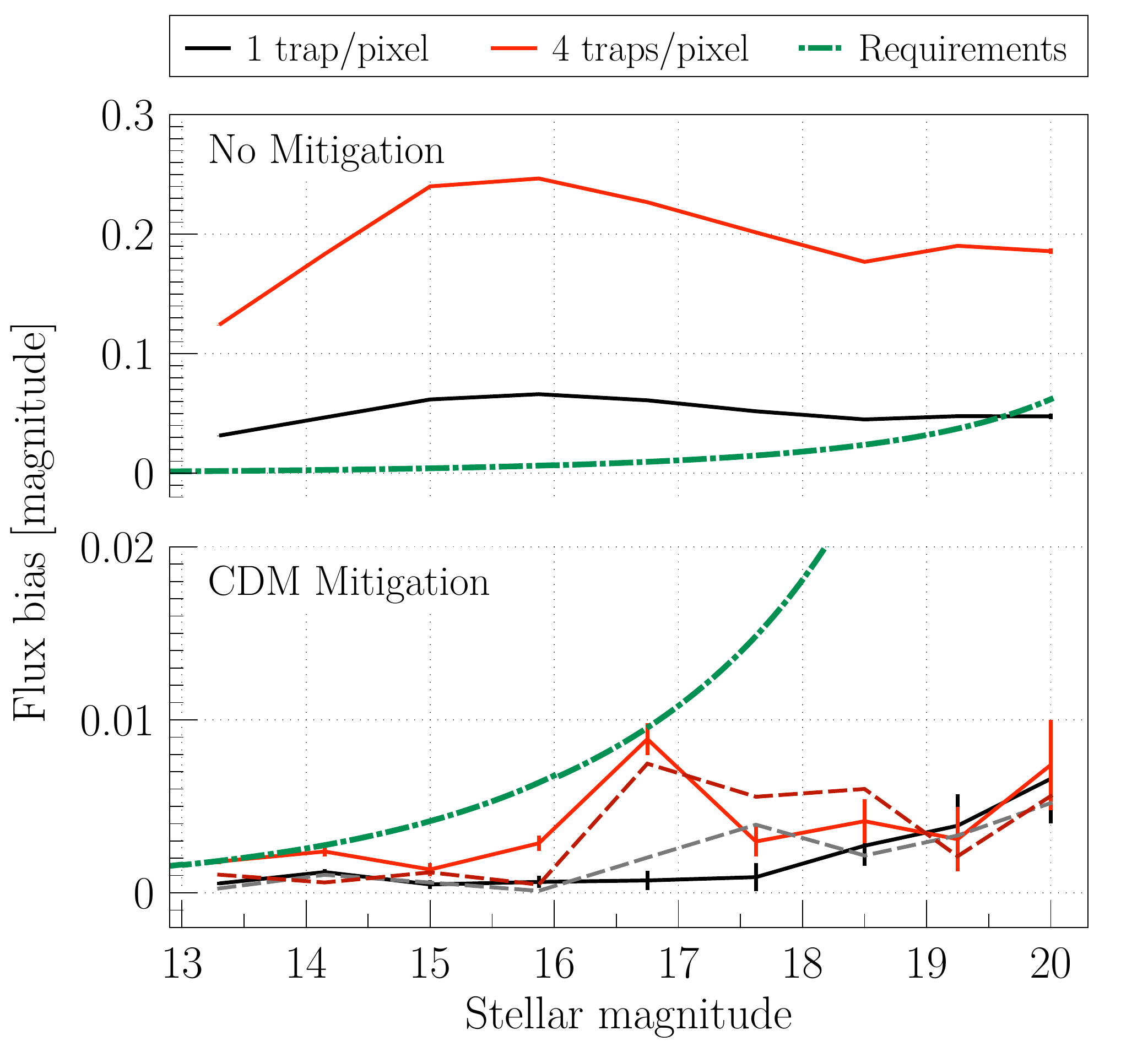}
\caption{Comparison between the image flux estimation bias (in units of magnitude) before (top) and after CTI mitigation (bottom) in the {\gaia} operating conditions considering the typical reference image for two different levels of trap density: 1~trap~pixel$^{-1}$ (black) and 4~traps pixel$^{-1}$ (red). These results are compared to the {\gaia} requirements \citep{jordi2010} for a transit across a single CCD (green dashed and dotted line). In the case of the CTI mitigation (bottom) we considered two different initial optimizations of the CDM parameters: fully optimized (continuous lines) and iteratively improved (dashed lines).}
\label{fig:fluxbiasRecovery}
\end{figure}

\subsection{Image flux bias recovery}\label{sect:fluxBiasRecovery}

This study focused on the estimation of the image location parameter as it is the most critical image parameter to be determined for astrometry.
However the accurate estimation of the integrated image counts is also important
as this forms the basis for the $G$-band photometry. The photometry, when
combined with the parallax measurements, will provide the absolute luminosities
of the stars observed by {\gaia}. In addition to this fundamental parameter the
multiple observations of each source constitute an all sky variability survey,
providing another treasure trove of astrophysical information. For both
applications high photometric accuracy, complementary to the astrometric
accuracy is required. We refer to \cite{jordi2010} for more details on
{\gaia}'s photometric capabilities.
As can be observed from Fig.~\ref{fig:damagedObservations}, CTI induces not only an image distortion but also an important charge loss that, if not properly taken into account, biases the image flux estimation. In order to judge the capability of our approach to CTI mitigation to achieve high photometric accuracy we show in Fig.~\ref{fig:fluxbiasRecovery} the image flux estimation bias (in units of magnitude) achieved with and without CTI mitigation using CDM. These biases are compared to the photometric performance predictions in \cite{jordi2010} including the intrinsic loss in photometric precision induced by CTI \citep{jordi2010}, where the numbers in their figure 19 have been translated to the photometric errors expected for a transit across a single CCD. The `safety margin' in equation (6) in \cite{jordi2010} was omitted.

Despite the rather strong bias induced by CTI in the image flux estimation (see Fig.~\ref{fig:fluxbiasRecovery} top), the presented CTI mitigation scheme allows to eliminate most of it (see Fig.~\ref{fig:fluxbiasRecovery} bottom). It thus allows an unbiased estimation of the image flux within the requirements if the estimation procedure is precise enough. In this context the current CDM performances are remarkable: for the highest trap density and when no mitigation is applied, the flux bias can reach $\sim$0.25 magnitudes at $G=15.875$, after mitigation we measure a flux estimation bias of 0.0029 magnitudes for the same magnitude, well below the requirement of 0.0067 magnitudes. It is also interesting to note that the image flux estimation is much less sensitive to the calibration of CDM than the image location estimation. Indeed Fig.~\ref{fig:fluxbiasRecovery} (bottom) shows that a similar level of performance is obtained for a fully optimized or an iteratively improved CDM calibration.

\section{Discussion}\label{sect:discussion}

\begin{figure}
\includegraphics[width=0.49\textwidth]{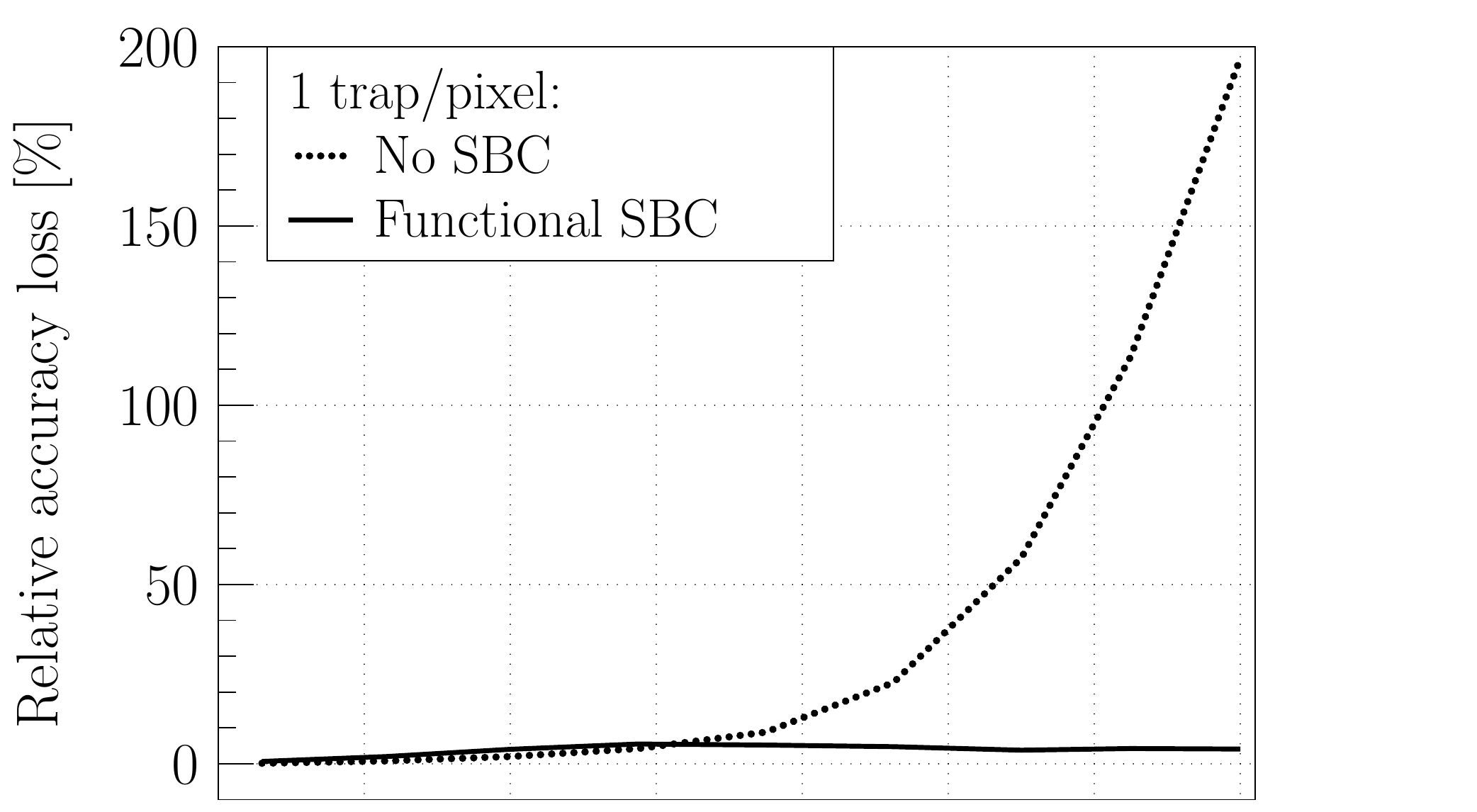}
\includegraphics[width=0.49\textwidth]{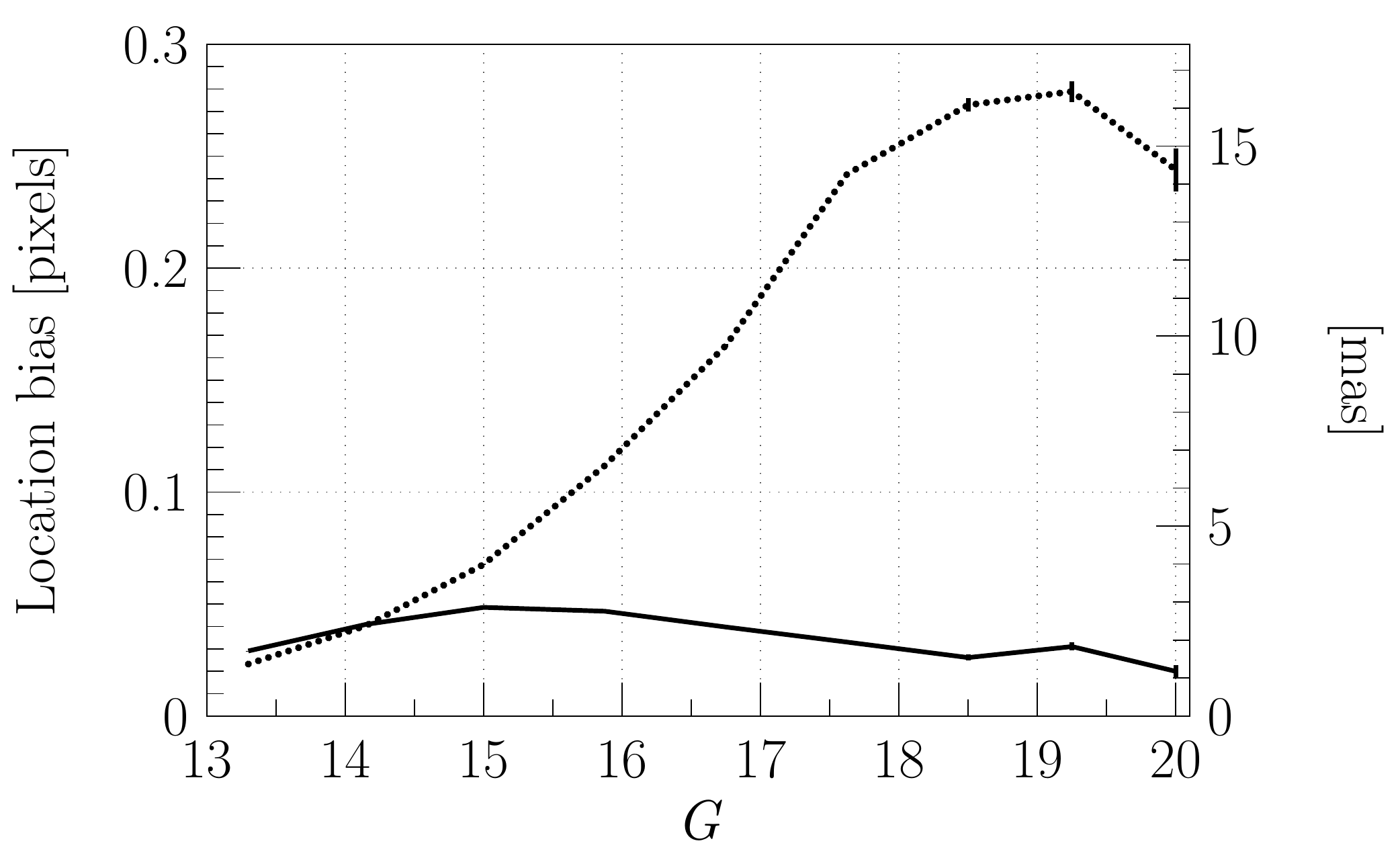}
\caption{Top: The relative intrinsic loss of accuracy induced by radiation damage as a function of $G$ and as computed in Section~\ref{sect:damagedCR}. The continuous line is the same as the black dashed line in Fig.~\ref{fig:crdamaged} and corresponds to the case where the SBC is functional. The dashed line is obtained using simulated transits for a CCD containing no SBC. Note the difference in ordinate scale range with Fig.~\ref{fig:crdamaged} (bottom).
\newline Bottom: Absolute image location bias as a function of $G$ and as computed in Section~\ref{sect:bias}. The continuous line is the same as the black continuous line in Fig.~\ref{fig:damagedBias} (typical image width, {\gaia} operating conditions, 1 trap pixel$^{-1}$) and corresponds to a functional SBC case. The dashed line was obtained for simulating star transits in the same conditions but with a CCD containing no SBC.}
\label{fig:noSBC}
\end{figure}

Throughout the paper we assumed a fully functional SBC.
Nevertheless the manufacturing of a SBC is a complex process and the CCDs of the Wield Field Camera channel of the Advanced Camera for Surveys on board {\hst} have been reported not to contain the SBC present in their design \citep{anderson2010}.
Fig.~\ref{fig:noSBC} shows the importance of the SBC CTI mitigation for achieving the {\gaia} requirements at low signal levels. If the SBC was not present (dashed line), the intrinsic loss of accuracy at low signal levels would reach 200\% instead of only 5\% in the presence of a SBC (continuous line and Fig.~\ref{fig:crdamaged}, bottom). The extra location bias induced by a missing SBC is also significant although there is no particular reason for which the presented CTI mitigation scheme would not be able to recover it.
In the case of {\gaia} the SBC has been demonstrated to be functional using experimental tests.
For instance, Fig.~5 in \cite{prodhomme2011} shows the effect of the SBC on the CTI-induced fractional charge loss as a function of signal level.
However a recent study by \cite{kohley2009} identified a non-functional SBC in the upper half of a {\gaia} CCD.
Based on more tested devices, Seabroke et al. (in prep.) will show that a significant number of the {\gaia} CCDs could be affected by this issue. Using the same methods as in this paper, they will evaluate the extra loss of accuracy and location bias induced by a non-functional SBC in the CCD upper half only; in contrast to
the no SBC case investigated here, Seabroke et al. will show that there is only at most 10\% extra loss of accuracy.

\section{Conclusions}\label{sect:conclusion}

We have presented a detailed characterization and evaluation of the impact of CCD radiation damage on the {\gaia} image location accuracy. The underlying principle of this study consisted of a systematic comparison between the computed theoretical limit to the image location accuracy, the Cram\'{e}r-Rao bound, and the actual performance of the {\gaia} image parameter estimation procedure under realistic {\gaia} operating conditions. The image location estimation bias and the associated standard errors were measured by applying the {\gaia} image parameter estimation procedure to a large set of synthetic data accounting for different stellar image widths, magnitudes, and background levels. We considered two different active trap densities; they are representative of two different levels of hardware mitigation in the presence of a radiation dose equivalent to the upper limit to the expected end-of-mission accumulated dose. The lowest active trap density constitutes the most realistic case because of a lower expected radiation dose and more importantly because of the planned injection of artificial charges in the CCD every second. In this context, a total of 41,472 synthetic two-dimensional {\gaia}-like observations have been generated using a detailed Monte-Carlo model of the CCD charge collection and transfer, and the radiation induced trapping. The dataset is readily available for the {\gaia} scientists to continue to test and further improve these critical steps in the {\gaia} data processing that are the image parameter estimation and the mitigation of the CTI effects. The main conclusions we can draw from this study are:  

\paragraph*{The {\gaia} image location estimation procedure is bias-free and efficient in the absence of radiation damage.} We showed that under realistic operating conditions, and from bright to faint magnitudes, the {\gaia} location estimator performs within the requirements at an accuracy close to the theoretical limit for CTI-free observations.

\paragraph*{The radiation damage effects induce an irreversible loss of accuracy that is independent of any image location estimator.}
It can only be avoided by the use of hardware CTI countermeasures that physically prevent the trapping. In the theoretical limit (i.e.\ perfect CTI calibration at the image processing level), the location accuracy loss is still acceptable when compared to the {\gaia} requirements: it can reach 6\% in the lowest trap density case and 24\% in the highest case. Due to the presence of a supplementary buried channel in each of the {\gaia} CCD pixels, the accuracy loss stops increasing for stars fainter than $G \sim 16$. 

\paragraph*{A CTI mitigation procedure is critical to achieve the {\gaia} requirements.}
We showed that if CTI is not taken into account in the image parameter estimation procedure, the resulting image location estimations are significantly biased. In the {\gaia} operating conditions, the most important bias is obtained for the widest type of stellar images at $G=15$: 0.05~pixels in the lowest trap density case and 0.2 pixels in the highest case. For comparison, at this magnitude the requirement on the image location accuracy for a G2V type star is 0.0045~pixels, at least an order of magnitude smaller than the measured bias.

\paragraph*{The CTI-induced image location bias varies significantly with the stellar image width and the background level.}
This is particularly relevant for experimental studies in which the image flux distribution and the background level cannot be absolutely known. At faint magnitudes small differences in the experimental setup can lead to significant differences in the measured CTI effects.

\paragraph*{In principle, a complete forward modelling approach to CTI mitigation allows for an accurate and bias-free estimation of the true image location from a damaged observation.}
We demonstrated that the forward modelling of a damaged observation using ideally calibrated models for both the PSF/LSF and the CTI effects provides a location estimate that on average never exceeds 0.003~pixels and does not deviate from zero within the error bars from bright to faint stars and for the two considered levels of radiation damage. The accuracy reached using this CTI mitigation scheme complies with the {\gaia} requirements. In the case of the lowest active trap density, this method even allows for the recovery of the theoretical limit to the image location accuracy in the presence of radiation damage.

\paragraph*{If calibrated well enough, the current best candidate for the charge distortion model (CDM) associated with the forward modelling approach allows significant image location and flux estimate bias recovery.}
In these favourable conditions (simple illumination history, 1~trap species, no serial CTI, well calibrated LSF model, and close to optimal CDM parameters) yet for a trap density level representative of the end-of-mission accumulated radiation dose, the {\gaia} image location accuracy is preserved. In the {\gaia} operating conditions and after CTI mitigation using the current best CDM at our disposal, the maximum measured location bias is $\sim$0.005~pixels for the lowest radiation level and $\sim$0.017~pixels for the highest.\newline

\section{Future work}\label{sect:futureWork}

Estimating the location of an image to milli-pixel accuracy is an extremely
challenging exercise, in which no detail must be neglected. This is especially
true in the presence of radiation damage as shown in this paper. The work
presented here is not the final word on the {\gaia} image parameter estimation
procedure. Indeed the estimation procedure must be tested and improved further
using synthetic and experimental data. In addition the elaboration and the
calibration of the charge distortion model is a key element in the success of
the presented CTI mitigation scheme. In this study, we have established
the level of agreement with the damaged observation that any CDM must achieve
to recover a bias-free image location estimation. We intend to test if the
current best CDM candidate is capable of reaching such agreement with
experimental data and improve it if necessary. Regarding the calibration of the
radiation damage parameters, the periodic charge injections will enable us to
monitor and characterize the radiation damage during the mission. In addition
the charge injections will act to reset the illumination history. We intend to
study what are the parameters that one can infer from the study of the CTI
effects on the charge injection signal, and how one can use these parameters to
initialize and calibrate the charge distortion model.

This study constitutes the first step in evaluating the impact of the CCD
radiation damage on the final astrometric accuracy of {\gaia}. It is indeed not
yet clear in detail how a biased and less precise estimation of the image
location, as induced by CTI, propagates into the astrometric parameters derived
by the {\gaia} astrometric global iterative solution (AGIS). In the follow-up
paper \citep{holl2011a} we investigate this particular question by using a small-scale
version of AGIS, AGISLab \citep{holl2009}, that will allow us to perform a careful error
propagation analysis for different cases (no CTI mitigation and optimal mitigation versus
CTI-free). The study presented here will be used to construct for each case a model that provides
location bias and uncertainty as function of magnitude, shape of the stellar
profile, illumination history (time since last charge injection), and mission
time (or trap density). These models will be used to disturb the image locations
(observation times) processed by AGIS. The study of the resulting astrometric
parameters will then allow us to characterize and evaluate the impact of CCD
radiation damage on {\gaia}'s astrometry. Also the effect of disturbing stars between a charge injection and a target
star will be assesed.

The future steps outlined are crucial ingredients in the successful radiation
damage mitigation strategy for {\gaia}, enabling the extraction of the best
scientific performance from this exciting and much anticipated mission.

\section*{Acknowledgments}
The work of TP and BH was supported by the European Marie-Curie research training network ELSA (MRTN-CT-2006-033481).  
LL acknowledges support by the Swedish National Space Board. AB acknowledges support by the Netherlands Research School for Astronomy (NOVA). The authors would like to kindly thank Carme Jordi for providing the {\gaia} photometric requirement values and Jos de Bruijne for the detailed feedback given on the conversion of the {\gaia} astrometric requirements. The authors would also like to thank the referee L.R.\ Bedin for his comments and suggestions that significantly improved the paper.

\bibliographystyle{mn2e}

\bibliography{references}

\begin{thebibliography}{}

\bibitem[\protect\citeauthoryear{{Anderson} \& {Bedin}}{{Anderson} \&
  {Bedin}}{2010}]{anderson2010}
{Anderson} J.,  {Bedin} L.~R.,  2010, PASP, 122, 1035

\bibitem[\protect\citeauthoryear{{Anderson} \& {King}}{{Anderson} \&
  {King}}{2000}]{anderson+king2000}
{Anderson} J.,  {King} I.~R.,  2000, PASP, 112, 1360

\bibitem[\protect\citeauthoryear{{Bastian} \& {Biermann}}{{Bastian} \&
  {Biermann}}{2005}]{bastian2005}
{Bastian} U.,  {Biermann} M.,  2005, Astronomy \& Astrophysics, 438, 745

\bibitem[\protect\citeauthoryear{{Bristow}}{{Bristow}}{2003}]{bristow2003}
{Bristow} P.,  2003, Technical report, {Application of Model Derived Charge
  Transfer Inefficiency Corrections to STIS Photometric CCD Data}

\bibitem[\protect\citeauthoryear{{Bristow}, {Kerber} \& {Rosa}}{{Bristow}
  et~al.}{2005}]{bristow2005}
{Bristow} P.,  {Kerber} F.,    {Rosa} M.~R.,  2005, in {A.~M. Koekemoer, P.
  Goudfrooij, \& L.~L. Dressel} ed., The 2005 HST Calibration Workshop {STIS
  Calibration Enhancement: Wavelength Calibration and CTI}

\bibitem[\protect\citeauthoryear{{Brown}}{{Brown}}{2009}]{swb2009}
{Brown} S.~W.,  2009, Technical report, Independent Analysis of Astriums
  Radiation Campaign 2.
DPAC

\bibitem[\protect\citeauthoryear{{de Bruijne}}{{de Bruijne}}{2009}]{JdB055}
{de Bruijne} J.,  2009, Technical report, Gaia astrometric performance: summer
  -2009 status.
ESTEC

\bibitem[\protect\citeauthoryear{{de Bruijne}}{{de Bruijne}}{2005}]{JdB2005Acc}
{de Bruijne} J.~H.~J.,  2005, in {C.~Turon, K.~S.~O'Flaherty, \&
  M.~A.~C.~Perryman} ed., The Three-Dimensional Universe with Gaia Vol.~576 of
  ESA Special Publication, {Accuracy Budget and Performances}.
pp 35--+

\bibitem[\protect\citeauthoryear{{Dolphin}}{{Dolphin}}{2009}]{dolphin2009}
{Dolphin} A.~E.,  2009, PASP, 121, 655

\bibitem[\protect\citeauthoryear{{Feynman}, {Spitale}, {Wang} \&
  {Gabriel}}{{Feynman} et~al.}{1993}]{feynman1993}
{Feynman} J.,  {Spitale} G.,  {Wang} J.,    {Gabriel} S.,  1993, JGR, 981,
  13281

\bibitem[\protect\citeauthoryear{{Georges}}{{Georges}}{2008}]{astrium2008}
{Georges} L.,  2008, Technical report, Tests Report of the Radiation Campaign
  2.
EADS Astrium

\bibitem[\protect\citeauthoryear{{Goudfrooij} \& {Kimble}}{{Goudfrooij} \&
  {Kimble}}{2002}]{goudfrooij2002}
{Goudfrooij} P.,  {Kimble} R.~A.,  2002, in {S.~Arribas, A.~Koekemoer, \&
  B.~Whitmore} ed., The 2002 HST Calibration Workshop : Hubble after the
  Installation of the ACS and the NICMOS Cooling System {Correcting STIS CCD
  Photometry for CTE Loss}.
pp 105--+

\bibitem[\protect\citeauthoryear{{Grant}, {Bautz}, {Kissel} \&
  {LaMarr}}{{Grant} et~al.}{2004}]{grant2004}
{Grant} C.~E.,  {Bautz} M.~W.,  {Kissel} S.~E.,    {LaMarr} B.,  2004, in
  {A.~D.~Holland} ed., Society of Photo-Optical Instrumentation Engineers
  (SPIE) Conference Series Vol.~5501 of Presented at the Society of
  Photo-Optical Instrumentation Engineers (SPIE) Conference, {A charge transfer
  inefficiency correction model for the Chandra advanced CCD imaging
  spectrometer}.
pp 177--188

\bibitem[\protect\citeauthoryear{{Hall}}{{Hall}}{1952}]{hall52}
{Hall} R.~N.,  1952, Physical Review, 87, 387

\bibitem[\protect\citeauthoryear{{Holl}, {Lindegren} \& {Hobbs}}{{Holl}
  et~al.}{2009}]{holl2009}
{Holl} B.,  {Lindegren} L.,    {Hobbs} D.,  2009, Technical report, AGISLab Ð A
  facility for experimental astrometric solutions.
DPAC

\bibitem[\protect\citeauthoryear{{Holl}, {Prod'homme}, {Lindegren} \&
  {Brown}}{{Holl} et~al.}{2011a}]{holl2011a}
{Holl} B.,  {Prod'homme} T.,  {Lindegren} L.,    {Brown} A.,  2011a, MNRAS

\bibitem[\protect\citeauthoryear{{Holl}, {Prod'homme}, {Lindegren} \&
  {Brown}}{{Holl} et~al.}{2011b}]{holl2011}
{Holl} B.,  {Prod'homme} T.,  {Lindegren} L.,    {Brown} A.~G.~A.,  2011b, in
  prep.

\bibitem[\protect\citeauthoryear{{Hopkinson}, {Short}, {Vetel}, {Zayer} \&
  {Holland}}{{Hopkinson} et~al.}{2005}]{hopkinson2005}
{Hopkinson} G.,  {Short} A.,  {Vetel} C.,  {Zayer} I.,    {Holland} A.,  2005,
  IEEE Transactions on Nuclear Science, 52, 2664

\bibitem[\protect\citeauthoryear{{Jordi}, {Gebran}, {Carrasco}, {de Bruijne},
  {Voss}, {Fabricius}, {Knude}, {Vallenari}, {Kohley} \& {Mora}}{{Jordi}
  et~al.}{2010}]{jordi2010}
{Jordi} C.,  {Gebran} M.,  {Carrasco} J.~M.,  {de Bruijne} J.,  {Voss} H.,
  {Fabricius} C.,  {Knude} J.,  {Vallenari} A.,  {Kohley} R.,    {Mora} A.,
  2010, Astronomy \& Astrophysics, 523, A48+

\bibitem[\protect\citeauthoryear{{Kohley}, {Raison} \&
  {Martin-Fleitas}}{{Kohley} et~al.}{2009}]{kohley2009}
{Kohley} R.,  {Raison} F.,    {Martin-Fleitas} J.~M.,  2009, in Society of
  Photo-Optical Instrumentation Engineers (SPIE) Conference Series Vol.~7439 of
  Presented at the Society of Photo-Optical Instrumentation Engineers (SPIE)
  Conference, {Gaia: operational aspects and tests of Gaia Flight Model CCDs}

\bibitem[\protect\citeauthoryear{{Lindegren}}{{Lindegren}}{1978}]{lindegren78}
{Lindegren} L.,  1978, in {F.~V.~Prochazka \& R.~H.~Tucker} ed., IAU Colloq.
  48: Modern Astrometry {Photoelectric astrometry - A comparison of methods for
  precise image location}.
pp 197--217

\bibitem[\protect\citeauthoryear{{Lindegren}}{{Lindegren}}{2003}]{lindegren200%
3LL046}
{Lindegren} L.,  2003, Technical report, Representation of LSF and PSF for
  GDAAS-2.
Lund Observatory

\bibitem[\protect\citeauthoryear{{Lindegren}}{{Lindegren}}{2006}]{lindegrenLL6%
8}
{Lindegren} L.,  2006, Technical report, Centroid definition for the Astro Line
  Spread Function.
Lund Observatory

\bibitem[\protect\citeauthoryear{{Lindegren}}{{Lindegren}}{2008}]{lindegren200%
8}
{Lindegren} L.,  2008, Technical report, A general Maximum-Likelihood algorithm
  for model fitting to a CCD sample data.
Lund Observatory

\bibitem[\protect\citeauthoryear{{Lindegren}}{{Lindegren}}{2009}]{lindegren200%
9LL084}
{Lindegren} L.,  2009, Technical report, Minimum-dimension LSF modelling.
Lund Observatory

\bibitem[\protect\citeauthoryear{{Lindegren}, {Babusiaux}, {Bailer-Jones},
  {Bastian}, {Brown}, {Cropper}, {H{\o}g}, {Jordi}, {Katz}, {van Leeuwen},
  {Luri}, {Mignard}, {de Bruijne} \& {Prusti}}{{Lindegren}
  et~al.}{2008}]{Gaia2008}
{Lindegren} L.,  {Babusiaux} C.,  {Bailer-Jones} C.,  {Bastian} U.,  {Brown}
  A.~G.~A.,  {Cropper} M.,  {H{\o}g} E.,  {Jordi} C.,  {Katz} D.,  {van
  Leeuwen} F.,  {Luri} X.,  {Mignard} F.,  {de Bruijne} J.~H.~J.,    {Prusti}
  T.,  2008, in {W.~J.~Jin, I.~Platais, \& M.~A.~C.~Perryman} ed., IAU
  Symposium Vol.~248 of IAU Symposium, {The Gaia mission: science, organization
  and present status}.
pp 217--223

\bibitem[\protect\citeauthoryear{{Lindegren} \& {Bastian}}{{Lindegren} \&
  {Bastian}}{2011}]{lindegren2011}
{Lindegren} L.,  {Bastian} U.,  2011, in {C. Turon, F. Meynadier and F. Arenou}
  ed., EAS Publications Series Vol.~45 of EAS Publications Series, {Basic
  principles of scanning space astrometry}.
pp 109--114

\bibitem[\protect\citeauthoryear{{Lindegren}, {Lammers} U.~{Hobbs}, {O'Mullane}
  \& {Bastian} U.~{Hernandez}}{{Lindegren} et~al.}{2011}]{lindegrenAgis2011}
{Lindegren} L.,  {Lammers} U.~{Hobbs} D.,  {O'Mullane} W.,    {Bastian}
  U.~{Hernandez} J.,  2011, A\& A, submitted

\bibitem[\protect\citeauthoryear{{Massey}, {Stoughton}, {Leauthaud}, {Rhodes},
  {Koekemoer}, {Ellis} \& {Shaghoulian}}{{Massey} et~al.}{2010}]{massey2010}
{Massey} R.,  {Stoughton} C.,  {Leauthaud} A.,  {Rhodes} J.,  {Koekemoer} A.,
  {Ellis} R.,    {Shaghoulian} E.,  2010, MNRAS, 401, 371

\bibitem[\protect\citeauthoryear{{Mignard}, {Bailer-Jones}, {Bastian},
  {Drimmel}, {Eyer}, {Katz}, {van Leeuwen}, {Luri}, {O'Mullane}, {Passot},
  {Pourbaix} \& {Prusti}}{{Mignard} et~al.}{2008}]{mignard2008}
{Mignard} F.,  {Bailer-Jones} C.,  {Bastian} U.,  {Drimmel} R.,  {Eyer} L.,
  {Katz} D.,  {van Leeuwen} F.,  {Luri} X.,  {O'Mullane} W.,  {Passot} X.,
  {Pourbaix} D.,    {Prusti} T.,  2008, in {W.~J.~Jin, I.~Platais, \&
  M.~A.~C.~Perryman} ed., IAU Symposium Vol.~248 of IAU Symposium, {Gaia:
  organisation and challenges for the data processing}.
pp 224--230

\bibitem[\protect\citeauthoryear{{Nelder} \& {Mead}}{{Nelder} \&
  {Mead}}{1965}]{nelder1965}
{Nelder} J.,  {Mead} R.,  1965, Computer Journal, 7, 308

\bibitem[\protect\citeauthoryear{{Paulet}}{{Paulet}}{2009}]{vpu}
{Paulet} P.,  2009, Technical report, Video processing unit (VPU)
  specification.
EADS Astrium

\bibitem[\protect\citeauthoryear{{Perryman}, {de Boer}, {Gilmore}, {H{\o}g},
  {Lattanzi}, {Lindegren}, {Luri}, {Mignard}, {Pace} \& {de Zeeuw}}{{Perryman}
  et~al.}{2001a}]{Gaia2001}
{Perryman} M.~A.~C.,  {de Boer} K.~S.,  {Gilmore} G.,  {H{\o}g} E.,  {Lattanzi}
  M.~G.,  {Lindegren} L.,  {Luri} X.,  {Mignard} F.,  {Pace} O.,    {de Zeeuw}
  P.~T.,  2001a, Astronomy \& Astrophysics, 369, 339

\bibitem[\protect\citeauthoryear{{Perryman}, {de Boer}, {Gilmore}, {H{\o}g},
  {Lattanzi}, {Lindegren}, {Luri}, {Mignard}, {Pace} \& {de Zeeuw}}{{Perryman}
  et~al.}{2001b}]{perryman2001}
{Perryman} M.~A.~C.,  {de Boer} K.~S.,  {Gilmore} G.,  {H{\o}g} E.,  {Lattanzi}
  M.~G.,  {Lindegren} L.,  {Luri} X.,  {Mignard} F.,  {Pace} O.,    {de Zeeuw}
  P.~T.,  2001b, A\& A, 369, 339

\bibitem[\protect\citeauthoryear{{Press}, B.P., {Teukolsky} \&
  {Vetterling}}{{Press} et~al.}{1992}]{press92}
{Press} W.,  B.P. F.,  {Teukolsky} S.,    {Vetterling} W.,  1992, Numerical
  recipes in Fortran.
Cambridge University Press

\bibitem[\protect\citeauthoryear{{Prod'homme}}{{Prod'homme}}{2011}]{prodhomme2%
010b}
{Prod'homme} T.,  2011, in {C. Turon, F. Meynadier and F. Arenou} ed., EAS
  Publications Series Vol.~45 of EAS Publications Series, {Radiation effects on
  Gaia CCDs, Modelling to mitigate the threat}.
pp 55--60

\bibitem[\protect\citeauthoryear{{Prod'homme}, {Brown}, {Lindegren} \&
  {Short}}{{Prod'homme} et~al.}{2011}]{prodhomme2011}
{Prod'homme} T.,  {Brown} A.,  {Lindegren} L.,    {Short} A.,  2011, MNRAS

\bibitem[\protect\citeauthoryear{{Prod'homme}, {Weiler}, {Brown}, {Short} \&
  {Brown}}{{Prod'homme} et~al.}{2010}]{prodhomme2010a}
{Prod'homme} T.,  {Weiler} M.,  {Brown} S.,  {Short} A.,    {Brown} A.,  2010,
  in Society of Photo-Optical Instrumentation Engineers (SPIE) Conference
  Series Vol.~7742 of Presented at the Society of Photo-Optical Instrumentation
  Engineers (SPIE) Conference, Comparison of a fast analytical model of
  radiation damage effects in ccds with experimental tests

\bibitem[\protect\citeauthoryear{{Rhodes} et~al.,}{{Rhodes}
  et~al.}{2007}]{rhodes2007}
{Rhodes} J.~D.,  et~al., 2007, ApJS, 172, 203

\bibitem[\protect\citeauthoryear{{Schrabback} et~al.,}{{Schrabback}
  et~al.}{2010}]{schrabback2010}
{Schrabback} T.,  et~al., 2010, A\&A, 516, A63+

\bibitem[\protect\citeauthoryear{{Seabroke}, {Holland} \& {Cropper}}{{Seabroke}
  et~al.}{2008}]{seabroke2008}
{Seabroke} G.,  {Holland} A.,    {Cropper} M.,  2008, in Society of
  Photo-Optical Instrumentation Engineers (SPIE) Conference Series Vol.~7021 of
  Presented at the Society of Photo-Optical Instrumentation Engineers (SPIE)
  Conference, {Modelling radiation damage to ESA's Gaia satellite CCDs}

\bibitem[\protect\citeauthoryear{{Shockley} \& {Read}}{{Shockley} \&
  {Read}}{1952}]{shockley52}
{Shockley} W.,  {Read} W.~T.,  1952, Physical Review, 87, 835

\bibitem[\protect\citeauthoryear{{Short}, {Prod'homme}, {Weiler}, {Brown} \&
  {Brown}}{{Short} et~al.}{2010}]{short2010}
{Short} A.,  {Prod'homme} T.,  {Weiler} M.,  {Brown} S.,    {Brown} A.,  2010,
  in Society of Photo-Optical Instrumentation Engineers (SPIE) Conference
  Series Vol.~7742 of Presented at the Society of Photo-Optical Instrumentation
  Engineers (SPIE) Conference, A fast model of radiation-induced electron
  trapping in ccds for implementation in the gaia data processing

\bibitem[\protect\citeauthoryear{{SIDC-team}}{{SIDC-team}}{2011}]{sidc}
{SIDC-team} 2011, Monthly Report on the International Sunspot Number, online
  catalogue

\bibitem[\protect\citeauthoryear{{Townsley}, {Broos}, {Garmire} \&
  {Nousek}}{{Townsley} et~al.}{2000}]{townsley2000}
{Townsley} L.~K.,  {Broos} P.~S.,  {Garmire} G.~P.,    {Nousek} J.~A.,  2000,
  ApJ, 534, L139

\bibitem[\protect\citeauthoryear{{Townsley}, {Broos}, {Nousek} \&
  {Garmire}}{{Townsley} et~al.}{2002}]{townsley2002}
{Townsley} L.~K.,  {Broos} P.~S.,  {Nousek} J.~A.,    {Garmire} G.~P.,  2002,
  Nuclear Instruments and Methods in Physics Research A, 486, 751

\bibitem[\protect\citeauthoryear{{van Leeuwen}}{{van
  Leeuwen}}{2007}]{vanLeeuwen2007b}
{van Leeuwen} F.,  2007, Technical report, Dealing with CCD radiation damage in
  the Gaia data processing.
DPAC

\bibitem[\protect\citeauthoryear{{van Leeuwen} \& {Lindegren}}{{van Leeuwen} \&
  {Lindegren}}{2007}]{vanLeeuwen2007a}
{van Leeuwen} F.,  {Lindegren} L.,  2007, Technical report, Outline of the
  Radiation Calibration Strategy for Astrometry, Photometry and Spectroscopy.
DPAC

\end{thebibliography}

Note that a large fraction of the {\gaia} internal technical reports we refer to in this paper are available at this http url: \url{http://www.rssd.esa.int/index.php?project=GAIA&page=Library_selected_papers}

\begin{table*}
  \caption{List of acronyms used in this paper.}
  \begin{tabular}{ll}
    \hline
    Acronym 	& Definition	\\
    \hline
    AC & ACross scan \\
    AF & Astrometric Field \\
    AGIS & Astrometric Global Iterative Solution \\
    AL & ALong scan \\
    CCD & Charge-Coupled Device \\
    CDM & Charge Distortion Model \\
    CEMGA & CTI Effects Models for GAia \\
    CI & Charge Injection \\
    CTI & Charge Transfer Inefficiency \\
    ELSA & European Leadership in Space Astrometry\\
    FWHM & Full Width Half Maximum\\
    {\hst} & Hubble Space Telescope\\
    L2 & Lagrangian Point 2 \\
    LSF & Line Spread Function \\
    mas & milli-arcsecond\\
    $\mu$as & micro-arcsecond\\
    ML & Maximum-Likelihood \\
    PSF & Point Spread Function \\
    RC & Radiation Campaign \\
    SBC & Supplementary Buried Channel \\
    TDI & Time-Delayed Integration \\
    \hline
  \end{tabular}
  \label{t:acronyms}
\end{table*}

\begin{table*}
\begin{center}
\resizebox{\textwidth}{!}{
\centering
 \begin{tabular}{|c|c|c|c|c|c|c|c|c|}
   \hline
   
\multicolumn{1}{|l|}{Reference image type} 	& \multicolumn{2}{|c|}{typical} 			& \multicolumn{2}{|c|}{typical} 	& \multicolumn{2}{|c|}{typical} 	& \multicolumn{2}{|c|}{typical} \\
   
\multicolumn{1}{|l|}{Background level (\electron pixel$^{-1}$ s$^{-1}$)} & \multicolumn{2}{|c|}{0} & \multicolumn{2}{|c|}{0} 		& \multicolumn{2}{|c|}{0.44698} 	& \multicolumn{2}{|c|}{0.44698} \\
   
\multicolumn{1}{|l|}{Readout noise (\electron)} & \multicolumn{2}{|c|}{0} 			& \multicolumn{2}{|c|}{4.35} 	& \multicolumn{2}{|c|}{4.35} 	& \multicolumn{2}{|c|}{4.35} \\
   
\multicolumn{1}{|l|}{Window size (pixels)} 	& \multicolumn{2}{|c|}{40} 				& \multicolumn{2}{|c|}{40} 		& \multicolumn{2}{|c|}{40} 		& \multicolumn{2}{|c|}{18} \\
   
\multicolumn{1}{|l|}{Trap density (traps pixel$^{-1}$)} 	& \multicolumn{2}{|c|}{0} 		& \multicolumn{2}{|c|}{0}		& \multicolumn{2}{|c|}{0} 		& \multicolumn{2}{|c|}{0} \\
   
   \hline
   
   Magnitude & Cram\'{e}r-Rao bound & $\sigma_{\kappa} \pm \upsilon_{\sigma_{\kappa}}$& \multicolumn{6}{|c|}{}\\
   
   (G-band)& ($10^{-3}$ pixel) & ($10^{-3}$ pixel)& \multicolumn{6}{|c|}{}\\  
    \hline
    13.3\phantom{0}\phantom{0}			& 1.67 & 1.668 $\pm$ 0.075 	& 1.67 & 1.672 $\pm$ 0.075 	& 1.68 & 1.655 $\pm$ 0.074 	& 1.67 & 1.662 $\pm$0.074 	\\
    14.15\phantom{0}		& 2.48 & 2.49 $\pm$ 0.11  	& 2.48 & 2.50 $\pm$ 0.11 	& 2.48 & 2.46 $\pm$ 0.11 	& 2.48 & 2.45 $\pm$0.11 	\\
    15.0\phantom{0}\phantom{0}		 	& 3.66 & 3.69 $\pm$ 0.17 	& 3.67 & 3.71 $\pm$ 0.17 	& 3.67 & 3.77 $\pm$ 0.17 	& 3.67 & 3.75 $\pm$0.17 	\\
    15.875 	& 5.48 & 5.95 $\pm$ 0.27 	& 5.50 & 6.00 $\pm$ 0.27 	& 5.50 & 5.45 $\pm$ 0.24 	& 5.51 & 5.63 $\pm$0.25 	\\
    16.75\phantom{0}	 	& 8.29 & 7.65 $\pm$ 0.34 	& 8.33 & 7.65 $\pm$ 0.34		& 8.34 & 8.46 $\pm$ 0.38 	& 8.30 & 8.34 $\pm$0.37		\\
    17.625 	& 12.40 & 12.65 $\pm$ 0.57 	& 12.54 & 12.73 $\pm$ 0.57 	& 12.57 & 12.63 $\pm$ 0.56 	& 12.61 & 12.70 $\pm$0.57	\\
    18.5\phantom{0}\phantom{0}		 	& 18.55 & 17.88 $\pm$ 0.80 	& 18.99 & 18.03 $\pm$ 0.81 	& 19.09 & 18.84 $\pm$ 0.84 	& 19.37 & 19.34 $\pm$0.86	\\
    19.25\phantom{0}	 	& 26.29 & 25.87 $\pm$ 1.16 	& 27.49 & 26.56 $\pm$ 1.19 	& 27.62 & 27.83 $\pm$ 1.24 	& 28.53 & 28.86 $\pm$1.29	\\
    20.0\phantom{0}\phantom{0}		 	& 37.06 & 37.12 $\pm$ 1.66 	& 40.26 & 39.99 $\pm$ 1.79	& 40.99 & 38.08 $\pm$ 1.70 	& 43.40 & 39.08 $\pm$1.75	\\
    \hline
    \hline

   \multicolumn{1}{|l|}{Reference image type} 	& \multicolumn{2}{|c|}{typical} 			& \multicolumn{2}{|c|}{narrow} 	& \multicolumn{2}{|c|}{typical} 	& \multicolumn{2}{|c|}{wide} \\
   
   \multicolumn{1}{|l|}{Background level (\electron pixel$^{-1}$ s$^{-1}$)} & \multicolumn{2}{|c|}{0} & \multicolumn{2}{|c|}{0.44698} 	& \multicolumn{2}{|c|}{0.44698} 	& \multicolumn{2}{|c|}{0.44698} \\
   
   \multicolumn{1}{|l|}{Readout noise (\electron)} & \multicolumn{2}{|c|}{0} 			& \multicolumn{2}{|c|}{4.35} 	& \multicolumn{2}{|c|}{4.35} 	& \multicolumn{2}{|c|}{4.35} \\
   
   \multicolumn{1}{|l|}{Window size (pixels)} 	& \multicolumn{2}{|c|}{telemetry$^\star$} 		& \multicolumn{2}{|c|}{telemetry}& \multicolumn{2}{|c|}{telemetry}& \multicolumn{2}{|c|}{telemetry} \\
   
   \multicolumn{1}{|l|}{Trap density (traps pixel$^{-1}$)} 	& \multicolumn{2}{|c|}{0} 		& \multicolumn{2}{|c|}{0}		& \multicolumn{2}{|c|}{0} 		& \multicolumn{2}{|c|}{0} \\
    
       \hline
   
   Magnitude & Cram\'{e}r-Rao bound & $\sigma_{\kappa} \pm \upsilon_{\sigma_{\kappa}}$& \multicolumn{6}{|c|}{}\\
   
   (G-band)& ($10^{-3}$ pixel) & ($10^{-3}$ pixel)& \multicolumn{6}{|c|}{}\\  
    \hline
13.3\phantom{0}\phantom{0}	 & 1.67 & 1.680 $\pm$ 0.075 	& 1.50 & 1.416  $\pm$  0.063 	& 1.67 & 1.658  $\pm$  0.074 & 1.95 & 1.920  $\pm$ 0.086\\
14.15\phantom{0} & 2.48 & 2.54 $\pm$ 0.11 		& 2.23 & 2.25  $\pm$  0.10 		& 2.48 & 2.46  $\pm$  0.11 & 2.88 & 2.83  $\pm$ 0.13\\
15.0\phantom{0}\phantom{0} & 3.66 & 3.69 $\pm$ 0.17 			& 3.30 & 3.52  $\pm$  0.16 		& 3.67 & 3.74  $\pm$  0.17 & 4.27 & 4.52  $\pm$ 0.20\\
15.875 & 5.48 & 5.98 $\pm$ 0.27 		& 4.95 & 5.47  $\pm$  0.24 		& 5.51 & 5.59  $\pm$  0.25 & 6.42 & 5.69 $\pm$ 0.25\\
16.75\phantom{0} & 8.29 & 7.69  $\pm$  0.34 	& 7.50 & 7.55  $\pm$  0.34 		& 8.38 & 8.45  $\pm$  0.38 & 9.85 & 9.80 $\pm$ 0.44\\
17.625 & 12.40 & 12.66  $\pm$  0.57 	& 11.39 & 11.13  $\pm$  0.50		& 12.70 & 12.94  $\pm$  0.58 & 14.94 & 15.55 $\pm$ 0.70\\
18.5\phantom{0}\phantom{0} & 18.55 & 17.99  $\pm$  0.80 	& 17.44 & 16.88  $\pm$  0.76 	& 19.49 & 19.39  $\pm$  0.87 & 22.99 & 25.09 $\pm$ 1.12\\
19.25\phantom{0} & 26.29 & 25.87  $\pm$  1.16 	& 25.48 & 26.33  $\pm$  1.18 	& 28.65 & 29.53  $\pm$  1.32 & 33.73 & 31.20 $\pm$ 1.40\\
20.0\phantom{0}\phantom{0} & 37.06 & 37.12  $\pm$  1.66 		& 38.43 & 39.73  $\pm$  1.78 	& 43.57 & 41.66  $\pm$  1.86 & 51.11 & 51.94 $\pm$ 2.32\\
    \hline
   \end{tabular}
   }
\end{center}
 \caption{Comparison between the theoretical and the actual limit to the image location accuracy in the absence of radiation damage as a function of $G$, sky background, readout noise, image width, and size of the telemetry windows in the along-scan direction.
The theoretical limit corresponds to the Cram\'{e}r-Rao bound, and the actual to the {\gaia} image location estimator standard errors $\sigma_{\kappa}$ with $\upsilon_{\sigma_{\kappa}}$ the statistical uncertainty (Eq.~\ref{eq:uncertStdev}).
  While the image width has a significant impact on those limits (e.g., for a 20\% increase in FWHM, one can note a $\sim$25\% decrease in accuracy for the faintest magnitude), the window size, readout noise, and background level only slightly affects the image location accuracy.
 \newline$^\star$ `telemetry size' refers to the size of the windows as they will be transmitted to the ground segment during the operational phase of {\gaia} \citep{vpu}: 12~pixels in the along-scan direction for $G<16$, and 6~pixels for $G>16$.}
 \label{tab:limits}
\end{table*}

\begin{table*}
\begin{center}
\resizebox{\textwidth}{!}{
\centering
 \begin{tabular}{|c|c|c|c|c|c|c|c|c|c|c|c|c|}
   \hline

   \multicolumn{1}{|l|}{Reference image type} 	& \multicolumn{3}{|c|}{typical} 			& \multicolumn{3}{|c|}{narrow} 	& \multicolumn{3}{|c|}{typical} 	& \multicolumn{3}{|c|}{wide} 	\\
   
   \multicolumn{1}{|l|}{Background level (\electron pixel$^{-1}$ s$^{-1}$)} & \multicolumn{3}{|c|}{0} & \multicolumn{3}{|c|}{0.44698}	& \multicolumn{3}{|c|}{0.44698}	& \multicolumn{3}{|c|}{0.44698}	\\
   
   \multicolumn{1}{|l|}{Readout noise (\electron)} & \multicolumn{3}{|c|}{4.35}			& \multicolumn{3}{|c|}{4.35} 	& \multicolumn{3}{|c|}{4.35}	 	& \multicolumn{3}{|c|}{4.35} 	\\
   
   \multicolumn{1}{|l|}{Window size (pixels)} 	& \multicolumn{3}{|c|}{telemetry}		& \multicolumn{3}{|c|}{telemetry}& \multicolumn{3}{|c|}{telemetry}& \multicolumn{3}{|c|}{telemetry}\\

   \multicolumn{1}{|l|}{Trap density (traps pixel$^{-1}$)} 	& 0 & 1 & 4 						& 0 & 1 & 4 						& 0 & 1 & 4  					& 0 & 1 & 4  		\\

   \hline
   
	Magnitude (G-band) & \multicolumn{12}{|c|}{Cram\'{e}r-Rao bound ($10^{-3}$ pixel)}\\
   
    \hline
13.3\phantom{0}\phantom{0} & 1.67 & 1.69 & 1.72 & 1.51 & 1.52 & 1.55 & 1.68 & 1.69 & 1.73 & 1.95 & 1.96 & 2.01\\
14.15\phantom{0} & 2.48 & 2.53 & 2.69 & 2.23 & 2.27 & 2.40 & 2.48 & 2.53 & 2.70 & 2.88 & 2.94 & 3.15\\
15.0\phantom{0}\phantom{0} & 3.67 & 3.81 & 4.28 & 3.30 & 3.43 & 3.87 & 3.67 & 3.82 & 4.31 & 4.27 & 4.45 & 5.02\\
15.875 & 5.50 & 5.79 & 6.61 & 4.95 & 5.25 & 6.10 & 5.51 & 5.82 & 6.67 & 6.42 & 6.73 & 7.64\\
16.75\phantom{0} & 8.33 & 8.72 & 9.97 & 7.53 & 7.98 & 9.23 & 8.41 & 8.85 & 10.06 & 9.97 & 10.43 & 11.68\\
17.625 & 12.54 & 13.08 & 14.89 & 11.44 & 12.01 & 13.69 & 12.74 & 13.35 & 15.01 & 15.11 & 15.70 & 17.46\\
18.5\phantom{0}\phantom{0} & 18.99 & 19.71 & 22.16 & 17.50 & 18.22 & 20.51 & 19.54 & 20.28 & 22.59 & 23.24 & 23.98 & 26.47\\
19.25\phantom{0} & 27.49 & 28.54 & 33.12 & 25.54 & 26.78 & 30.51 & 28.72 & 29.93 & 33.84 & 34.01 & 35.50 & 39.57\\
20.0\phantom{0}\phantom{0} & 40.26 & 43.27 & 53.28 & 38.51 & 40.27 & 45.96 & 43.64 & 45.43 & 51.19 & 51.43 & 53.54 & 60.32\\
    \hline
  
   \end{tabular}
   }
\end{center}
 \caption{Comparison between the Cram\'{e}r-Rao bounds computed for different image widths, background levels, and radiation damage levels. This comparison allows to characterize the intrinsic loss of precision and ultimately accuracy induced by radiation damage. This loss is relatively more important for the narrowest image, and increase with the trap density as expected.}
 \label{tab:cr}
\end{table*}

\begin{table*}
\begin{center}
\resizebox{\textwidth}{!}{
\centering
 \begin{tabular}{|c|c|c|c|c|c|c|}
   \hline

   	\multicolumn{1}{|l|}{Reference image type} 	 			& \multicolumn{2}{|c|}{narrow} 	& \multicolumn{2}{|c|}{typical} 	& \multicolumn{2}{|c|}{wide} 	\\
   
   	\multicolumn{1}{|l|}{Background level (\electron pixel$^{-1}$ s$^{-1}$)}	& \multicolumn{2}{|c|}{0.44698}	& \multicolumn{2}{|c|}{0.44698}	& \multicolumn{2}{|c|}{0.44698}	\\
   
   	\multicolumn{1}{|l|}{Readout noise (\electron)} 			& \multicolumn{2}{|c|}{4.35} 	& \multicolumn{2}{|c|}{4.35}	 	& \multicolumn{2}{|c|}{4.35} 	\\
   
   	\multicolumn{1}{|l|}{Window size (pixels)} 				& \multicolumn{2}{|c|}{telemetry}& \multicolumn{2}{|c|}{telemetry}& \multicolumn{2}{|c|}{telemetry}\\
	
	\multicolumn{1}{|l|}{Trap density (traps pixel$^{-1}$)} 			& 1 & 4 							& 1 & 4  						& 1 & 4  		\\
	
	   	\multicolumn{1}{|l|}{Mitigation} 							& \multicolumn{2}{|c|}{none}		& \multicolumn{2}{|c|}{none}		& \multicolumn{2}{|c|}{none}\\

   \hline
   
	Magnitude (G-band) & \multicolumn{6}{|c|}{Location bias $\langle\delta_{\kappa}\rangle_{G} \pm \upsilon_{\langle\delta_{\kappa}\rangle}$ ($10^{-3}$ pixel)}\\
    
    \hline
    13.3\phantom{0}\phantom{0} 	& 24.85 $\pm$ 0.11 & 93.48 $\pm$ 0.27 	& 29.06 $\pm$ 0.11 	& 110.84 $\pm$ 0.22	& 35.64 $\pm$ 0.13 & 137.18 $\pm$ 0.19\\
    
	14.15\phantom{0} 			& 35.79 $\pm$ 0.14 & 137.10 $\pm$ 0.22 	& 40.93 $\pm$ 0.17 	& 156.83 $\pm$ 0.29	& 49.68 $\pm$ 0.20 & 188.40 $\pm$ 0.28\\
	
	15.0\phantom{0}\phantom{0} 	& 43.54 $\pm$ 0.24 & 166.65 $\pm$ 0.66 	& 48.53 $\pm$ 0.23 	& 180.64 $\pm$ 0.47	& 56.48 $\pm$ 0.27 & 207.34 $\pm$ 0.54\\
	
	15.875 						& 44.06 $\pm$ 0.44 & 161.29 $\pm$ 1.11 	& 46.81 $\pm$ 0.43 	& 170.11 $\pm$ 0.82	& 51.19 $\pm$ 0.48 & 187.06 $\pm$ 0.74\\
	
	16.75\phantom{0} 			& 37.70 $\pm$ 0.56 & 139.54 $\pm$ 1.12 	& 39.75 $\pm$ 0.60 	& 143.66 $\pm$ 0.93	& 43.03 $\pm$ 0.64 & 157.81 $\pm$ 0.97\\
	
	17.625 						& 31.26 $\pm$ 0.81 & 120.41 $\pm$ 1.18 	& 33.02 $\pm$ 0.87 	& 126.87 $\pm$ 1.04	& 33.66 $\pm$ 0.99 & 139.18 $\pm$ 1.20\\
	
	18.5\phantom{0}\phantom{0} 	& 26.09 $\pm$ 1.07 & 108.10 $\pm$ 1.29 	& 26.11 $\pm$ 1.28 	& 114.60 $\pm$ 1.48	& 33.79 $\pm$ 1.58 & 130.81 $\pm$ 1.50\\
	
	19.25\phantom{0} 			& 28.67 $\pm$ 1.64 & 105.25 $\pm$ 1.92 	& 31.02 $\pm$ 1.81 	& 109.66 $\pm$ 2.03	& 25.92 $\pm$ 2.14 & 119.33 $\pm$ 2.20\\
	
	20.0\phantom{0}\phantom{0} 	& 24.92 $\pm$ 2.51 & 84.61 $\pm$ 2.73 	& 19.97 $\pm$ 2.86 	& 87.34 $\pm$ 3.15	& 24.16 $\pm$ 3.57 & 95.08 $\pm$ 3.71\\
    \hline
    \hline
     \multicolumn{1}{|l|}{Mitigation}  			& \multicolumn{2}{|c|}{ideal}		& \multicolumn{2}{|c|}{ideal}		& \multicolumn{2}{|c|}{ideal}\\
   \hline
   
	Magnitude (G-band) & \multicolumn{6}{|c|}{Location bias $\langle\delta_{\kappa}\rangle_{G} \pm \upsilon_{\langle\delta_{\kappa}\rangle}$ ($10^{-3}$ pixel)}\\
    
    \hline
13.3\phantom{0}\phantom{0} 	& 0.01  $\pm$ 0.10 	& 0.02  $\pm$ 0.12 & -0.05  $\pm$ 0.11 & 0.18  $\pm$ 0.12 & 0.09  $\pm$ 0.13 & -0.19  $\pm$ 0.13\\
14.15\phantom{0} 	& -0.16  $\pm$  0.14 & 0.05  $\pm$  0.16 & 0.15  $\pm$  0.16 & 0.11  $\pm$  0.20 & -0.17  $\pm$  0.20 & -0.15  $\pm$  0.20\\
15.0\phantom{0}\phantom{0} 		& 0.27  $\pm$  0.22 	& 0.55  $\pm$  0.36 & -0.32  $\pm$  0.25 & 0.29  $\pm$  0.33 & -0.25  $\pm$  0.28 & 0.11  $\pm$  0.32\\
15.875 	& -0.12  $\pm$  0.36 & -0.30  $\pm$  0.52 & 0.12  $\pm$  0.38 & 1.03  $\pm$  0.47 & -0.33  $\pm$  0.45 & 0.44  $\pm$  0.47\\
16.75\phantom{0} 	& 0.00  $\pm$  0.50 	& 0.51  $\pm$  0.63 & -0.14  $\pm$  0.58 & 0.40  $\pm$  0.62 & -0.07  $\pm$  0.65 & 0.81  $\pm$  0.70\\
17.625 	& -1.63  $\pm$  0.77 & 0.25  $\pm$  0.87 & 0.22  $\pm$  0.85 & 0.00  $\pm$  0.90 & -1.82  $\pm$  0.97 & 1.04  $\pm$  1.10\\
18.5\phantom{0}\phantom{0} 	& -0.54  $\pm$  1.07 & -1.94  $\pm$  1.24 & 0.90  $\pm$  1.26 & -1.52  $\pm$  1.42 & 2.04  $\pm$  1.58 & -0.38  $\pm$  1.51\\
19.25\phantom{0} 	& -0.28  $\pm$  1.66 & -1.98  $\pm$  1.91 & -2.09  $\pm$  1.86 & -0.67  $\pm$  1.95 & 1.30  $\pm$  2.15 & -0.75  $\pm$  2.18\\
20.0\phantom{0}\phantom{0} 		& 4.84 $\pm$ 2.44  &  0.30  $\pm$  2.91 & -3.08  $\pm$  2.96 & -2.06  $\pm$  3.18 & -2.63 $\pm$ 3.53  &  6.64  $\pm$  3.58\\
    
    \hline
    \hline
     \multicolumn{1}{|l|}{Mitigation}  							& \multicolumn{2}{|c|}{CDM}		& \multicolumn{2}{|c|}{CDM}		& \multicolumn{2}{|c|}{CDM}\\
	
   \hline
   
	Magnitude (G-band) & \multicolumn{6}{|c|}{Location bias $\langle\delta_{\kappa}\rangle_{G} \pm \upsilon_{\langle\delta_{\kappa}\rangle}$ ($10^{-3}$ pixel)}\\
       
    \hline
    13.3\phantom{0}\phantom{0} 	& -1.64  $\pm$  0.11	& -6.55  $\pm$  0.17 & -0.79  $\pm$  0.11 & -2.53  $\pm$  0.18 & 0.89  $\pm$  0.13 & 5.41  $\pm$  0.17\\
	14.15\phantom{0} 	& 0.13  $\pm$  0.16 	& 3.67  $\pm$  0.33 & 2.07  $\pm$  0.17 & 7.06  $\pm$  0.33 & 3.04  $\pm$  0.20 & 12.49  $\pm$  0.34\\
	15.0\phantom{0}\phantom{0} 		& -0.71  $\pm$  0.27	& -0.69  $\pm$  0.48 & 0.35  $\pm$  0.25 & -1.16  $\pm$  0.36 & 3.42 $\pm$  0.28 & 3.94  $\pm$  0.36\\
	15.875 	& -1.18  $\pm$  0.32	& -7.15  $\pm$  0.49 & -3.61  $\pm$  0.37 & -13.56  $\pm$  0.45 & -3.60  $\pm$  0.43 & -18.86  $\pm$  0.48\\
	16.75\phantom{0} 	& 1.21  $\pm$  0.50	& 22.88  $\pm$  1.14 & -1.04  $\pm$  0.57 & 11.65  $\pm$  1.02 & -2.98  $\pm$  0.63 & 7.38  $\pm$  1.04\\
	17.625 	& -3.11  $\pm$  0.76	& 5.06  $\pm$  0.96 & -1.23  $\pm$  0.85 & -1.95  $\pm$  0.99 & -9.33  $\pm$  0.96 & -3.82  $\pm$  1.19\\
	18.5\phantom{0}\phantom{0} 	& 8.54  $\pm$  1.09	& -4.25  $\pm$  1.24 & 2.16  $\pm$  1.25 & 4.73  $\pm$  1.50 & 9.13  $\pm$  1.54 & -10.50  $\pm$  1.60\\
	19.25\phantom{0} 	& 2.12  $\pm$  1.62	& 35.19  $\pm$  1.95 & 1.65  $\pm$  1.82 & -3.24  $\pm$  2.08 & -9.38  $\pm$  2.03 & 23.92  $\pm$  2.20\\
	20.0\phantom{0}\phantom{0} 		& 23.62  $\pm$  2.39	& 25.12  $\pm$  2.89 & 5.06  $\pm$  2.80 & 17.92  $\pm$  3.27 & 20.99  $\pm$  3.50 & 11.74  $\pm$  3.84\\
    \hline

   \end{tabular}
   }
\end{center}
 \caption{Summary of the measured image location biases in the {\gaia} operating conditions for different stellar image widths, radiation levels, and different levels of mitigation: (i) `none' corresponds to no mitigation (Section \ref{sect:bias}), (ii) `ideal' to the presented forward modelling approach associated to an ideal CDM and LSF model and calibration (Section \ref{sect:idealFwdTest}), and (iii) `CDM' to the presented forward modelling approach including the current implementation of CDM (Section \ref{sect:locBiasRecovery}). Note that in the latter case, the optimization of the CDM corresponds to the fully optimized case as described in Sections \ref{sect:initialization} and \ref{sect:locBiasRecovery}. This optimization was performed for the typical reference image only, the same CDM parameters were used for the two other reference images.}
 \label{tab:bias}
\end{table*}

\begin{table*}
\begin{center}
\resizebox{\textwidth}{!}{
\centering
 \begin{tabular}{|c|c|c|c|c|c|c|}
   \hline

   	\multicolumn{1}{|l|}{Reference image type} 	 			& \multicolumn{2}{|c|}{narrow} 	& \multicolumn{2}{|c|}{typical} 	& \multicolumn{2}{|c|}{wide} 	\\
   
   	\multicolumn{1}{|l|}{Background level (\electron pixel$^{-1}$ s$^{-1}$)}	& \multicolumn{2}{|c|}{0.44698}	& \multicolumn{2}{|c|}{0.44698}	& \multicolumn{2}{|c|}{0.44698}	\\
   
   	\multicolumn{1}{|l|}{Readout noise (\electron)} 			& \multicolumn{2}{|c|}{4.35} 	& \multicolumn{2}{|c|}{4.35}	 	& \multicolumn{2}{|c|}{4.35} 	\\
   
   	\multicolumn{1}{|l|}{Window size (pixels)} 				& \multicolumn{2}{|c|}{telemetry}& \multicolumn{2}{|c|}{telemetry}& \multicolumn{2}{|c|}{telemetry}\\
	
	\multicolumn{1}{|l|}{Trap density (traps pixel$^{-1}$)} 			& 1 & 4 							& 1 & 4  						& 1 & 4  		\\
	
   	\multicolumn{1}{|l|}{Mitigation} 							& \multicolumn{2}{|c|}{none}		& \multicolumn{2}{|c|}{none}		& \multicolumn{2}{|c|}{none}\\

   \hline
   
	Magnitude (G-band) & \multicolumn{6}{|c|}{Standard errors $\sigma_{\kappa} \pm \upsilon_{\sigma_{\kappa}}$ ($10^{-3}$ pixel)}\\
    
    \hline
13.3\phantom{0}\phantom{0} & 1.80  $\pm$  0.08 	& 4.26  $\pm$  0.19 	& 2.00  $\pm$  0.09 & 3.50  $\pm$ 0.16 & 2.08  $\pm$ 0.09 & 3.08  $\pm$ 0.14\\
14.15\phantom{0} & 2.31  $\pm$  0.10 	& 3.55  $\pm$  0.16 	& 2.67  $\pm$  0.12 & 4.56  $\pm$ 0.20 & 3.16  $\pm$ 0.14 & 4.43 $\pm$ 0.20\\
15.0\phantom{0}\phantom{0} & 3.75  $\pm$  0.17 		& 10.50 $\pm$  0.47 	& 4.29  $\pm$  0.19 & 7.43  $\pm$ 0.33 & 4.40  $\pm$ 0.20 & 8.59  $\pm$ 0.38\\
15.875 & 7.10  $\pm$  0.32 	& 17.58  $\pm$  0.79 & 6.59  $\pm$  0.29 & 12.91  $\pm$ 0.58 & 7.62  $\pm$ 0.34 & 11.69  $\pm$ 0.52\\
16.75\phantom{0} & 9.02  $\pm$  0.40 	& 17.78  $\pm$  0.80 & 9.48  $\pm$  0.42 & 14.71  $\pm$ 0.66 & 10.05  $\pm$ 0.45 & 15.36  $\pm$ 0.69\\
17.625 & 13.04  $\pm$  0.58 	& 18.63  $\pm$  0.83 & 13.60  $\pm$  0.61 & 16.47  $\pm$ 0.74 & 15.79  $\pm$ 0.71 & 18.95  $\pm$ 0.85\\
18.5\phantom{0}\phantom{0} & 16.96  $\pm$  0.76 	& 20.39  $\pm$  0.91 & 20.49  $\pm$  0.92 & 23.48  $\pm$ 1.05 & 24.99  $\pm$ 1.12 & 23.72  $\pm$ 1.06\\
19.25\phantom{0} & 26.44  $\pm$  1.18 	& 30.43  $\pm$  1.36 & 29.02  $\pm$  1.30 & 32.12  $\pm$ 1.44 & 33.86  $\pm$ 1.51 & 34.77  $\pm$ 1.56\\
20.0\phantom{0}\phantom{0} & 37.29  $\pm$  1.67 		& 43.24  $\pm$  1.93 & 45.83  $\pm$  2.05 & 49.85  $\pm$ 2.23 & 56.45  $\pm$ 2.52 & 58.65  $\pm$ 2.62\\
    \hline
    \hline
     \multicolumn{1}{|l|}{Mitigation}  			& \multicolumn{2}{|c|}{ideal}		& \multicolumn{2}{|c|}{ideal}		& \multicolumn{2}{|c|}{ideal}\\
	
   \hline
   
	Magnitude (G-band) & \multicolumn{6}{|c|}{Standard errors $\sigma_{\kappa} \pm \upsilon_{\sigma_{\kappa}}$ ($10^{-3}$ pixel)}\\    
    
    \hline
13.3\phantom{0}\phantom{0} 	& 1.57   $\pm$  0.07 & 1.91   $\pm$  0.09 & 1.77   $\pm$  0.08 & 2.01   $\pm$  0.09 & 1.87   $\pm$  0.08 & 2.05   $\pm$  0.09\\
14.15\phantom{0} 	& 2.24   $\pm$  0.10 & 2.55   $\pm$  0.11 & 2.49   $\pm$  0.11 & 3.09   $\pm$  0.14 & 3.13   $\pm$  0.14 & 3.12   $\pm$  0.14\\
15.0\phantom{0}\phantom{0} 		& 3.45   $\pm$  0.15 & 5.90   $\pm$  0.26 & 3.90   $\pm$  0.17 & 5.01   $\pm$  0.22 & 5.18   $\pm$  0.23 & 4.37   $\pm$  0.20\\
15.875 	& 5.70   $\pm$  0.25 & 8.20   $\pm$  0.37 & 5.98   $\pm$  0.27 & 7.49   $\pm$  0.34 & 7.49   $\pm$  0.34 & 7.10   $\pm$  0.32\\
16.75\phantom{0} 	& 7.92   $\pm$  0.35 & 9.72   $\pm$  0.43 & 9.15   $\pm$  0.41 & 11.04   $\pm$  0.49 & 9.76   $\pm$  0.44 & 10.28   $\pm$  0.46\\
17.625 	& 12.19   $\pm$  0.55 & 13.36   $\pm$  0.60 & 13.51   $\pm$  0.60 & 17.43   $\pm$  0.78 & 14.26   $\pm$  0.64 & 15.39   $\pm$  0.69\\
18.5\phantom{0}\phantom{0} 	& 16.89   $\pm$  0.76 & 18.50   $\pm$  0.83 & 19.86   $\pm$  0.89 & 23.80   $\pm$  1.06 & 22.42   $\pm$  1.00 & 24.94   $\pm$  1.12\\
19.25\phantom{0} 	& 26.21   $\pm$  1.17 & 30.35   $\pm$  1.36 & 29.43   $\pm$  1.32 & 34.44   $\pm$  1.54 & 30.76   $\pm$  1.38 & 34.00   $\pm$  1.52\\
20.0\phantom{0}\phantom{0} 		& 38.64   $\pm$  1.73 & 45.00   $\pm$  2.01 & 46.82   $\pm$  2.09 & 56.58   $\pm$  2.53 & 50.36   $\pm$  2.25 & 55.76   $\pm$  2.49\\
    \hline
    \hline
     \multicolumn{1}{|l|}{Mitigation}  							& \multicolumn{2}{|c|}{CDM}		& \multicolumn{2}{|c|}{CDM}		& \multicolumn{2}{|c|}{CDM}\\
	
   \hline
   
	Magnitude (G-band) & \multicolumn{6}{|c|}{Standard errors $\sigma_{\kappa} \pm \upsilon_{\sigma_{\kappa}}$ ($10^{-3}$ pixel)}\\    
    
    \hline
    13.3\phantom{0}\phantom{0} 	& 1.67  $\pm$  0.07 & 2.72  $\pm$  0.12 & 1.77  $\pm$  0.08 & 2.79  $\pm$  0.12 & 2.03  $\pm$  0.09 & 2.65  $\pm$  0.12\\
    14.15\phantom{0} 	& 2.47  $\pm$  0.11 & 5.21  $\pm$  0.23 & 2.72  $\pm$  0.12 & 5.15  $\pm$  0.23 & 3.18  $\pm$  0.14 & 5.38  $\pm$  0.24\\
	15.0\phantom{0}\phantom{0} 		& 4.33  $\pm$  0.19 & 7.54  $\pm$  0.34 & 3.92  $\pm$  0.18 & 5.69  $\pm$  0.25 & 4.42  $\pm$  0.20 & 5.71  $\pm$  0.26\\
	15.875 	& 5.02  $\pm$  0.22 & 7.74  $\pm$  0.35 & 5.86  $\pm$  0.26 & 7.10  $\pm$  0.32 & 6.77  $\pm$  0.30 & 7.55  $\pm$  0.34\\
	16.75\phantom{0} 	& 7.91  $\pm$  0.35 & 17.99  $\pm$  0.80 & 8.95  $\pm$  0.40 & 16.05  $\pm$  0.72 & 9.96  $\pm$  0.45 & 16.51  $\pm$  0.74\\
	17.625 	& 12.02  $\pm$  0.54 & 15.12  $\pm$  0.68 & 13.30  $\pm$  0.59 & 15.69  $\pm$  0.70 & 15.11  $\pm$  0.68 & 18.74  $\pm$  0.84\\
	18.5\phantom{0}\phantom{0} 	& 17.23  $\pm$  0.77 & 19.66  $\pm$  0.88 & 20.12  $\pm$  0.90 & 23.65  $\pm$  1.06 & 24.31  $\pm$  1.09 & 25.32  $\pm$  1.13\\
	19.25\phantom{0} 	& 25.55  $\pm$  1.14 & 30.77  $\pm$  1.38 & 28.33  $\pm$  1.27 & 32.95  $\pm$  1.47 & 32.16  $\pm$  1.44 & 34.73  $\pm$  1.55\\
	20.0\phantom{0}\phantom{0} 		& 37.83  $\pm$  1.69 & 45.73  $\pm$  2.05 & 45.74  $\pm$  2.05 & 51.67  $\pm$  2.31 & 55.35  $\pm$  2.48 & 60.64  $\pm$  2.71\\
    \hline
   \end{tabular}
   }
\end{center}

 \caption{Summary of the measured image location estimator standard errors in the {\gaia} operating conditions for different stellar image widths, radiation levels, and different levels of mitigation: (i) `none', (ii) `ideal', and (iii) `CDM' (see Table~\ref{tab:bias}).}
 \label{tab:stdErr}
\end{table*}

\label{lastpage}

\end{document}